\documentclass{aa}

\usepackage{natbib}
\usepackage{hyperref}

\usepackage{graphicx}
\usepackage{txfonts}
\usepackage{color}
\usepackage[none]{hyphenat}

\def\galex{{\it GALEX\/}} 
\def\spitzer{{\it Spitzer\/}} 
\def\herschel{{\it Herschel\/}}

\def\otelodeep{{OTELO-Deep}}
\def\otelocustom{{OTELO-custom}}
\def\oteloint{{OTELO$_{\rm Int}$}}

\def\otelotwo{{OTELO-II}}

\def\acs606{\textit{F606W}}
\def\acs814{\textit{F814W}}
\def\bandu{\textit{u}}
\def\bandg{\textit{g}}
\def\bandr{\textit{r}}
\def\bandi{\textit{i}}
\def\bandz{\textit{z}}
\def\ztrue{$z_{\rm\, true}$}
\def\bandj{\textit{J}}
\def\bandh{\textit{H}}
\def\bandk{\textit{K$_{\rm s}$}}
\def\irac36{${\rm 3.6\mu m}$}
\def\irac45{${\rm 4.5\mu m}$}
\def\irac58{${\rm 5.8\mu m}$}
\def\irac80{${\rm 8.0\mu m}$}
\def\mips24{${\rm 24\mu m}$}
\def\pep100{${\rm 100\mu m}$}
\def\pep160{${\rm 160\mu m}$}
\def\her250{${\rm 250\mu m}$}
\def\her350{${\rm 350\mu m}$}
\def\her500{${\rm 500\mu m}$}
\def\irac{{\rm IRAC\/}}
\def\mips{{\rm MIPS\/}}
\def\pacs{{\rm PACS\/}}
\def\spire{{\rm SPIRE\/}}
\def\deep{\hbox{DEEP2}}
\def\otelo{\hbox{OTELO}}
\def\chandra{\hbox{Chandra}}
\def\acs{\hbox{HST-ACS}}
\def\cfhtls{\hbox{CFHTLS}}
\def\wirds{\hbox{WIRDS}}
\def\pep{\hbox{PEP}}

\def\lephare{\textit{LePhare}}
\def\zred{{z}}
\def\zs{z$_{\rm\,DEEP2}$}
\def\zp{z$_{\rm\,phot}$}
\def\zreal{z$_{\rm\,real}$}
\def\zrel{z$_{\rm\,reliable}$}

\def\zpY{{\tt z\_BEST\_deepY}}
\def\zpN{{\tt z\_BEST\_deepN}}
\def\disp{|$\Delta$\zred|/(1+\zs)}
\def\dispnobar{$\Delta$\zred/(1+\zs)}
\def\aa{${\rm \AA}$}
\def\lya{Ly$\alpha$}
\def\ha{H$\alpha$}
\def\hb{H$\beta$}
\def\mgii{MgII}
\def\nii{[NII]$\lambda$6548,6583}
\def\ciii{CIII]$\lambda$1908}
\def\oii{[OII]$\lambda$3726,3729}
\def\oiii{[OIII]$\lambda$4959,5007}
\def\nev{[NeV]$\lambda$3426}
\def\sii{[SII]$\lambda$6717,6731}
\def\niii{NIII]$\lambda$1749,1752}
\def\oiiip{OIII]$\lambda$1665}
\def\heii{HeII$\lambda$1640}
\def\civ{CIV$\lambda$1548,1551}
\def\nv{NV$\lambda$1238,1242}

\begin{document}

\title{The OTELO survey}
\subtitle{I. Description, data reduction, and multi-wavelength catalogue}
\author{\'Angel Bongiovanni\inst{1,2,3}
\and Marina Ram\'on-P\'erez\inst{1,2}
\and Ana Mar\'ia P\'erez Garc\'ia\inst{4,5,3}
\and Jordi Cepa\inst{1,2,3}
\and Miguel Cervi\~no\inst{1,6,4}
\and Jakub Nadolny\inst{1,2}
\and Ricardo P\'erez Mart\'inez\inst{5,3}
\and Emilio Alfaro\inst{6}
\and H\'ector O. Casta\~neda\inst{12}
\and Jos\'e A. de Diego\inst{11}
\and Alessandro Ederoclite\inst{10,3}
\and Mirian Fern\'andez-Lorenzo\inst{6}
\and Jes\'us Gallego\inst{8}
\and J. Jes\'us Gonz\'alez\inst{11}
\and J. Ignacio Gonz\'alez-Serrano\inst{7,3}
\and Maritza A. Lara-L\'opez\inst{19}
\and Iv\'an Oteo G\'omez\inst{15,16}
\and Carmen P. Padilla Torres\inst{22}
\and Irene Pintos-Castro\inst{14}
\and Mirjana Povi\'c\,\inst{6,13}
\and Miguel S\'anchez-Portal\inst{17,18,3}
\and D. Heath Jones\inst{20}
\and Joss Bland-Hawthorn\inst{21}
\and Antonio Cabrera-Lavers\inst{9}
}

\institute{Instituto de Astrof\'isica de Canarias (IAC), E-38200 La Laguna, Tenerife, Spain 
\and Departamento de Astrof\'isica, Universidad de La Laguna (ULL), E-38205 La Laguna, Tenerife, Spain 
\and Asociaci\'on Astrof\'isica para la Promoci\'on de la Investigaci\'on, Instrumentaci\'on y su Desarrollo, ASPID, E-38205 La Laguna, Tenerife, Spain 
\and Centro de Astrobiolog\'ia (CSIC/INTA), 28850 Torrej\'on de Ardoz, Madrid, Spain 
\and ISDEFE for European Space Astronomy Centre (ESAC)/ESA, P.O. Box 78, E-28690, Villanueva de la Ca\~nada, Madrid, Spain 
\and Instituto de Astrof\'isica de Andaluc\'ia, CSIC, E-18080, Granada, Spain  
\and Instituto de F\'isica de Cantabria (CSIC-Universidad de Cantabria), E-39005 Santander, Spain 
\and Departamento de F\'isica de la Tierra y Astrof\'isica \& Instituto de F\'isica de Part\'iculas y del Cosmos (IPARCOS), Facultad de CC. F\'isicas, Universidad Complutense de Madrid, E-28040, Madrid, Spain 
\and Grantecan S.~A., Centro de Astrof\'isica de La Palma, Cuesta de San Jos\'e, E-38712 Bre\~na Baja, La Palma, Spain 
\and Centro de Estudios de F\'isica del Cosmos de Arag\'on, Plaza San Juan 1, Planta 2, E-44001 Teruel, Spain 
\and Instituto de Astronom\'ia, Universidad Nacional Aut\'onoma de M\'exico, 04510 Ciudad de M\'exico, Mexico 
\and Departamento de F\'isica, Escuela Superior de F\'isica y Matem\'aticas, Instituto Polit\'ecnico Nacional, 07738 Ciudad de M\'exico, Mexico 
\and Ethiopian Space Science and Technology Institute (ESSTI), Entoto Observatory and Research Center (EORC), Astronomy and Astrophysics Research Division, PO Box 33679, Addis Abbaba, Ethiopia 
\and Department of Astronomy \& Astrophysics, University of Toronto, Canada 
\and Institute for Astronomy, University of Edinburgh, Royal Observatory, Blackford Hill, Edinburgh, EH9  3HJ, UK 
\and European Southern Observatory, Karl-Schwarzschild-Str. 2, 85748, Garching, Germany  
\and European Southern Observatory, Alonso de C\'ordova 3107, Vitacura Casilla 19001, Santiago, Chile  
\and Joint ALMA Observatory, Alonso de C\'ordova 3107, Vitacura Casilla 763 0355, Santiago, Chile 
\and Dark Cosmology Centre, Niels Bohr Institute, University of Copenhagen, Juliane Maries Vej 30, 2100 Copenhagen {\O}, Denmark 
\and English Language and Foundation Studies Centre, University of Newcastle, Callaghan NSW 2308, Australia 
\and Sydney Institute of Astronomy, School of Physics, University of Sydney, NSW 2006, Australia 
\and INAF, Telescopio Nazionale Galileo, Apartado de Correos 565, E-38700 Santa Cruz de la Palma, Spain 
}

\date{Received 25 April 2018 / Accepted 19 September 2018}

  \abstract
   {The evolution of galaxies through cosmic time is  studied observationally by means of extragalactic surveys. 
   The usefulness of these surveys is greatly improved by increasing the cosmological volume, in either depth or area, 
   and by observing the same targets in different wavelength ranges. A multi-wavelength approach using different
   observational techniques can compensate for observational biases.}
   {The OTELO survey aims to provide the deepest narrow-band survey to date in terms of minimum detectable flux
   and emission line equivalent width in order to detect the faintest extragalactic emission line systems. In this way, 
   OTELO data will complements other broad-band, narrow-band, and spectroscopic surveys.}
   {The red tunable filter of the OSIRIS instrument on the 10.4 m Gran Telescopio Canarias (GTC) is used to scan 
   a spectral window centred at 9175 \AA , which is free from strong sky emission lines, with a sampling interval 
   of 6 \AA\ and a bandwidth of 12 \AA\ in the most deeply explored EGS region. Careful data reduction using 
   improved techniques for sky ring subtraction, accurate astrometry, photometric calibration, and source extraction 
   enables us to compile the OTELO catalogue. This catalogue 
   is complemented with ancillary data ranging from deep X-ray to far-infrared, including high resolution HST images, 
   which allow us to segregate the different types of targets, derive precise photometric redshifts, and obtain the
   morphological classification of the extragalactic objects detected.}
   {The OTELO multi-wavelength catalogue contains 11\,237 entries and is 50\% complete at AB magnitude 26.38. Of these 
   sources, 6600 have photometric redshifts with an uncertainty $\delta\,$\zp\ better than 0.2\,(1+\zp). A total 
   of 4336 of these sources correspond to preliminary emission line candidates, which are complemented by 81 candidate 
   stars and 483 sources that qualify as absorption line systems. The OTELO survey data products were released to the 
   public on 2019.}   
   {}

   \titlerunning{The OTELO survey}
   \authorrunning{Bongiovanni et al.}
   \maketitle

\section{Introduction}
\label{sec:intro}

Extragalactic surveys are an essential tool for studying galaxy evolution. Considerable amounts of observing time have been 
invested, mainly in the last few decades, in gathering deeper and larger datasets, enriched with observations covering wide
wavelength ranges, through the use of different instruments covering the same areas of sky.

Imaging surveys using broad-band filters, with passbands of the order of 100 nm or more, detect the faintest objects, per 
unit observing time and telescope aperture, at the price of losing spectral resolution. However, the development of 
such photometric redshift tools as \lephare\ \citep{arnouts99,ilbert06}, Zebra \citep{feldmann06}, BPZ \citep{benitez04}, 
and other SED-fitting facilities has somewhat alleviated this limitation. Moreover, high spatial resolution broad-
band surveys allow the determination of galaxy morphologies, an essential parameter for studying galaxy evolution. The large number 
of existing or planned surveys of this kind makes it difficult to provide a reasonably complete compilation, but the Hubble Deep 
Field \citep{williams96}, including its southern, flanking, and deep extensions, SDSS \citep{york00}, 
COSMOS \citep{scoville07}, VVDS \citep{lefevre04a}, and CANDELS \citep{grogin11}  give a hint of the importance of broad-band surveys.

The first attempts at obtaining better spectral resolutions in extragalactic surveys were based on slitless blind 
spectroscopy. KISS \citep{wegener03}, UCM \citep{gallego93}, CUYS \citep{bongio05}, and more recently PEARS \citep{straughn09} 
are  representative examples. They are hampered by spectra overlapping in crowded fields, especially in the 
case of extended sources. These limitations has been overcome with the advent of multi-plexing spectroscopic techniques 
(via multiple slits, integral field units, and image slicers). In this case, broad-band surveys provide the slit or fibre positions 
required for spectroscopic surveys, either blind or with target pre-selection, using the same broad-band or other 
ancillary data. The spectral resolution provided and the rich physical information that can then be derived compensate
for the lower limiting magnitude, with respect to imaging, that can be reached with these kinds of surveys. 
Worth mentioning are SDSS \citep{york00}, GAMA \citep{driver11}, z-COSMOS \citep{lilly07}, \deep\ \citep{newman13}, and 
VVDS-CFDS \citep{lefevre04b}. For a more detailed compilation of spectroscopic surveys of galaxies at  $z\leqslant 2$ see \cite{hayashi18}.

Mid-band surveys, with filter passbands of the order of ten to a few tens of nm,  possibly with some overlapping of 
contiguous filters covering a relatively wide spectral band, represent an intermediate situation between the depth
achieved in imaging, and the spectral coverage and resolution achieved in spectroscopy. They are advantageous when the 
number of sources in the field is so large that the amount of  time invested in observing through a large
number of filters is comparable to, or lower than, what should be spent in gathering spectroscopic information 
\citep{benitez14}. This situation can be achieved by either by the depth or the angular coverage of the survey. COMBO-17
\citep{wolf03}, ALHAMBRA \citep{moles08}, J-PAS \citep{benitez14}, and SHARDS \citep{perezgonzalez13}, are some recent examples of this kind of survey.

Narrow band imaging surveys use passbands of the order of 10 nm or lower and are usually designed to reach the 
maximum depth in a wavelength range restricted by the filter response. They target mainly emission line candidates, 
identified using colour--magnitude diagrams \citep[see for example][]{thompson95,pascual07,ota10}. Different redshift
ranges are explored, defined by the emission line detected, and the wavelength range is defined by the filter. For a
complete review of the narrow-band surveys performed so far, see \cite{hayashi18}.

One particular type of narrow-band imaging survey uses tunable filters (TFs) instead of standard fixed-cavity 
interference filters \citep[see for example][]{glazebrook04}. TFs define narrower passbands, of the order of 1 nm up 
to a few nm \citep{atherton81}. This allows the study of lower equivalent width (EW) emission features because the passband of the filter is
related to the EW of the emission lines that can be detected. This effect can be estimated using 
the contrast parameter defined in \cite{thompson95} and is explicitly acknowledged in, for example, the ongoing fixed-cavity 
standard narrow-band survey HSC-SSP \citep{hayashi18}, which uses narrow band filters of 113 and 135 \AA. A practical example 
of the lower EW bound reached in \otelo\ can be seen in \cite{ramonperez19}, hereafter referred as \otelotwo. 
However, this advantage is usually at the 
price of requiring several images at different wavelengths with some overlapping between them \citep[see for
example][]{jonesbland01} to increase emission line identification and improve flux accuracy \citep{lara10}. Another 
advantage of TF surveys is that they allow the detection of the faintest emission line targets with low continuum, which are 
probably missed in broad-band, and hence  spectroscopic, surveys. \cite{jonesbland01} pointed out that there is 
little overlap between emission-line selected galaxies (hereafter, emission line source or ELS) found in broad-band 
selected redshift surveys and TF surveys. The effect of the  bandpass width and  transmission profile of narrow-
band filters on the finding of \lya\ emitter (LAE) candidates at redshift \zred\ $\sim$ 6.5, was studied by \cite{dediego13} in 
a pilot survey to test the performance of TFs to find this and other emission-line candidates. They 
anticipated that fixed-cavity standard narrow-band filter surveys underestimate the number counts of LAEs and other emitters,
when the observed EW $\lesssim 60$ \AA. Such bias can be largely mitigated using TFs such as that of the \otelo\ survey.
TTF \citep{bland98}, CADIS \citep{hippelein03}, and more recently GLACE \citep{sanchezportal15} 
are examples of narrow-band surveys using TFs. GLACE has been  conducted mainly by members of the \otelo\ team and benefits from \otelo\ observing 
strategies, whereas \otelo\ uses certain GLACE data analysis approaches.

This is the first of a series of papers devoted to the OSIRIS Tunable Filter Emission Line Object (OTELO) 
survey,\footnote{\tt http://research.iac.es/proyecto/otelo} a pencil-beam probe designed for finding faint ELSs at different 
comoving volumes up to redshift \zred\ $\sim6.5$ through the exploitation of the red TF of the OSIRIS instrument on the 
GTC. The data gathering and reduction and the construction of the OTELO multi-wavelength
catalogue are described here. This article includes a first selection of ELS candidates and a study of their properties. 
The second article of the series (\otelotwo) and subsequent contributions about this survey set forth the techniques 
adopted for the study of pre-selected collections of ELSs based on \otelo\ low-resolution spectra, and a science case example as a demonstration of the survey potential.  In the calculations carried out this paper we assume a standard $\Lambda$-cold dark matter cosmology with
$\Omega_\Lambda$=0.69, $\Omega_{\rm m}$=0.31, and $H_0$=67.8 km s$^{-1}$ Mpc$^{-1}$, as extracted from \cite{planck15}.

\section{The OTELO survey}
\label{sec:otelo}

\noindent \otelo\ is a very deep, 2D spectroscopic (resolution $R\sim700$) blind survey, 
defined on a window of 230 \aa, centred at 9175 \aa. The first pointing of 
\otelo\ targets a region of the Extended Groth Field  embedded in Deep field 3 
of the Canada--France--Hawaii Telescope Legacy Survey\footnote{\tt http://www.cfht.hawaii.edu/Science/CFHTLS} 
(\cfhtls) and the deepest pointing of GALEX 
in imaging and spectroscopy. 
\otelo\ obtains {\it pseudo-spectra} (i.e.\ conventional spectra affected by the distinctive TF response,
as further explained in Section \ref{sec:products}) of all emission line sources in the 
field, sampling different cosmological 
volumes between \zred=0.4 and 6, thereby providing valuable data for tackling a wide variety 
of science projects, which include the evolution of star formation density up to redshift
$\sim$1.5, an approach to the demographics of low-luminosity emission-line galaxies and detailed
studies of emission-line ellipticals in the field, high-$z$ QSO, Lyman-$\alpha$ emitters, and Galactic
emission stars \citep{cepa13}. Such pseudo-spectra were obtained by means of the red tunable filter 
(RTF) of the OSIRIS instrument at GTC. For further details of
the OSIRIS instrument see \cite{cepa03}.

\subsection{Technical description}
\label{sec:tfprinciples}

Modern TFs or etalons are kinds of Fabry-Perot interferometers in which the cavity is formed by 
barely separated (by a few microns) plane--parallel plates (unlike their high-resolution 
counterparts), covered with multilayer, high-reflectivity coatings. The 
spacing between plates can be accurately changed by means of a stack of piezo-electric transducers 
actuating on one of these plates. 

In the case of an etalon in a parallel beam, with identical coating reflectivity 
\textit{R} and finite absorbance plates, 
the intensity transmission coefficient as a function of wavelength $\lambda$ is given by 
the Airy formula \citep{hecht01}:

\begin{equation}
\tau=\left(\frac{T}{1-R}\right)^2\left[1+\frac{4\ R}{(1-R^2)} \sin^2\left(\frac{\delta}{2}\right)\right]^{-1},
\label{eq:airyformula}
\end{equation}

\noindent where \textit{T} is the transmission coefficient of the coatings,
\textit{R} is the mean reflection coefficient, \textit{d} the plate spacing, and 
$\delta = 4\pi\mu d \cos \theta / \lambda$\ 
is the phase difference between interfering waves for a given incidence angle $\theta$, and 
a refractive index $\mu$ of the medium (=1 if air) between plates.  

Equation \ref{eq:airyformula} defines not only the transmission profile shape (Airy function) 
but also the periodicity of its maxima, which occurs when $\delta = 2\pi m$, with $m\in\mathbb{N}$. 
Hence, for the given parameters the 
transmission of the TF is at maximum (i.e.\ constructive interference) if 
the space between the reflectors is an \textit{m}-tuple of an allowed state with the 
same energy as the photon, $E_{\gamma}=hc/\lambda$. Therefore, the interference
condition remains at

\begin{equation}
m\lambda = 2\mu d \cos \theta.
\label{eq:interf}
\end{equation}

On the basis of this approximation finding the peak transmission, $T/{(1-R)}^2$, is trivial and
the wavelength spacing between consecutive orders, or free spectral range (FSR), is:

\begin{equation}
\Delta\lambda = \frac{\lambda}{m}.
\label{eq:fsr}
\end{equation}

Assuming that reflectivity is high enough, or $(1-R)^2 \ll R$, we can solve Equation
\ref{eq:airyformula} for $\tau=0.5$ to obtain an expression for the FWHM (or bandwidth) of 
the transmission peak given by:

\begin{equation}
\delta\lambda\,_{\rm FWHM}=\frac{\lambda\ (1-R)}{m\pi\ R^{1/2}}.
\label{eq:fwhm}
\end{equation}

Within a given \textit{m} and for small $\theta$, the TF transmission 
profile for a single maximum can then be approximated, again from Equation \ref{eq:airyformula}, 
by the expression:

\begin{equation}
\tau:\tau\left(\lambda\right) \simeq \left[1 + \left(\frac{2 \left(\lambda -\lambda_0\right)}{\delta\lambda\,_{\rm FWHM}}\right)^2\right]^{-1},
\label{trans}
\end{equation} 

\noindent where $\lambda_0$ is the wavelength at maximum transmission. 

From the above equations it is clear that $\delta\lambda\,_{\rm FWHM}$ depends only on the 
order of interference for
a given illuminating wavelength. We then, in practice, require  a mid-band 
filter (known as an order-sorter) of width $\lesssim {\rm FSR}$ that allows us to isolate 
an individual transmission profile corresponding to \textit{m}. Under this assumption,
a useful expression for the TF {\it effective} passband width can be obtained by
integrating Equation \ref{trans} analytically with respect to $\lambda$ in the interval
defined by ${\rm FSR}$, which yields 

\begin{equation}
\delta\lambda\,_{\rm e} \simeq \frac{\pi}{2}\ \delta\lambda\,_{\rm FWHM}.
\label{eq:effwidth}
\end{equation}

Equation \ref{eq:interf} provides the key control tool of 
a TF. The central wavelength $\lambda_0$ of such a device can be tuned by changing the cavity 
spacing \textit{d}. For a low-resolution TFs, if \textit{d} varies by only a few 
nm, $\lambda_0$ slips in the FSR domain, while the order of interference \textit{m} 
(and hence the bandwidth) can be changed by varying the cavity spacing in the order of 
microns. An additional consequence, often called the \textit{phase effect}, is also noticeable: 
the filter-transmitted wavelength will be progressively shifted to the blue as the incident 
angle $\theta$ with respect to the optical axis of the TF increases. The projection of this 
axis on to the detector plane defines the optical centre of the TF.

Theoretically, in the particular case of the OSIRIS instrument, the incident angle $\theta$ 
should be related to the radial distance \textit{r} to the optical centre by means of the 
ratio between the GTC and the OSIRIS collimator mirror focal lengths. However, for the OSIRIS 
RTF, the dependence referred to of the output wavelength on the radial distance is really 
given by \cite{gonzalez14}:

\begin{equation}
\lambda = \lambda_0 - 5.04\ r^2 + a_3(\lambda)\ r^3,
\label{eq:radep}
\end{equation}

\noindent where $\lambda_0$ is the tuned central wavelength in \AA, \textit{r} is the 
distance to the optical centre in arcminutes, and $a_3(\lambda)$ is an additional term given by

\begin{equation}
a_3(\lambda) = 6.0396 - 1.5698 \times 10^{-3} \lambda + 1.0024 \times 10^{-7} \lambda^2.
\label{eq:third}
\end{equation}
 
This empirical parametrization of the output wavelength on radial distance is in accordance with the 
fact that, in general, the performance of a Fabry-Perot interferometer is highly dependent 
in turn on the properties of the cavity coatings. As demonstrated in \cite{gonzalez14}, 
chromatic dispersion caused by multilayered thick coatings of the RTF gives rise to an 
anomalous phase effect driven by Equation \ref{eq:radep}. This expression, as well as 
Equation \ref{trans}, are used hereafter for modelling the RTF's behaviour.

\subsection{Survey design and observations}
\label{sec:observations}

\begin{figure*}[!ht]
\centering
\includegraphics[width=1.0\textwidth]{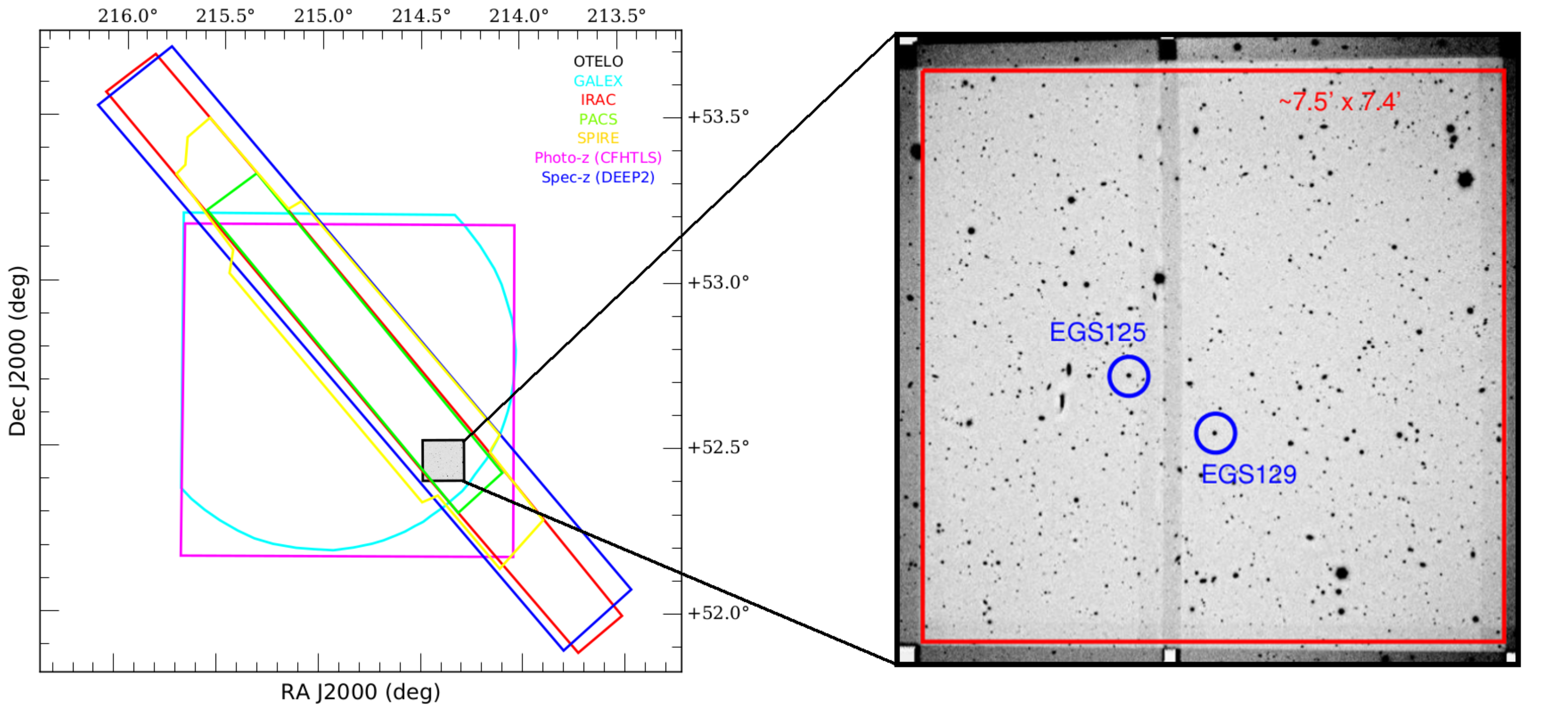}
\caption[Spatial distribution of the sources in the auxiliary catalogues]
{Spatial distribution boundaries of the UV, optical, IR, and spectroscopic redshift 
data used to build OTELO's multi-wavelength catalogue (left panel). In black: 
the \otelo\ raw field, 
cyan: GALEX, red: \irac, green: \pacs, yellow: \spire, magenta: photo-z (\cfhtls), blue: 
spec-z (\deep). The right panel expands the surveyed area, showing the coadded RTF data 
of \otelo\ with its weight map (see Section \ref{sec:coadd_ext} for details) overlapped in 
order to highlight the survey sensitivity footprint. The red polygon defines the survey's final
area (7.5\arcmin$\times$7.4\arcmin). Blue circles indicate the position of the flux 
calibration stars (Sec.\ \ref{sec:standards}).}
\label{catalogues}
\end{figure*}

As mentioned above, the first pointing of the \otelo\ survey is located at the 
southwest edge of the most deeply explored region of the EGS, specifically centred
at RA = 14 17 33, Dec = +52 28 22 (J2000.0), subtending almost 56 square arcmin.
This choice benefits not only from the plethora of observational and derived data products 
created and/or compiled by the Team of the 
AEGIS\footnote{All-wavelength Extended Groth strip International Survey; {\tt http://aegis.ucolick.org}} 
survey, as well as recently acquired information from the \herschel\ Space Telescope, 
but these ancillary data are also an imperative requirement for obtaining the products described
in this paper. The fine selection of pointing coordinates was partially determined
by the position of isolated flux calibration sources, as accurate flux calibration in physical 
units  is necessary for every individual RTF observation. Figure \ref{catalogues} indicates 
the position of the first \otelo\ field relative to the main data contributors of AEGIS. Details 
of these contributions are given in Table 6 and are discussed in Sections \ref{sec:ancillary}
and \ref{sec:addancillary}. 

According to the science goals of \otelo, the strategy of the survey consists 
of the tomography of 36 
slices equally distributed in the (central) wavelength range between 9070 and 9280 \aa. An RTF width 
$\delta\lambda\,_{\rm FWHM} = 12$ \aa\ was adopted, scanning every 6 \aa\ (i.e.\  
$\delta\lambda\,_{\rm FWHM}/2$). This sampling 
represents almost the best compromise between a photometric accuracy of $\sim$20\% in the 
deblendence of the \ha\ from \nii\ emission lines (as demonstrated in thorough simulations
by \citealt{lara10}) and a reasonable observing time span.

A total of 108 dark hours, under a guaranteed gime (GT) agreement\footnote{Defined between
the OSIRIS Instrument Team and the Instituto de Astrof\'isica de Canarias.}, distributed over
four campaigns between 2010 and 2014, were dedicated to obtain the RTF data. Table \ref{table_observations}
contains a summary of the observing log. These observations were performed under quite uniform
seeing conditions, with a global mean of $0.83\pm0.08$ arcseconds, as averaged directly on the 
scientific images. 

Additional time ($\sim$1 h) was devoted, with the same instrument
(see Section \ref{sec:standards}), to obtaining low-resolution 
spectra of two colour-selected F8 sub-dwarf stars (EGS125 \& EGS129 in the right panel of
Figure \ref{catalogues}) in the \otelo\ field, and an STIS Next Generation Spectral 
Library\footnote{\tt https://archive.stsci.edu/prepds/stisngsl} 
spectro-photometric standard HD126511 ($V$=8.359, Sp.~type G5), all under photometric conditions.

\begin{figure}[h]
\centering
\includegraphics[angle=-90,width=\linewidth]{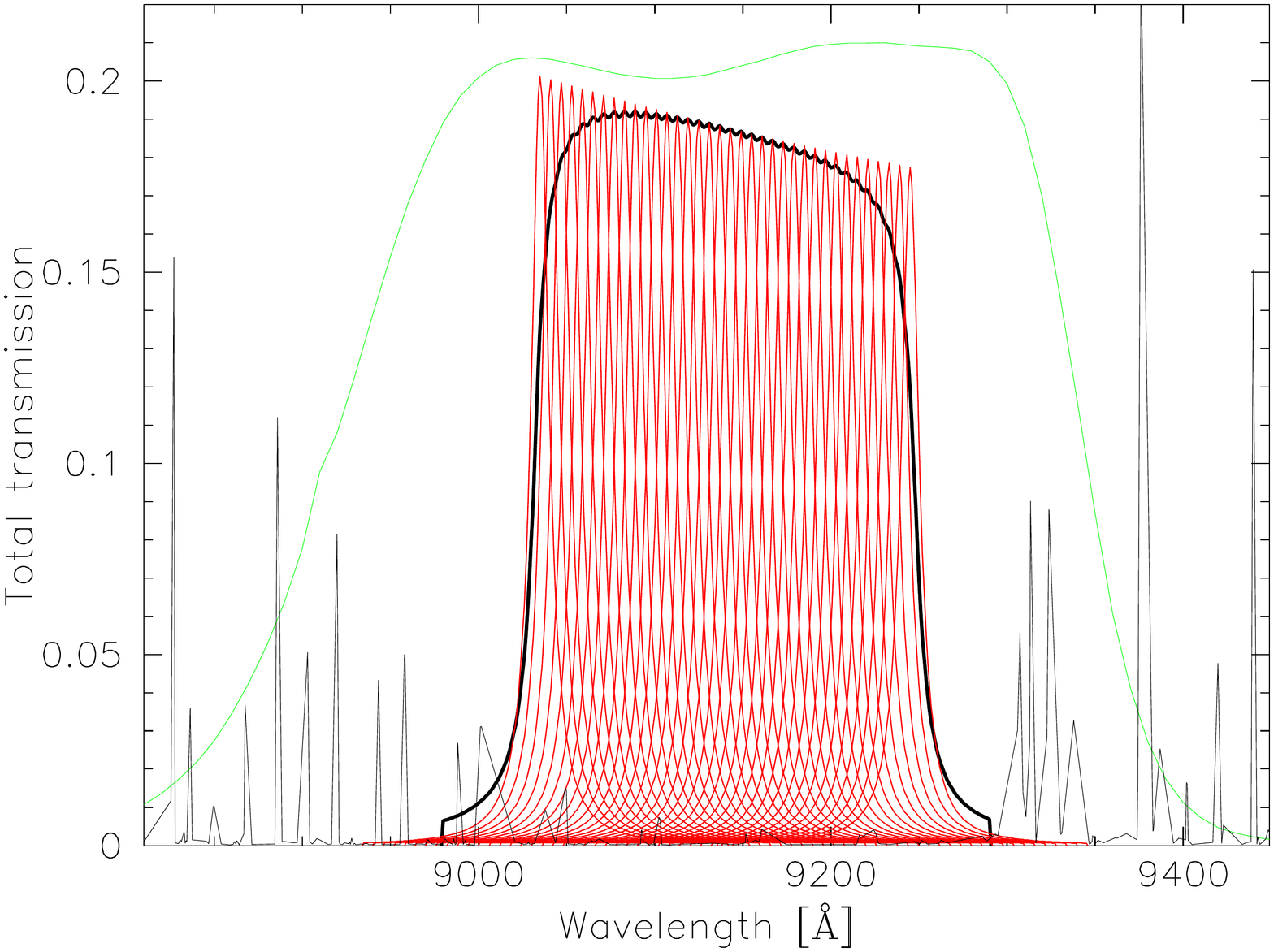}
\caption[OTELO response]{Characteristic transmission of the RTF for the 36 slices 
that define the OTELO survey spectral scan (red solid lines), as dictated by
Equation \ref{trans}. The thick solid line 
corresponds to the synthesis of OTELO RTF responses (hereafter \otelocustom) a kind of 
intermediate-band ($\sim$230 \AA) 
filter affected by the telescope, the OSIRIS camera, and the QE
of the detectors. Note the location of those responses, avoiding the Meinel bands represented 
by the airglow spectrum (thin black line) from \cite{hanuschik03}. 
The green line represents the transmission (scaled to the total
efficiency of the system) of the order sorter OS 911/42 
used to acquire RTF data.}
\label{airy}
\end{figure}

Data for each wavelength was obtained from six exposures of 1100 s each, nominally dithered 
18\arcsec\ in a cross-shaped symmetrical pattern with 
a recurrence to the initial pointing, in order to fix bad pixels, residual cosmic rays and fill 
in the gap between detectors 
of the OSIRIS mosaic. This pattern also facilitates the identification of diametric ghost images (see 
Appendix B in \citealt{jones02} for a succinct description of ghost families), as well 
as the modelling of fringes. Observations were distributed into observing blocks of two single exposures, 
resulting in a total of 216 RTF science mosaics. A filter, named OS 911/42 (hereafter referred as OS), 
centred at 911 nm and with  a bandwidth of 42 nm, was used as 
order sorter. Figure \ref{airy} shows the Airy profile corresponding to each wavelength
slice or $\lambda_0$, and the order-sorter transmission between Meinel bands. The RTF 
tuning during the observations was found to  be stable at the nominal accuracy of 1 \AA, as 
expected. 

\begin{table*}[ht]
\caption[Summary of observations]{\otelo\ survey: Log summary of the red tunable filter of OSIRIS 
(RTF) \& long-slit (LS) observation campaigns. $N_{\rm obs}$ indicates the number of individual
observing blocks of $\sim$1 hour each. In the case of RTF observing mode, to each block corresponds
a single tuning.}
\label{table_observations}
\centering     
\begin{center}\begin{tabular}{c c c c c c c}
\hline   \\         
Observing & Dates & $N_{\rm obs}$ & Exposure & Wavelength & Mean seeing & $\sigma_{\rm seeing}$  \\  
Mode &  & & time [ks] & range [\aa] & [\arcsec] & [\arcsec]  \\    
\hline      \\[1pt]                        
RTF & April 11 \,--\, July 7, 2010 & 36 & 39.6 & 9250 \,--\, 9280 & 0.83 & 0.06 \\ 
RTF & May 4 \,--\, August 2, 2011 & 38 & 41.8 & 9208 \,--\, 9244 & 0.84 & 0.08 \\ 
RTF & May 4 \,--\, Aug. 10, 2013 & 58 & 63.8 & 9154 \,--\, 9202 & 0.82 & 0.09 \\ 
RTF & March 1 \,--\, June 5, 2014 & 84 & 92.4 & 9070 \,--\, 9148 & 0.82 & 0.07 \\ 
LS & April 13, 2011 & 8 & 3 & 4800 \,--\, 9500 & <0.9 & \,--\, \\ 
\hline  
\end{tabular}\end{center}
\end{table*}

\subsection{Survey products}
\label{sec:products}
The data reduction process described below affords 
astrometry-corrected and flux-calibrated images of each RTF slice. Coaddition of these
images is used to obtain a deep RTF image, as well as a raw source list in a sort of 
photometric catalogue of RTF integrated fluxes. This catalogue is enhanced by 
measuring and cross-correlating ancillary information cited in Section 
\ref{sec:observations}. Individual RTF frames are also stacked to obtain 
representative frames of each slice, used only to obtain source cutouts
for illustrative purposes. 

\begin{figure}[h]
\centering
\includegraphics[angle=-90,width=\linewidth]{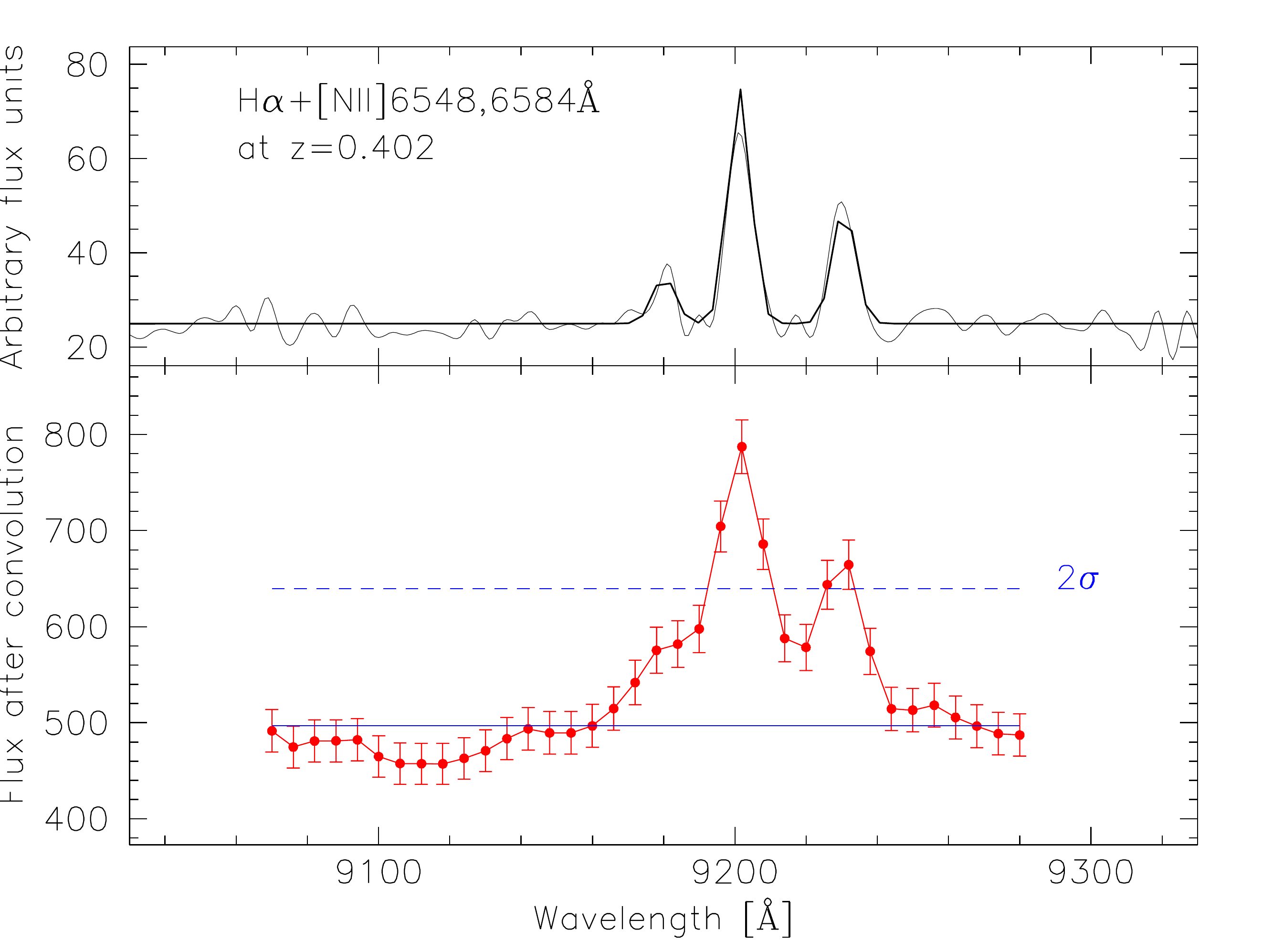}
\caption[Example of a pseudo-spectrum]{Example of a synthetic pseudo-spectrum as
typically provided by the \otelo\ survey. The top panel shows a flat continuum (thick
line) with \ha+[NII]$6549,6584\AA$ emission lines as Gaussian profiles (FWHM=6$\AA$). A
$\log_{10}$([NII]$6584\AA$/\ha) = -0.3 ratio was chosen for this example. The observed
\ha\ and [NII]$6584\AA$ EWs are 15.5 and 7.7 $\AA$ respectively. Poissonian
noise was added to this spectrum (thin line) and then convolved with the transmission 
curves represented in Figure \ref{airy} to mimic the \otelo\ survey scan. The
result is shown in the bottom panel. Flux errors were computed by propagation, 
taking into account the continuum deviation at the input spectrum multiplied by the 
minimum pixel area, and the nominal readout noise. The dashed line represents 
2$\sigma_c$ deviations above the fitted pseudo-continuum.}
\label{pseudospectra}
\end{figure}

Secondly, the \otelo\ survey provides a pseudo-spectrum for each object detected
in the deep RTF image. Formally, a pseudo-spectrum is a wavelength convolution of 
the source SED by the RTF response sequence defining the scan. Unlike the spectra 
obtained from diffractive devices, we denoted as pseudo-spectra the vectors obtained 
from TF scans, properly calibrated in flux and wavelength. An example of 
the synthetic pseudo-spectra as provided by \otelo\ can be seen in Figure 
\ref{pseudospectra}, and their further analysis is a subject of the \otelotwo\
paper. 

\begin{figure*}[!ht]
\centering
\includegraphics[angle=0,width=1.0\textwidth]{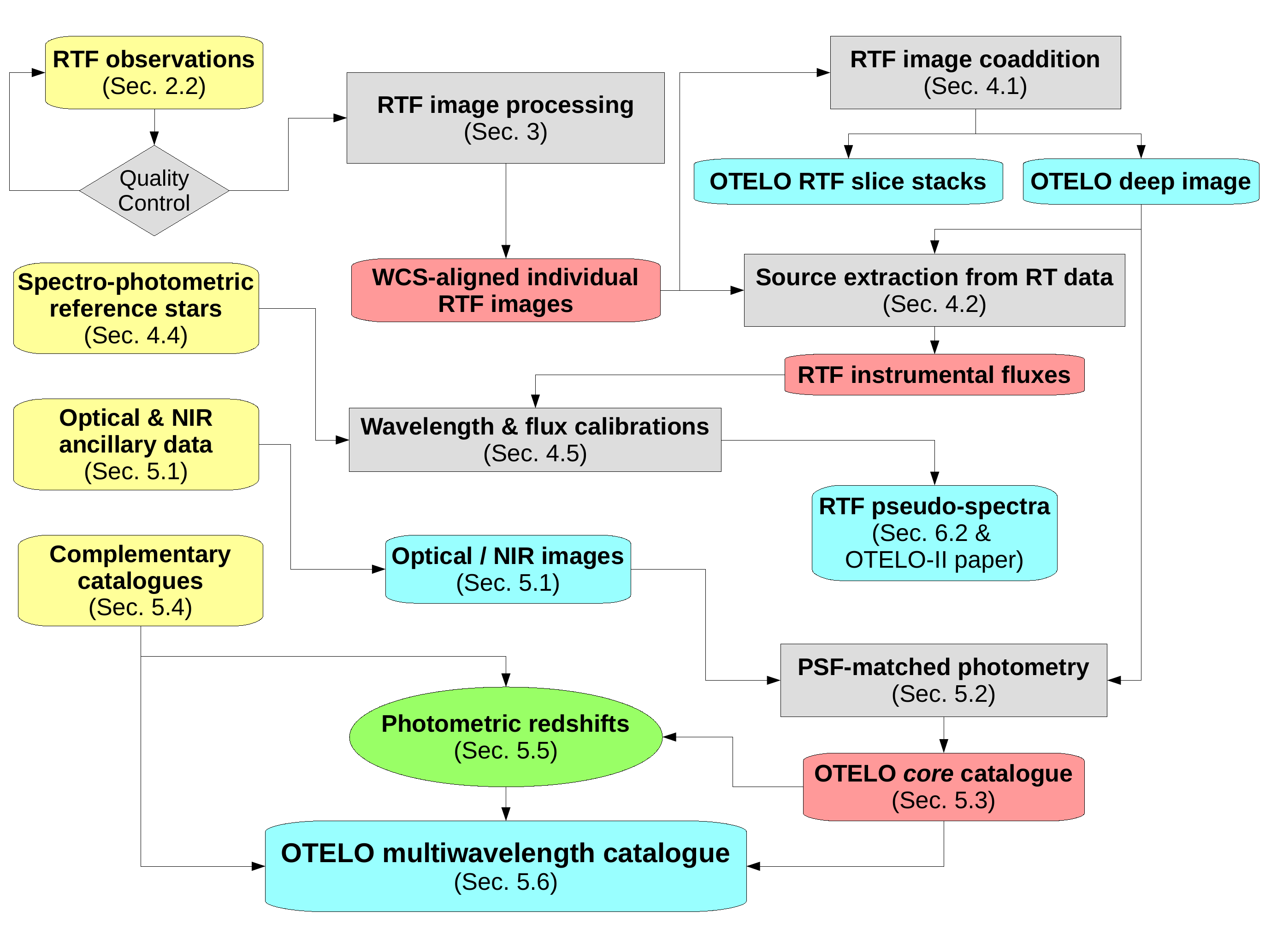}
\caption[]{Data flow chart of the \otelo\ survey. Sharp corner grey boxes enclose processes, 
and the round corner ones represent input data (yellow), or intermediate (red) and final (blue) 
products. The green ellipse denotes both a process and a product. Where appropriate, the relevant 
section of this article is quoted in parentheses.
}
\label{oteloflow}
\end{figure*}

The main processes and products outlined in this work, with a reference to
the relevant section, are mapped in the data flow diagram of Figure \ref{oteloflow}.

\section{RTF data reduction}
\label{sec:data_reduction}

\subsection{First steps}
\label{sec:firststeps}

Image mosaics from the OSIRIS instrument consist of two 1049$\times$2051 pixel$^2$ 
frames at binning 2$\times$2 (giving a pixel scale of 0.254 \arcsec/pixel), with a 
projected gap of about 38 pixel between them along the longest 
axis\footnote{\tt http://www.gtc.iac.es/instruments/osiris/osiris.php}. The effect of
the latter can be appreciated as the slight, vertical shadow 
in the middle of the right side image of Figure \ref{catalogues}. Individual science, 
and all auxiliary, frames of \otelo\ were bias- and 
overscan-subtracted before trimming
(according to the unvignetted FoV) using the {\tt IRAF ccdproc} task. Cosmic-ray
removal was carried out on each frame by means of the {\tt IRAF} implementation
of the stand-alone procedure {\tt lacos-im} \citep{dokkum01}, which identifies traces of 
these events by Laplacian filtering. 

OSIRIS detectors are affected by bad pixels, mostly on the right edges and in a few columns
near the upper middle part of CCD2. Bad-pixel masks for each observing epoch were obtained 
using a set of reduced OS night-sky flats following a standard procedure. A number of 
flats with low ($\sim$6000--12\,000) and high counts ($\sim$33\,000) were median-combined
apart. The normalized ratio of such low-to-high intensity level flats was used as a 
bad-pixel mask. Using {\tt IRAF}'s \texttt{ccdmask} task, we identified those pixels for which 
this ratio was greater than 15\% and created boolean masks. Based on these masks, bad 
pixels in the science frames (which do not exceed 0.37\% of each trimmed frame area) 
were finally corrected using the {\tt IRAF} \texttt{fixpix} task, which performs an interpolation 
of neighbouring pixels.

Flat-fielding is not a straightforward step in conventional TF data reduction because 
sensitivity variations across the field radially depends on wavelength for a given 
tuning and, in the particular case of this survey, a non-negligible fringing component 
is present in all frames. In this case, pixel-to-pixel fluctuation maps in each mosaic 
component, observing epoch, and TF tuning were obtained from a combination of bias-reduced, 
defringed, and bad-pixel fixed TF dome flats. These flats were then corrected by illumination 
using a mode-scaled sky flat obtained with the OSIRIS order sorter OS 911/42. 
Low-frequency fluctuation maps were obtained from airglow emission maps 
representative of each slice: each bias-corrected science frame was object-masked up to 
2-$\sigma$ object counts above the median background level, where $\sigma$ is the standard 
deviation of the local background, and each slice sextuple was median (sigma-clipped) 
combined. The fluctuations were measured in concentric radii, each 
50 pixels around the optical centre. These measures were normalized and then used to 
generate a low-order surface, which constitutes an analogy to a large-scale night-sky 
flat. Both small- and large-scale maps were combined into a super-flat used for reducing
individual mosaic components. The mean background homogeneity of science frames after applying
this procedure is better than 4\%.   

\subsection{Ring subtraction and defringing}
\label{sec:rings}

Sky subtraction from astronomical images would formally imply control over the 
physical conditions of sky brightness, its gradients, the behaviour of detectors during 
the integration, and the optical properties of the telescope and instruments 
involved. This task, both impractical and intractable \citep{blanton11}, tends to focus on 
relatively simple approximations that depend on the resulting superposition of the main 
observational effects, and on the angular extension of the sources of interest. 

\begin{figure*}[htb]
\centering 
  \begin{tabular}{ccc@{}c}
    \multicolumn{4}{@{}c}{\includegraphics[angle=-90,width=\textwidth]{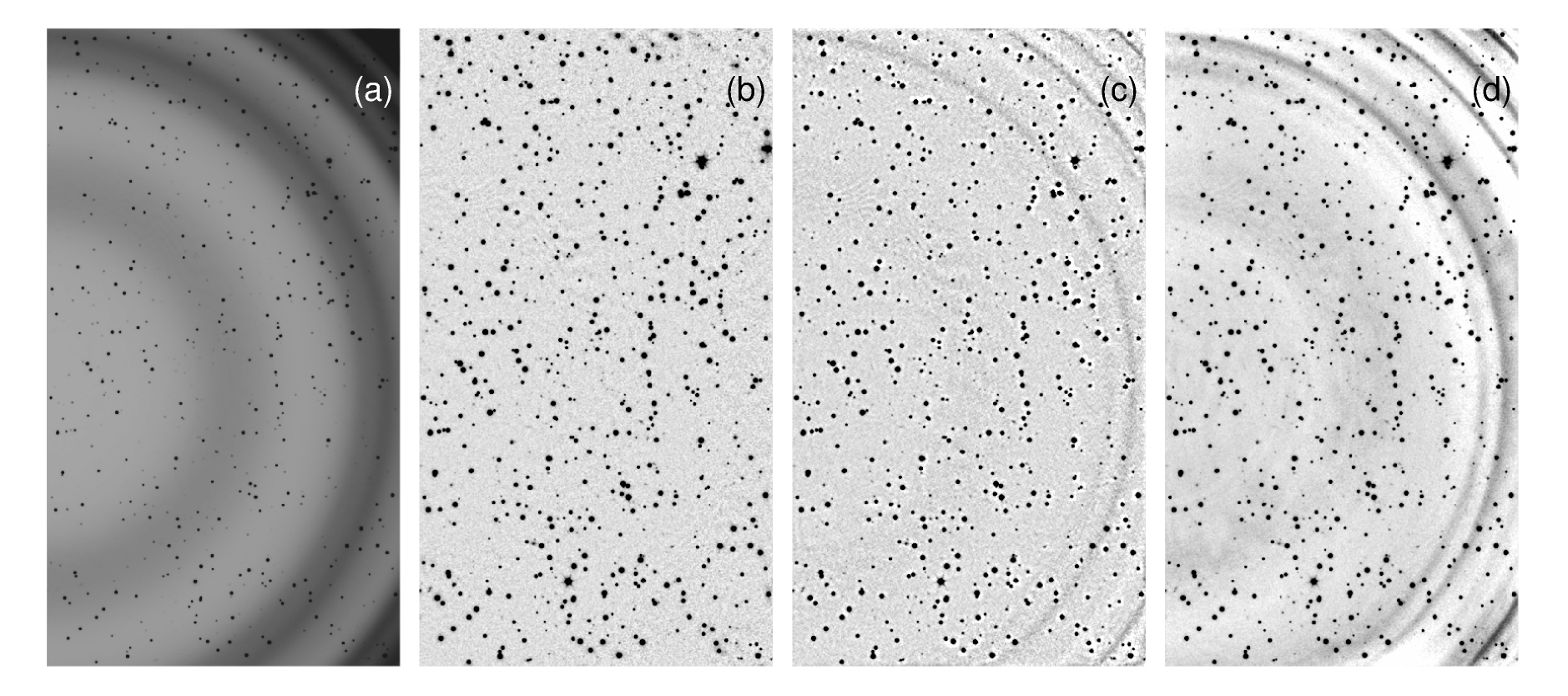}} \\
    \ \includegraphics[angle=-90,width=.226\textwidth]{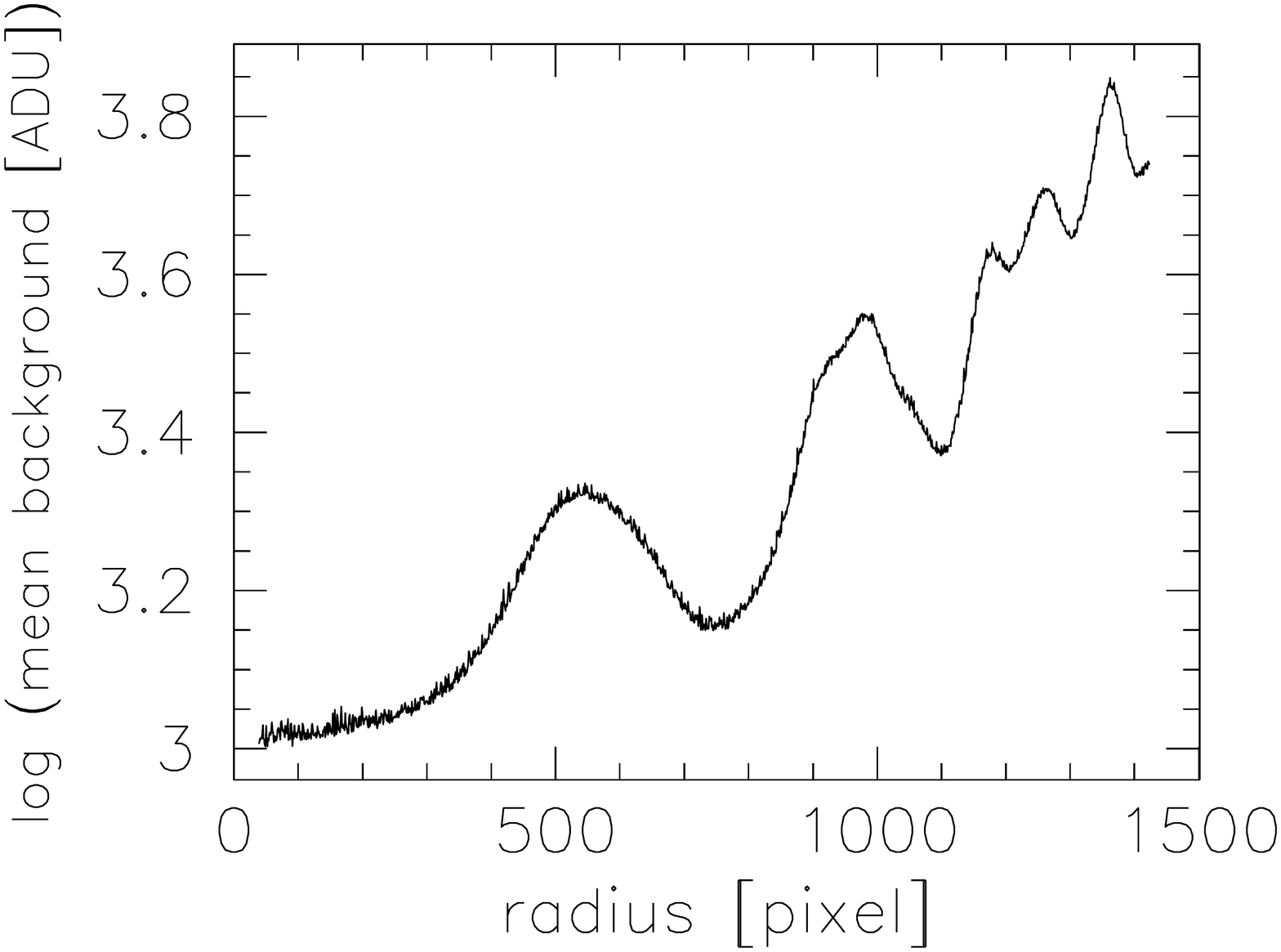} &
    \includegraphics[angle=-90,width=.226\textwidth]{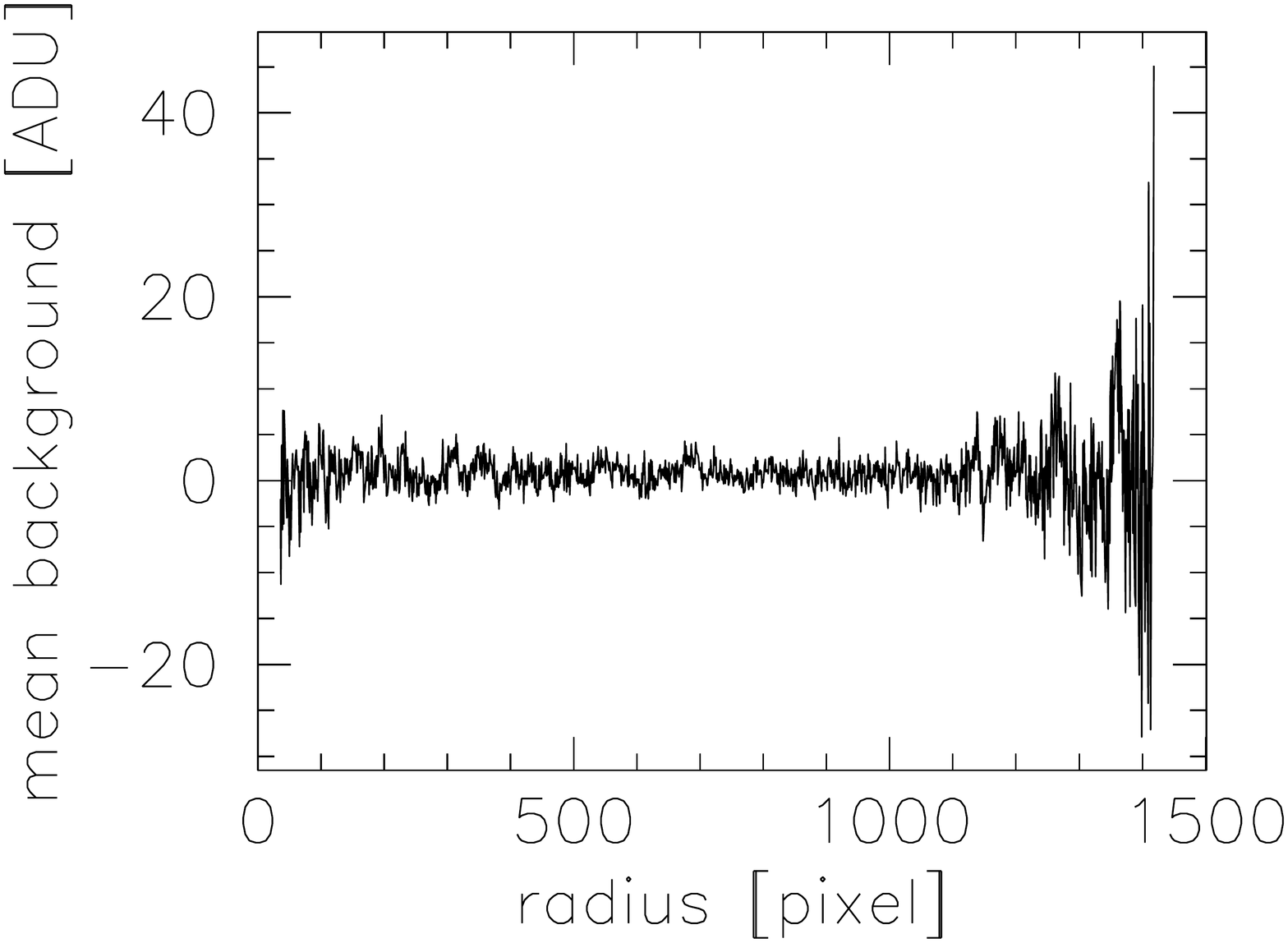} &
    \includegraphics[angle=-90,width=.226\textwidth]{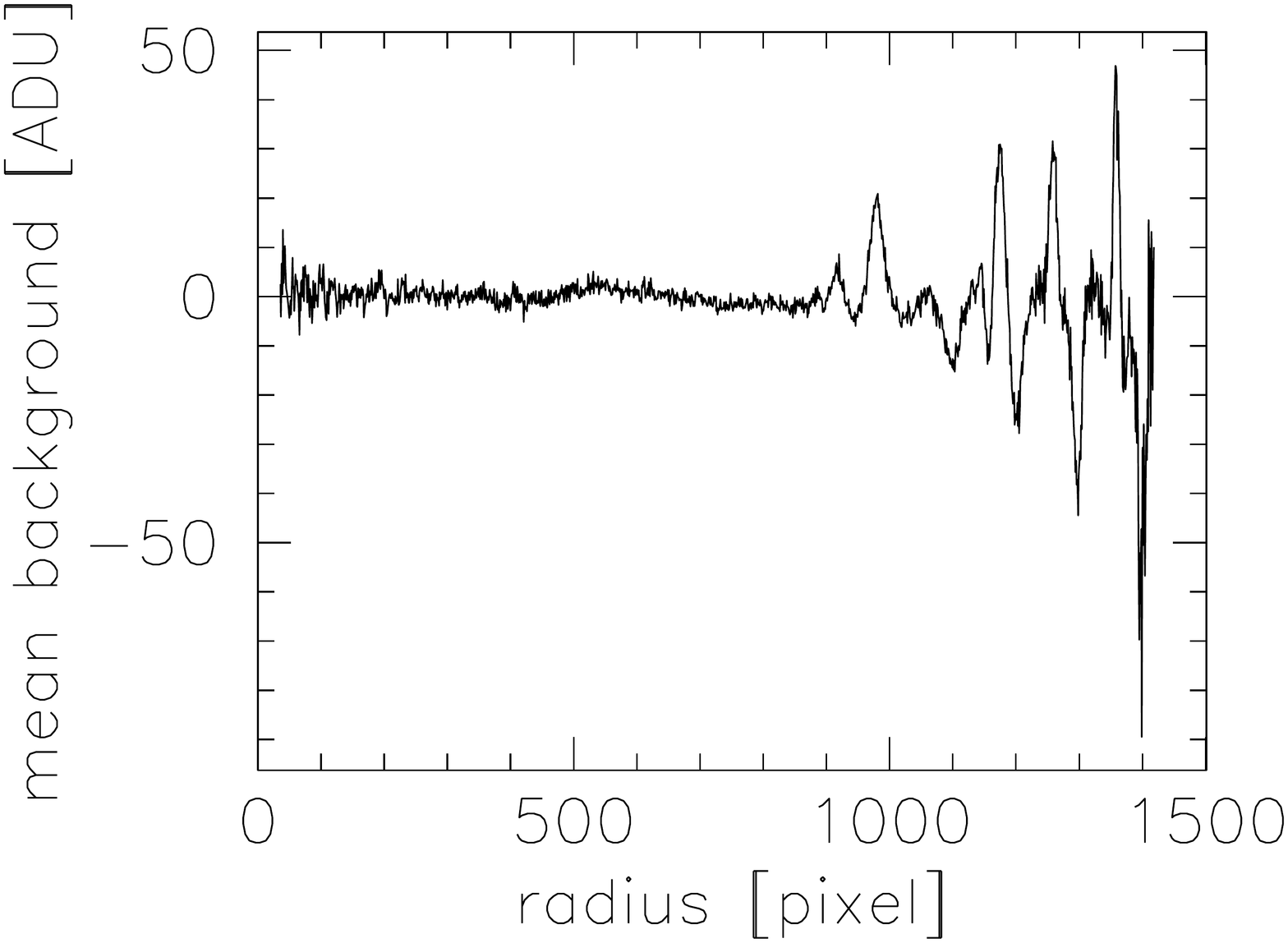} &
    \includegraphics[angle=-90,width=.226\textwidth]{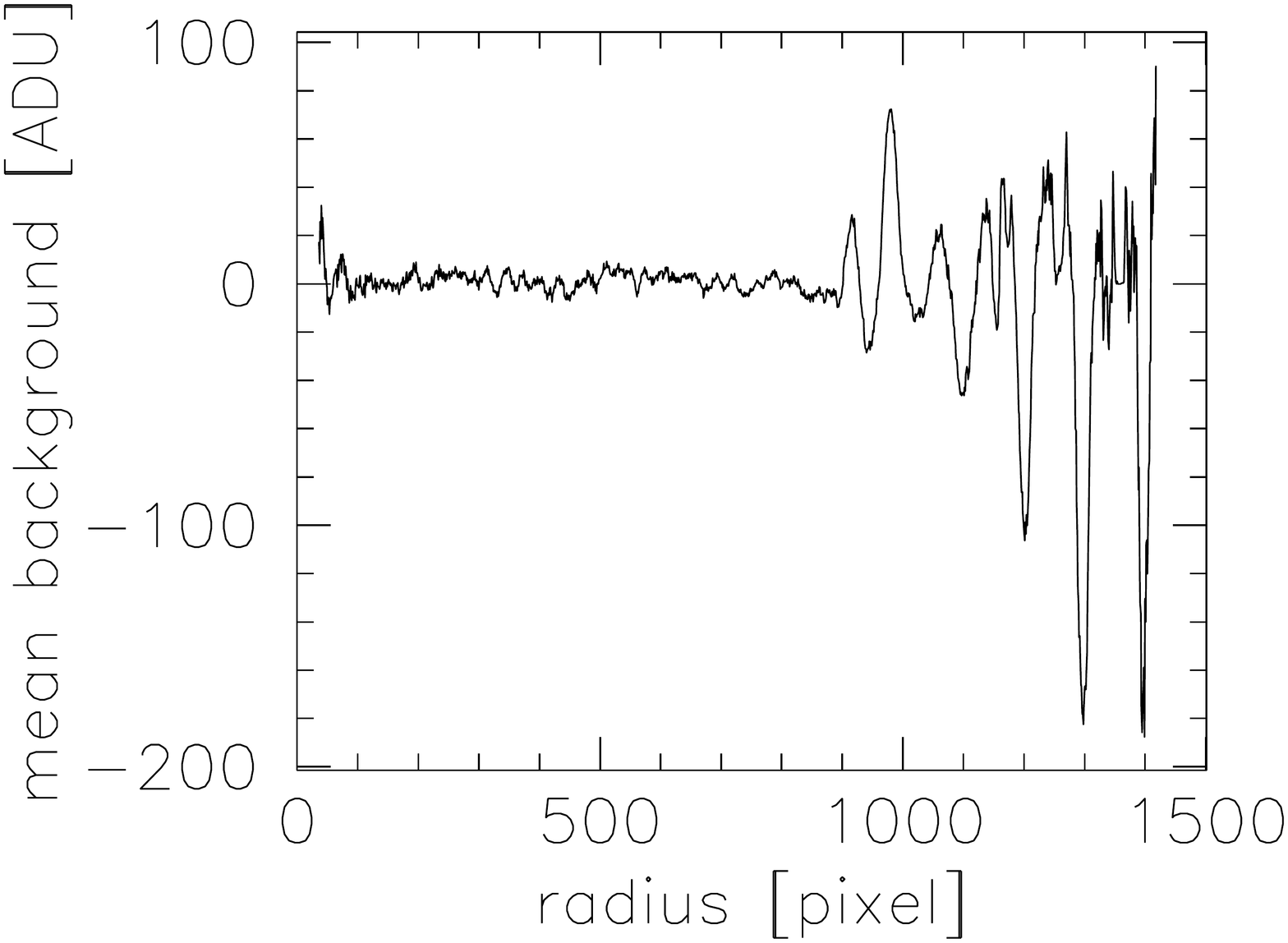} \\
  \end{tabular}
\caption[Ring subtraction example]{An example of sky-ring subtraction tests described in the
text. Panel (a) shows one selected raw image of the CCD2 corresponding to a central
wavelength of 9130${\rm \AA}$, with 500 artificial stars added. Arc patterns (bluer as 
the curvature radius grows) corresponding to equal wavelengths of the airglow band spectrum are 
noticeable. Panels (b), (c), and (d) show sky-ring subtracted and
defringed images by using, respectively, the {\tt ringsub} algorithm (adopted in this work), 
{\tt tringSub}, and the azimuthally averaged approach. All images are scaled to
the median. Below each panel is shown the mean radial spectrum (1 pixel sampling) of the 
original image background or residual, as appropriate.}
\label{ringsub}
\end{figure*}

As set out above, TF observations show a radial distribution of the transmitted spectral 
features bluewards of the central (i.e.\ tuned) wavelength. In practice, spectral 
calibration lines or airglow emission illuminating the exit pupil of the telescope appear 
on images as concentric rings around the projection of the TF's optical axis. The latter 
are particularly strong in the NIR, as shown in panel (a) of Figure \ref{ringsub}. 
Although the OTELO survey sampled a spectral region between Meinel bands, airglow OH emission 
bands, even with minor strengths, severely affect observations with long exposure 
times. As a side effect, considerable fringing usually accompanies the most intense airglow 
bands. Thus, an appropriate subtraction of these effects not only ensures the photometric 
integrity of all the sources, especially the faintest ones, but also prepares the individual science 
frames for a residual-free image stacking. 

With these aims, we explored several technical approaches to these problems in the
literature. Evidently, airglow contribution to observations can be removed on the fly 
from data to levels $\ll$1\% by using nod-and-shuffling \citep{glazebrook01} or similar 
techniques (i.e.\ of the ON--OFF type). But the FOV reduction and/or a prohibitive 
increase in observing time made us discard these strategies from the very beginning of 
the survey, although the OSIRIS instrument was designed to be used also in these observing 
modes. Apart from this, there are different approaches for ring subtraction in TF images 
obtained in the usual mode. \cite{jones02} include a complete review of reasonable 
alternatives to remove night-sky rings. They finally lean towards a simple but effective 
method for those cases in which objects of interest are much smaller than the ring structure 
(as in the case of the OTELO survey): a background map is created by median filtering of 
regularly shifted (by only a few pixels) copies of the individual image to be reduced. The 
result is then optionally smoothed and subtracted from the original frame, leaving little 
or no night-sky residual, according to the authors. This procedure is a part of the 
{\tt TFRed} collection of {\tt IRAF} tasks for TF data reduction \citep{jones01}, 
identified as {\tt tringSub}. It was included in the OSIRIS Offline Pipeline 
Software\footnote{Available at \tt http://gtc-osiris.blogspot.com.es} after some improvements.

\cite{veilleux10} model the sky background by obtaining an azimuthally 
averaged sky spectrum and then sweeping it around the known position of the optical axis. 
Prior filtering of sources and cosmic rays are performed, and  constitute a part of 
the Maryland-Magellan TF Data Reduction Pipeline. On the other hand, \cite{weinzirl15} 
describe a technique based on an image transformation to polar coordinates with the 
purpose of subtracting the airglow emission, this time as straight patterns, and then 
applying an inverse transformation to restore image geometry. 

Even though this approach is qualitatively similar to that of Veilleux et al. (2010), we do 
not test its performance in order to avoid flux-conserving issues in the 
projection/reprojection processes.

For this work, we opted to combine the advantages of the cited algorithms by performing 
the sky spectrum subtraction on each individual image with an automatic, two-step approach. 
A first cleaned 
image, useful only for source mapping, is obtained by subtracting a background model resulting 
from  median-filtering eight offset ($\sim$10 pixel) copies of the input image by using 
the {\tt tringSub} task described above. An object mask for each individual image 
is then constructed by flagging positive features 2-$\sigma$ above residual background using 
the {\tt IRAF objmask} task. This initial guess for the 
background-subtracted image is also used 
to create a fringing model. Once the original image is defringed and the sources on 
it - traced by the object mask - are replaced by neighbouring background values, we create 
a series of replicas of the resulting image by rotating it around the optical centre at 
optimized angles. Such angles result from an equilibrium between maximum pixel sampling
and the maintenance of the sky ring ellipticity ($<$0.3\% in our case; J. Cepa 2017,
priv.\ comm.) effects below a pixel in the radial direction. A median combination of these 
replicas before trimming to the original image size results in a sky background base image. 
The final background model is obtained by fitting a surface on to the base image. The procedure 
ends after subtraction of this model from the original image.

\begin{figure}[h]
\centering
\includegraphics[angle=-90,width=\linewidth]{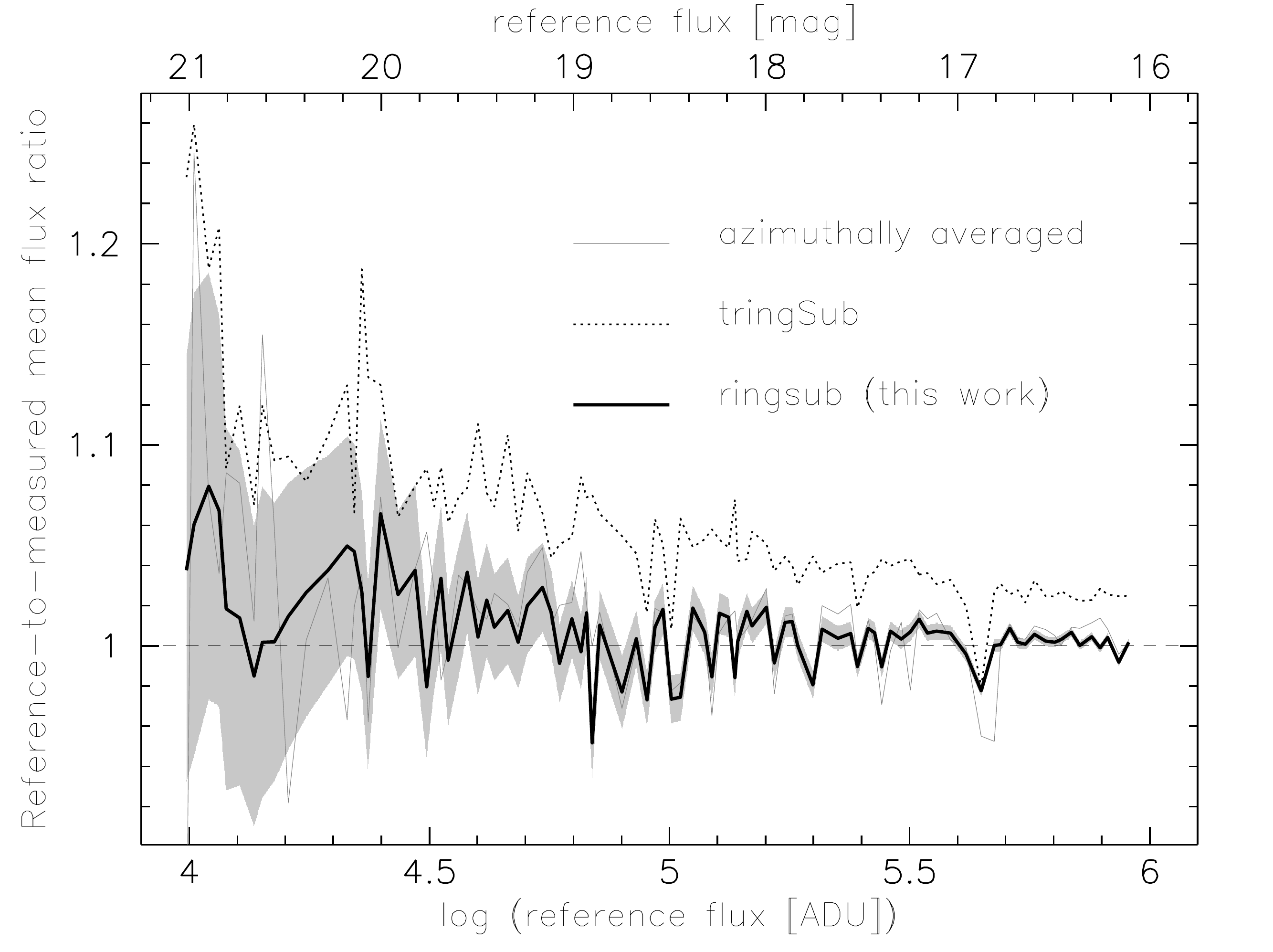}
\caption[]{
Running mean of the reference-to-measured flux ratio of 500 artificial stars added 
to an example raw RTF image before sky-ring subtraction using the {\tt ringsub} procedure, 
the {\tt tringSub} task, and the azimuthally averaged spectrum approach. Consistently with
data given in Table \ref{ringsub_table}, the first and last show a similar flux recovery 
behaviour, whilst the {\tt tringSub} approach leads to a mean flux overestimation about 6\%.
The shaded region represents the error of measured fluxes corresponding to the {\tt ringsub}
routine. The error distributions of the other two procedures resemble this one. See text for 
details. 
}
\label{ringsubcomparison}
\end{figure}

A complete procedure, named {\tt ringsub}, was written in a single parametric {\tt IRAF} script 
with zero user interaction and is available on request to the authors. The
{\tt ringsub} performance was tested and compared with {\tt tringSub} and the azimuthally averaged
algorithms. Science frames with large sky background fluctuations (i.e.\ around $\lambda_{0}=9140 \AA$) 
were selected and catalogues created of 500 artificial stars (using the {\tt IRAF 
starlist} task) uniformly distributed in position and brightness, according the instrumental flux 
range of the tested images. These stars were added to
the images using the {\tt IRAF mkobject} task and their fluxes recovered using the {\tt SExtractor} 
(v. 2.19.5) application \citep{bertin96} in standard mode after subtracting the sky background. 
Figure \ref{ringsub} shows a representative example of an \otelo\ image (CCD2) before sky subtraction 
with the mock stars added. Additional panels show the results of subtraction and the mean background
in each case. The use of {\tt ringsub} produces the smallest background residuals. 
If the reference-to-measured mock star flux ratio for each sky background subtraction approach is compared,
it can be seen from the statistics presented in Table \ref{ringsub_table} that the {\tt ringsub} 
algorithm yields the results nearest to the unity with the smallest dispersion. The running mean of 
each flux ratio as a function of the mock reference flux is represented in Figure \ref{ringsubcomparison}.
It shows that maximum departure of the {\tt ringsub} ratio from unity at low flux regime is between 
$\sim$2 and 4\%, which is in any case a fraction of the in quadrature error of the measured flux. This
gives an idea of the real performance of the adopted sky subtraction routine. 

Finally, as the sky-subtracted model is essentially a high-order surface fitting, the images obtained
so far must be reduced by additive fringing. For each slice we median-stack the maximum 
number of object-masked science frames. We then arithmetically subtract this fringe model 
from the sky-subtracted frames with the same central wavelength.

\begin{table}[ht]
\vspace*{2mm}
\caption[Statistics of sky background comparison test]{
Mean ($\mu$), standard deviation ($\sigma$) and selected quartiles ($Q_{\rm 1,\,3}$) of the 
reference-to-measured ratio of all mock sources added to the example image in Figure \ref{ringsub},
corresponding to the sky subtraction algorithms compared in this work.
}
\vspace*{-5mm}
\label{ringsub_table}      
\centering   \small                                   
\begin{center}\begin{tabular}{c c c c c}       \hline   \\                   
 Algorithm & $\mu_{\rm ratio}$ & $\sigma_{\rm ratio}$ & $Q_{\rm 1}$ & $Q_{\rm 3}$ \\   
\hline      \\[1pt]                       
    Azimuthally & 1.011 & 0.077 & 0.999 & 1.028 \\
    averaged &  &  &  & \\
    {\tt tringSub} & 1.063 & 0.059 & 1.031 & 1.074 \\
    {\tt ringsub} & 1.012 & 0.044 & 1.001 & 1.021 \\
    (this work) &  &  &  & \\
\hline                
\end{tabular} \end{center}
\end{table}
\normalsize

\subsection{Astrometry}
\label{sec:astrometry}
Astrometry calibration is necessary not only for referencing targets in a celestial 
coordinate system but also to put all the individual science images in a common spatial 
frame for mosaic assembly and coadding (see below) with accuracies of a few tenths of a 
pixel. Relative astrometry of individual science frames was referred to a custom catalogue 
whose construction was based on the \cfhtls\
Deep Field 3, 25\% best seeing (D3-25), \bandz-band data, whose internal root mean square (RMS) 
astrometric errors are 0.075 and 0.074 
arcseconds\footnote{From the Final \cfhtls\ Release Explanatory Document, available at 
\tt http://terapix.iap.fr/cplt/T0007/doc/T0007-doc.html} in equatorial coordinates ($\alpha,\delta$). 
Concerning external errors, from a comparison with 2MASS astrometric positions, the D3-25 source 
catalogue used has an accuracy of 0.23\arcsec\ in each coordinate.
Selecting all non-saturated, point-like sources (i.e.\ in this case with {\tt CLASS\_STAR} $>$ 0.95) 
up to magnitude AB=23 (i.e.\ around the \cfhtls-provided limiting magnitude of point-like sources with 
a SNR$\gtrsim$10 in science exposures) from this catalogue, we consolidated a reference 
catalogue with 892 entries with a resulting maximal RMS internal positional error of 
0.03\arcsec.

The reference catalogue was cross-matched using the {\tt IRAF ccxymatch} task,
with the list of sources extracted from each  OTELO 
mosaic component. An equal number of astrometry solutions was obtained by adopting a 
fourth-order polynomial geometry and the non-standard 
{\tt TNX}\footnote{\tt http://iraf.noao.edu/projects/ccdmosaic/tnx.html} World Coordinate
System (WCS), which has proven to be the best projection for an accurate modelling of the 
geometric distortion present in the OSIRIS instrument 
images,\footnote{From the OSIRIS Instrument User Manual; available at 
\tt http://www.gtc.iac.es/instruments/osiris}, which, in turn, is
implemented in the {\tt IRAF ccmap} task.  
The internal mean accuracy of the individual WCS solutions obtained (RMS 
residuals) is  $0.043 \pm 0.007$ arcseconds in the  standard coordinate 
($\xi,\eta$) space. Taking into account the in quadrature  upper error limit of the reference 
catalogue, internal deviations of the WCS are below 0.068\arcsec\ in both coordinates. This 
translates into a $\sim$0.27 pixel plate scale. WCS-based image registration for
mosaic assembly and subsequent coaddition  is therefore feasible within this sub-pixel regime. 

Individual components of each mosaic were then warped and referenced to each other 
(i.e.\ registered) using the {\tt IRAF mscred.mscimage} task, according to the corresponding 
high-order polynomial coefficients of the astrometric solution and conserving the instrumental 
flux per area unit. Before proceeding with mosaic assembly we created image weight maps, 
expressed in units of relative inverse variance per pixel. Weight maps influence flux error
determinations and prevent false detections due to satellite trails, diffraction spikes,
and certain instrumental signatures coming from detectors. Twenty of 216 RTF science frames
are affected by satellite trails which were represented by zero-weight traces in those maps. 
Each mosaic and its resulting weight map was afterwards assembled at sub-pixel accuracy using the 
{\tt SWarp} (v. 2.38.0) application \citep{bertin02}. 

\section{RTF data measurement}
\label{sec:tfmeasuring}

The \otelo\ survey is conceived as a blind, magnitude-limited spectral
tomography. This involves the creation of a deep detection map, hereafter \otelodeep, 
resulting from the weighted combination of the RTF registered science frames. This image is 
utlized not only to maximize the detection of all real sources in the field, keeping the 
false-positive statistics under control, but also to be used as a source of photometric 
data integrated over all the slices.

\subsection{Image coaddition}
\label{sec:coadd_ext}

There are several methods of tackling the image coaddition problem. Apart from such 
approaches as Lucky Imaging or Fourier-based methods for combining  stacks of images 
\citep{homrighausen11}, the pixelwise statistics techniques stand out among the commonly
found approach of PSF homogenization (see \citealt{zackay17} for a recent review of these
techniques). To a first approximation, we disregard  image convolution to the worst 
seeing before coadding because it alters the information contained in the image, degrades the 
PSF of almost the whole input image set,  amplifies the background noise at high frequencies, 
creates correlated artefacts. Instead, we proceed by using a two-step coaddition
scheme. First, for each slice we combined up to five of the six images dithered far enough apart 
(to ensure the rejection of diametric ghost images) with the best mean FWHM. Thus, nearly 83\% 
of all RTF science frames were coadded in the corresponding slices. The remaining ones did not 
contribute to \otelodeep\, but were naturally  taken into account in the flux 
measurements described in Section \ref{sec:products}. For this step we obtained the 36 
image stacks by using the named {\tt clipped-mean} algorithm described in \cite{gruen14} 
and implemented in {\tt SWarp}. This
algorithm has been specifically designed for rejecting artefacts present in individual
contributors to resulting stacks. The PSF differences of selected images for each slice 
stack are below the canonical requirement established by its authors (i.e.\ $\sim$10\%).
The main configuration parameters adopted for {\tt SWarp} runs are given in Table \ref{swarp}. 

The results obtained were compared with the median coadding for selected slice stacks,  
this being the most popular artefact-free model for image coadding. As expected, the instrumental 
flux recovered is quite similar in both cases, but the measured SNR is 20\% less in 
median-combined stacks than when using the clipped-mean approach. Moreover, taking into account the 
discrete number of individual frames for slices, median combining is not so efficient 
at discarding diametric ghosts and other residual artefacts of extended bright sources as 
the alternative used here. It is worth noting that slice stacks are  useful only for
producing the \otelodeep\ image  and for a data cube representation of
a given source.

The combination of the  slice images obtained must conserve the intrinsic flux variation 
of the sources over the RTF scan. For this reason, and as a final step, the resulting 36 
stacks were simply averaged using {\tt SWarp} again to obtain the \otelodeep\ 
image. All coadding products include their corresponding weight maps. In particular, we 
used the local variance in the weight map of the \otelodeep\ image to define the 
highest sensitivity survey area (i.e.\ the region of 7.5\arcmin$\times$7.4\arcmin, or 
1754$\times$1734 pixel$^2$, represented in the right panel of Figure \ref{catalogues}).
After this, and as a requirement of the source extraction procedure, all science frames 
(whether stacked or not) where trimmed to the same size as the \otelodeep\ image.

\begin{table}[ht]
\vspace*{2mm}
\caption[SWarp input]{Main configuration parameters in {\tt SWarp} to obtain
the slice representatives that contribute to \otelodeep\ image.}      
\vspace*{-5mm}
\label{swarp}     
\begin{center}\begin{tabular}{c c}         
\hline   \\                   
Parameter & Value   \\
\hline   \\                     
{\tt WEIGHT\_TYPE} & MAP\_WEIGHT\\
{\tt COMBINE\_TYPE} & CLIPPED\\
{\tt RESAMPLE} & Y \\ 
{\tt RESAMPLE\_TYPE} & LANCZOS3 \\ 
{\tt SUBTRACT\_BACK} & N \\ 
\hline      \\[1pt]                       
\end{tabular}\end{center}
\end{table}

\subsection{Source extraction and instrumental fluxes from RTF data}
\label{sec:sextraction}

Sources detected in \otelodeep\ were flux-measured on the image itself
and on each RTF frame by using {\tt SExtractor} in dual-mode. This
choice conforms to the recognized performance and ease of use of this detection tool,
particularly in the case of faint, extended sources (see \citealt{masias12} for a 
review of source detection approaches). Under this
scheme, the thresholding and final detection (segmentation) map on the \otelodeep\ is 
translated to each RTF image to be analysed. For this purpose, it was necessary to select 
the most appropriate 
configuration parameters for the {\tt SExtractor} runs, taking into account the peculiarities of the 
\otelo\ survey. The configuration parameters adopted and which differ from default ones are 
given in Table \ref{sexRTF}. The main parameters are justified in what follows. 

From the astrometric analysis and quality control of the \otelodeep\ image, the plate
scale is fixed at 0.254 \arcsec/pixel and  {SEEING\_FWHM} was set at 0.8\arcsec.

The detection and analysis thresholds adopted can be defined as multiples of the local background 
variance. High threshold values result in the missing of real fainter fluxes, but lower ones will
increase the false detection rate in the final source catalogue because of correlated
noise peaks in the \otelodeep\ image. This issue is discussed in Section 
\ref{sec:completeness}.

As demonstrated in \otelotwo, the emission line likelihood of a source is quantified
from the analysis of pseudo-spectra and the parameters derived from cross-correlation with 
ancillary data. The detection threshold was fixed to the maximum variance required  
in the \otelodeep\ background in order to recovery sources whose pseudo-spectra contain 
at least two adjacent slices with a flux $\gtrsim$\,2$\sigma_{c}$ above the 
pseudo-continuum, where $\sigma_c$ is defined as the standard 
deviation of the pseudo-continuum counts. An example of a pseudo-spectrum that dovetails this 
requirement (concretely, the [NII]$6584\AA$ emission line) is represented in
Figure \ref{pseudospectra}. The criteria that lead to the practice of this hypothesis for ELS 
selection are specified in Section \ref{sec:elgselection}. 

The detection/analysis threshold was obtained by isolating three  regions - background 
residual only - of 30$\times$30 pixels on the 
\otelodeep. Such cutouts were extracted from the slice images used to obtain the
\otelodeep\ image to create sets of 36 stamps each. For each set we then added
point-like artificial sources (as described in Section \ref{sec:rings}) with SNR $\simeq$ 3
on selected pairs of slice image regions, leaving intact the remaining slices of each collection.
The flux of each artificial source was carefully scaled to 3$\sigma$ above the background
of the selected slice cutout. This procedure was repeated six times in each collection. After 
this, the 36 cutouts of each collection and realization were averaged as \otelodeep. 
The detection/analysis threshold relative to the background of each averaged image was 
decreased in successive steps of 0.1 units until recovering the mock source flux. By
linearly fitting the input SNR against the recovery thresholds, we finally
obtained the detection/analysis threshold that exactly satisfies the previous 
hypothesis. The values found after this procedure are in agreement with the 
detection (= analysis) threshold adopted, for example, by \cite{jones02} and \cite{galametz13}
for faint source extraction.

From this procedure, we also determined the minimum area above the threshold that a true 
detection should have. The {\tt SExtractor} manual suggests setting from 1 to 5 pixels.
We fixed it consistently at 4 pixel, which is equivalent to a circular area with radius 
$\sim$0.5*{\tt SEEING\_FWHM}. 

Depending on count peaks and neighbouring fluctuations in a raw detection, {\tt SExtractor} 
 hierarchically splits the object into smaller (child) ones. The deblending threshold is set as 
powers of 2 (default value is 32) and constitutes the allowed number of levels in this 
object hierarchy, whilst the minimum flux ratio between the objects at the extremes of a 
decomposition is defined by {\tt DEBLEND\_MINCONT}. After educated tests
we adopt the deblending parameters found by \cite{annunziatella13} from their analysis
of source extraction software. In the same way we proceeded with the background estimation
parameters (i.e.\ mesh gridding map and background smoothing factor), except that we leaned
towards a local estimate of the background around a given detection rather than a global 
one in order to take into account the sky noise gradient on images with
the radius to the optical centre. Image filtering after background fluctuations was
done by means of a `top-hat' function,  optimized to  faint, low-surface 
brightness source detection. 

Instrumental fluxes measured in the \otelodeep\ image were directly converted into AB 
magnitudes. These are referred to below as \oteloint\ magnitudes. Using the effective gain and exposure 
time, and the estimation of the zero-point magnitude corresponding to the synthetic spectral 
response of the \otelodeep\ image, we obtained a {\tt MAG\_ZEROPOINT} of 30.504 mag. 

Once the configuration parameters of {\tt SExtractor} were obtained, the RTF data flow
passed from the image to the catalogue domain: the standard Kron ({\tt AUTO}), isophotal
({\tt ISO}) and aperture ({\tt APER}: 2\arcsec\ and 3\arcsec\ in diameter) instrumental fluxes, $F_{\rm ADU}$, 
of the 11237 raw sources detected on \otelodeep\ and their errors were measured in the 216 individual
RTF frames, apart from position, source image geometry (including isophotal area, $A_{ISO}$), and 
the corresponding extraction flags. Flux measurement uncertainties were determined by means 
of the expression

\begin{equation}
\label{eq:flux_err_tspect}
\Delta F_{\rm ADU} = \sqrt{A_{\tt ISO,\ APER}\ \sigma^2 + F_{\rm ADU} / g_{\rm \,eff}}, 
\end{equation}

\noindent where $A_{\tt ISO,\ APER}$ are the aperture areas (isophotal or apertures respectively) 
in pixels, $\sigma$ is the source local background RMS, and $g_{\rm \,eff}$ is the effective 
gain in e$^-$ ADU$^{-1}$, depending on the measured image (\otelodeep\ or individual 
RTF frame).

As described in Section \ref{sec:fluxcal}, the individual instrumental fluxes must be first converted 
into physical units and  an effective wavelength assigned to them before generating the provided 
pseudo-spectra.

\begin{table}[ht]
\vspace*{2mm}
\caption[SExtractor input]{Main configuration parameters in {\tt SExtractor} for
instrumental flux measurement of RTF data using \otelodeep\ as detection image
in {\it dual mode}.}       
\vspace*{-5mm}
\label{sexRTF}     
\centering     \small                               
\begin{center}\begin{tabular}{c c}          
\hline   \\                  
Parameter & Value   \\  
\hline   \\                     
{\tt DETECT\_MINAREA} & 4 pixels\\
{\tt THRESH\_TYPE} & RELATIVE\\ 
{\tt DETECT\_THRESH} & 0.73$\sigma$ \\ 
(={\tt ANALYSIS\_THRESH}) &  \\ 
{\tt FILTER\_NAME} & {\tt tophat\_3.0\_3x3.conv} \\ 
{\tt DEBLEND\_NTHRESH} & 64 branches \\ 
{\tt DEBLEND\_MINCONT} & 0.001 fraction\\ 
{\tt CLEAN} & Y \\
{\tt CLEAN\_PARAM} & 1.0 \\
{\tt WEIGHT\_TYPE} & MAP\_WEIGHT \\
{\tt PIXEL\_SCALE} & 0.254 \arcsec/pixel\\
{\tt MAG\_ZEROPOINT} & 30.504 \\ 
{\tt BACK\_TYPE} & LOCAL \\ 
{\tt BACK\_SIZE} & 64$\times$64 pixels \\ 
{\tt BACK\_FILTERSIZE} & 8 pixels\\ 
\hline      \\[1pt]                        
\end{tabular}\end{center}
\end{table}

\subsection{Completeness and contamination}
\label{sec:completeness}

\begin{figure}[h]
\centering
\includegraphics[angle=-90,width=\linewidth]{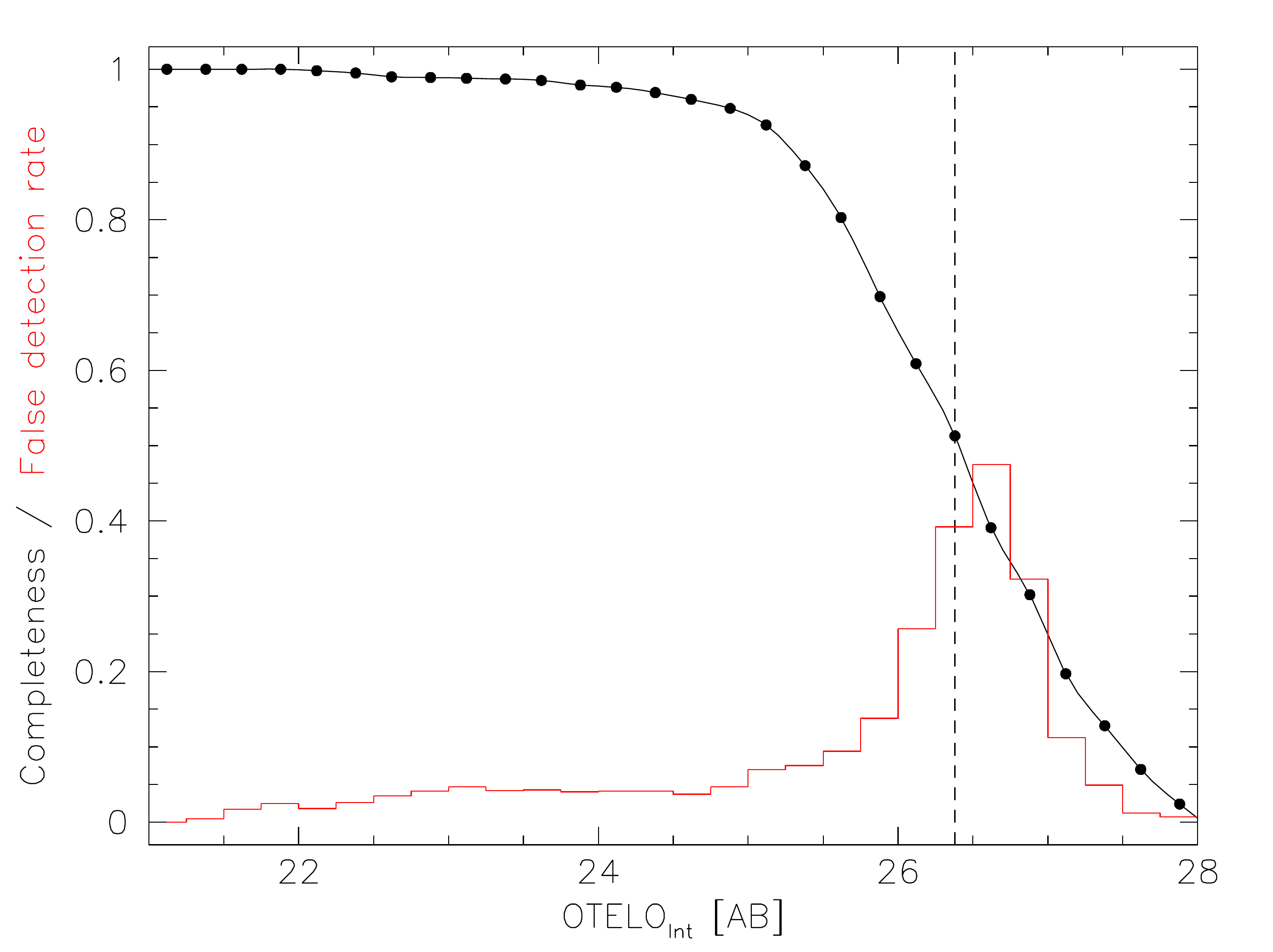}
\caption[OTELO-deep completeness]{Detection completeness obtained from averaging the
ratio of recovered-to-catalogued artificial sources in random-shoot trials on the
\otelodeep\ image as described in text. A spline black curve joins averages over \oteloint\
fluxes in 0.25 magnitude bins. The dashed line at \oteloint=26.38 [AB] corresponds to the
50\% completeness magnitude. The histogram traces the false detection rate (FDR)
for the \otelodeep\ image, corresponding to an estimate of the spurious from correlated 
noise.}
\label{NBcompleteness}
\end{figure}

The completeness profile of source detections was obtained by recovering and comparing fluxes
of artificial sources added to the \otelodeep\ image in a similar procedure to that described in
Section \ref{sec:rings}. Five independent catalogues of mock sources, with a power-law
distribution limited to magnitude 28 (i.e.\ an upper bound of the \otelodeep\ limiting magnitude
or AB = 27.8: see Section \ref{sec:catalogue}), were randomly dropped into the \otelodeep\ image 
by using the {\tt IRAF mkobject} task in an equal number of realizations. Mock data photometry was
performed identically to the procedure described above for the observed \otelo\ sources.
The completeness estimation is defined by
the average over the five realizations of catalogued-to-recovered source number as a
function of the binned \oteloint\ flux. Figure \ref{NBcompleteness}
shows that \oteloint\ data are complete up to $\sim$24 magnitudes and 50\% completeness flux is
reached at \oteloint = 26.38 [AB]. As is evident, this completeness estimation does not take into 
account the fraction of lost sources close to bright star imprints in the image, or
 detections that could be favoured by gravitational lenses.

As pointed out above, the nature of the so-called false-positives or false detections
in a deep astronomical image is diverse. Despite having taken actions to reduce the risk of 
such fake objects by replacing bad pixels, adopting the clipped-mean algorithm for slice stacking, 
and building individual weight maps, the highest frequency and most spatially homogeneous source 
of false detections (FD) is constituted by the correlated noise spikes.

Using {\tt SExtractor} with default parameters in \otelodeep\ and its negative image 
(i.e.\ source masked \otelodeep\ $\times$ -1), it is possible to obtain a rapid estimate of false 
source count statistics. This negative image is not only a fair statistical representation of the 
residual background of the coadded image to be measured but also retains 
fringing residuals, particularly the imprints of the incomplete correction of the scattered 
haloes from bright sources in the background subtraction or {\it dithering holes}. In our case,
the asymmetry of the background model obtained from {\tt SExtractor} discourages
this approach. Instead of this, we created a set of images that mimic the \otelodeep\ background,
which differ in the random noise pattern. Each simulation was created starting
from the \otelodeep\ background model mentioned above and the corresponding background 
variance map, by means of a custom {\tt IRAF} script. The effective gain and readout noise 
of  \otelodeep, as well as the sky photon map which explain the background variance 
one, are the other inputs of the task. A comparison of the pixel distributions of model with
each mock background image gives a mean Kolmogorov--Smirnov probability of 0.98. Such statistically
identical but independent images were then measured using {\tt SExtractor} with exactly the same
configuration used for \otelodeep. All the detections were regarded as spurious and compared in number
per magnitude bin with all the \otelodeep\ detections. The false detection rate (FDR) is defined 
as the ratio FD/(FD+TD), where TD are the true detections per bin. All FRDs obtained from each simulation
were averaged (with a mean absolute deviation of 0.01) and plotted as a function of the \oteloint\ 
in Figure \ref{NBcompleteness}. The FDR provides a measure for discriminating between spurious and correlated 
noise, and therefore a lower bound - close to the total - of the FD in \otelo. 

In summary, from the raw 10487 sources up to AB = 26.5 (i.e.\ the upper limit of the simulation
bin that contains the 50\% completeness flux of \oteloint), the potential FDs amount to 1150 objects. 
For fainter magnitudes, nearly 69\% of the sources are possibly spurious. Thus, the total number of 
FDs is close to 1650 raw entries in the
catalogue. To put it another way, in the 22 $<$ \oteloint $<$ 25 range, the probability that
an object qualifies as spurious is about 4\%. Between the latter threshold and the 
50\% completeness magnitude, this probability doubles every 0.5 magnitude.

\subsection{Flux calibration stars}
\label{sec:standards}

The colour-selected F8 sub-dwarf stars in the \otelo\ FoV and the secondary 
spectro-photometric standard used to calibrate them in flux, all referred to in 
Section \ref{sec:observations}, were  reduced in the standard way using {\tt IRAF.noao} 
spectral reduction packages. All targets were observed 
in OSIRIS long-slit mode with a red grism at resolution $\sim$500. A slit of
1.5\arcsec\ was used (with seeing conditions better than 0.9\arcsec) to achieve 
the maximum flux accuracy.
According to $r$(AB) magnitude of the targets, total integration times were
generous enough to reach an SNR$\simeq$10 between 5500 and 9500 $\AA$.

\begin{figure}[h]
\centering
\includegraphics[angle=-90,width=\linewidth]{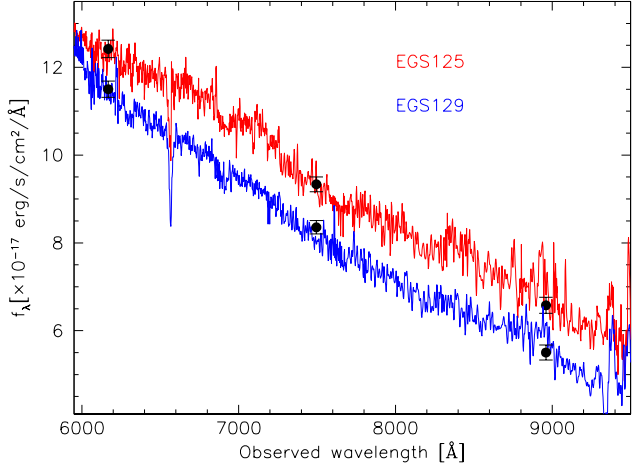}
\caption[Calibration stars]{Spectra of the calibration stars present in the FoV of \otelo\ 
EGS pointing, as is highlighted in Figure \ref{catalogues}. These spectra were used in turn to 
estimate the total efficiency of the system and its deviations depending on the particular 
conditions of each  RTF observation. As a check on flux consistence, overlapped black 
dots correspond to PSF fluxes in \bandr-, \bandi-, and \bandz-band from SDSS-DR12 for these stars.}  
\label{calstars}
\end{figure}

Bias frames were combined and subtracted from science spectra using the {\tt imcombine}
and {\tt ccdproc} tasks. High-count flat-field images were combined and the result
corrected by the fitting of the continuum lamp spectrum ({\tt flatcombine} and
{\tt response}). The sky spectrum (sky flats from science images) was averaged
and a sky flat was interactively fitted by a spline function using the  
{\tt illumination} task. After correcting all science frames by illumination,
wavelength calibration was carried out using {\tt transform}. To this end, line 
identification of Ne and HgAr lamp exposures obtained with the same rotator angle as the 
science exposures was dumped on to a database ({\tt identify}) and interactively analysed 
using {\tt fitcoords}. Once the sky background were subtracted from individual 2D spectra,
they were combined and collapsed in the spatial direction. A dereddened sensitivity 
curve for flux calibration was obtained from the standard star (HD126511) spectrum and
the instrumental flux of the \otelo\ calibration stars were converted into physical ones.
Figure \ref{calstars} shows the reduced spectra resulting from this procedure. The mean flux 
error of spectra is better than 6\%. The consistency of the  flux density obtained with 
the SDSS-DR12 photometry would make it possible to apply the spectro-photometric flux calibration 
procedure used in the SDSS to the case of TF observations that contain such spectral-type
stars in the FoV.

\subsection{Wavelength and flux calibrations}
\label{sec:fluxcal}

The first step in the conversion of instrumental to physical fluxes consists in 
deriving the total efficiency $\epsilon\equiv\epsilon\,(\lambda_{\rm ob},{\rm i},{\rm CCD})$ of the system 
(telescope, optics, and detector), defined as the ratio of the measured-to-reference
flux $F_{\rm m}(\lambda_{\rm ob},{\rm i},{\rm CCD})/F_{\rm r}(\lambda_{\rm ob},{\rm CCD})$ using the 
two on-purpose calibration 
stars, one for each detector. As the efficiency depends on the observed wavelength, the 
$i$-science frame and the detector of the OSIRIS mosaic (CCD=1,2), isophotal fluxes 
$F_{\rm m}(\lambda_{\rm ob},{\rm i},{\rm CCD})$ were measured in each $i$-science frame, accompanied by precise 
wavelength determinations at the position of both stars with 
respect to the optical centre using Equations \ref{eq:radep} and \ref{eq:third}. It is necessary to 
emphasize that small variations in the $i$-telescope pointings and the effects of dithering 
on RTF observations are taken into account in the observed wavelength calculation,
not only for calibration stars but for all remaining sources in the field. 

The reference fluxes are measured at the observed wavelength by convolving the corresponding
spectra obtained from the process described in Section \ref{sec:standards} with the Airy profile
approximation given by Equation \ref{trans}, by setting $\lambda_0=\lambda_{\rm ob}$ and integrating.
The measured instrumental fluxes in counts are then converted into physical ones 
(erg\,s$^{-1}$\,cm$^{-2}$) for each calibration star by using:

\begin{equation}
\label{eq:eff}
F_{\rm \,m}(\lambda_{\rm ob},i)=\frac{g\,K(\lambda_{\rm ob},i)\,E_{\gamma}(\lambda_{\rm ob})}{t\,A_{\rm tel}}F_{\rm ADU}(\lambda_{\rm ob},i),
\end{equation}

\noindent where $g=0.95$ e$^-$ADU$^{-1}$ is the CCD gain, $E_{\gamma}(\lambda_{\rm ob})$ is the 
energy of a photon in erg, $t$ is the exposure time in seconds, $A_{\rm tel}$ is the effective 
collection area of the telescope in cm$^2$, and $K(\lambda_{\rm ob})$ is the 
correction for atmospheric extinction, given by

\begin{equation}
K(\lambda_{\rm ob})=10^{0.4\,k(\lambda_{\rm ob}) \, \left\langle \chi \right\rangle},
\end{equation}

\noindent which depends on the extinction coefficient $k(\lambda_{\rm ob})$ and the mean airmass 
of the observation. In our case, we estimated  $k(\lambda_{\rm ob})$ by fitting the extinction 
curve of 
La Palma\footnote{\tt http://www.ing.iac.es/Astronomy/observing/\\manuals/ps/tech\_notes/tn031.pdf} 
in the wavelength range defined in the survey. The uncertainty in the efficiency is defined by 
the sum in quadrature of the errors in the measured and the reference flux (Sec.\ 
\ref{sec:standards}). 

\begin{figure}[h]
\centering
\includegraphics[angle=-90,width=\linewidth]{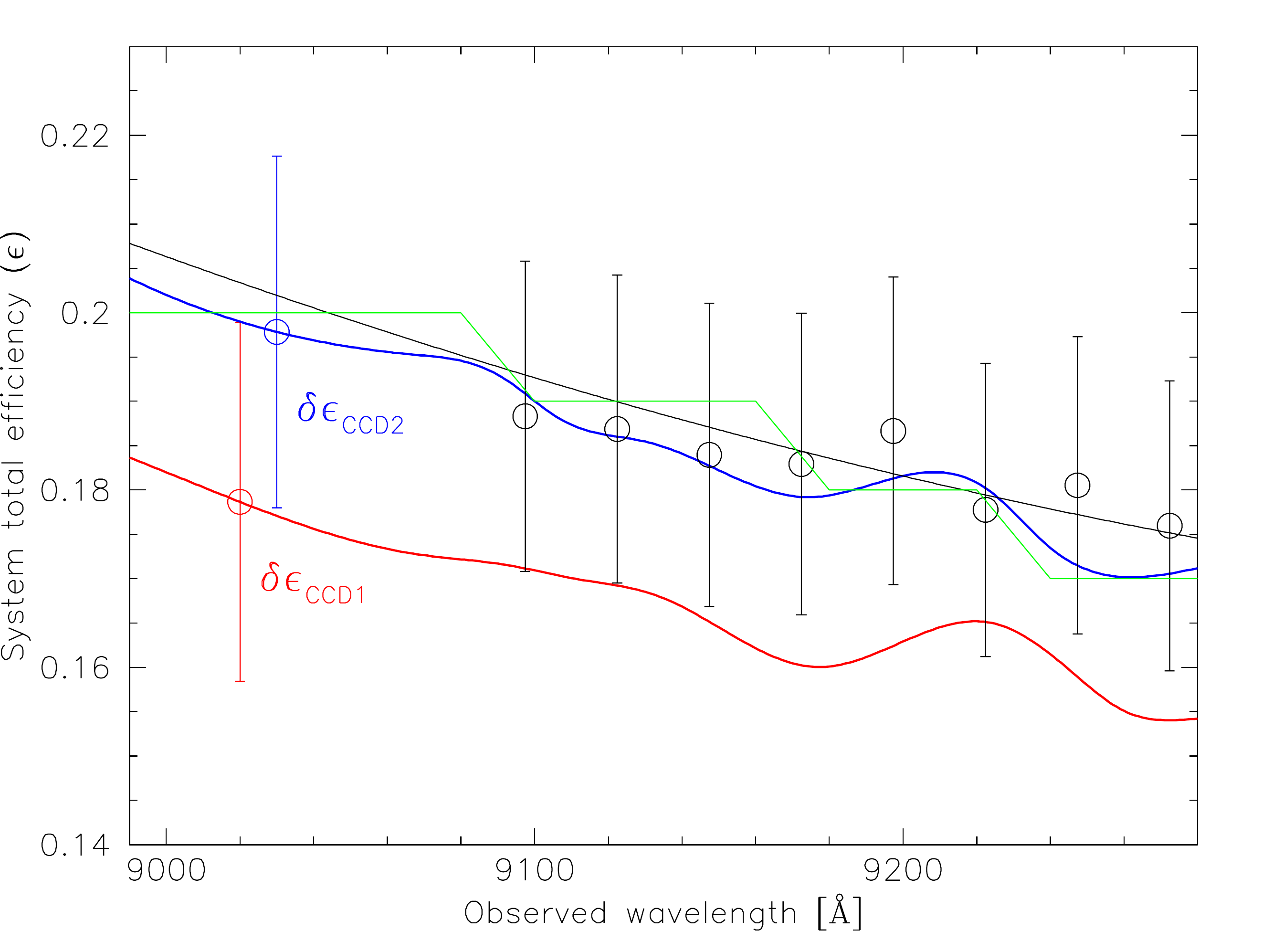}
\caption[OTELO efficiency]{Spline fitting of the total efficiency of the telescope + RTF + OSIRIS 
camera system for CCD1 (red curve) and CCD2 (blue curve), based on the OTELO survey calibration stars and 
averaged for each slice. The black continuous line is the polynomial fitting
from the GTC \cite{cabrera14} of the values represented by the open black circles, whilst the 
green steped one represents the efficiency tabulated in the OSIRIS SNR 
Calculator.\footnote{\tt http://gtc-phase2.gtc.iac.es/science/\\OsirisETC/html/Calculators.html} 
These spectra were in turn used 
 to estimate the total efficiency of the system and its deviations depending on the 
particular conditions of each individual RTF observation. Overlapping black dots correspond to 
PSF fluxes in \bandr-, \bandi-, and \bandz-band from SDSS-DR12 for these stars. The flux differences with 
those from SDSS are within the mean error.}
\label{efficiency}
\end{figure}

The total efficiency of the system (telescope + RTF + OSIRIS camera) has recently been extensively 
sampled, but only using the CCD2 to
obtain spectro-photometric data for such calibrations. Figure \ref{efficiency} shows the general
trend of our efficiency estimates from measurements in each individual science frame. 
They are in accordance with those obtained by \cite{sanchezportal15} for the \ha\ imaging of a galaxy 
cluster at \zred\ = 0.395 using the RTF in the framework of the GLACE survey, as well as with the 
efficiency estimates compiled by \cite{cabrera14} for the same device. A systematic 
differential sensitivity of a factor $\sim$1.12 between both detectors in favour of CCD2 was
noted. The behaviour of our efficiencies was fitted by spline and conveniently sampled
to perform the calibration at the observed wavelength of each source and RTF tuning as 
$\epsilon(\lambda,{\rm CCD})$. 

Once  the $\epsilon$-spectrum for each detector becomes available, the next step is to convert the
instrumental flux of each source $s$, measured with a given CCD, to a vphysical flux density in 
CGS units (ergs\,s$^{-1}$\,cm$^{-2}$\,\AA\ $^{-1}$) by means of the expression:

\begin{equation}
\label{eq:flux}
f(\lambda_{\rm ob},CCD)_s = 
\frac{g\,K(\lambda_{\rm ob})\,E_{\gamma}(\lambda_{\rm ob})}{t\,A_{\rm tel}\,\delta\lambda\,_{\rm e}\,\epsilon(\lambda_{\rm ob},CCD)}F_{\rm ADU}(\lambda_{\rm ob},s),
\end{equation}

\noindent where $\epsilon(\lambda_{\rm ob},{\rm CCD})$ is the total efficiency at
$\lambda_{\rm ob}$, $\delta\lambda\,_{\rm e}$ is the effective passband width in \aa\ 
(Eq.\ \ref{eq:effwidth}), and the 
remaining terms are as in Equation \ref{eq:eff}. Estimation of flux error takes into account the 
efficiency error in quadrature, depending on the detector and the source flux measurement 
uncertainty as described in Section \ref{sec:sextraction}.

\subsection{RTF outputs}
\label{sec:tunableproducts}

Two products result from the RTF data reduction: a raw set of 11\,237 objects detected 
in the \otelodeep\ image and an equal number of calibrated pseudo-spectra. As described in
Section \ref{sec:catalogue}, this source list is complemented with ancillary data
to produce the \otelo\ catalogue. Even though this catalogue contains integrated fluxes expressed 
in different parameters (Kron, isophotal, apertures, and more sophisticated ones, as described
below), we adopted the isophotal flux measured in individual RTF frames for pseudo-spectra building 
as the best approximation to corrected aperture flux in crowded fields. When  isophotal 
flux pseudo-spectra of the standard stars of \otelo\ are compared with the convolution of their spectra, 
smaller deviations ($\leq$4\%) than those using any other photometric parameter are revealed.

The procedure for constructing the \otelo\ pseudo-spectra is outlined below. For each source 
detected, we have a vector f$_{\rm i}$($\lambda_{\rm ob}$) of $i=216$ effective wavelengths 
and physical fluxes with their errors. We should group them into $N = 36$ wavelength windows or 
cells of $\delta\lambda\,_{\rm FWHM}/2$ = 6 \AA\ width (i.e.\ the scan step) and combine the 
individual fluxes in each window. In practice, this is possible as long 
as the mean angular distance of the source to the optical centre is smaller than the size of
the Jacquinot spot (i.e.\ 
a nearly monochromatic region over which the change in wavelength does not exceed by a
factor $\sim\sqrt{2}$ $\delta\lambda\,_{\rm FWHM}$), or $\lesssim$1\arcmin\ for the \otelo\ 
observing design. 

As a consequence of the observing strategy described in Section \ref{sec:observations},
 concerning the dithering pattern (which, in practice, also includes  small telescope pointing
deviations) and the wavelength change bluewards with the distance to the optical
centre for a given nominal RTF tune (Eq.\ \ref{eq:radep}), the wavelength distribution
of the vector f$_{\rm i}$($\lambda_{\rm ob}$) not only moves bluewards as the mean distance of the 
source to the optical centre increases, but the f$_{\rm i}$($\lambda_{\rm ob}$) obtained from observations
could be best distributed into more than $N = 36$ wavelength windows. In other words, as the source 
is farther from the optical centre, the f$_{\rm i}$($\lambda_{\rm ob}$) fluxes obtained at the same 
central wavelength tuning could correspond to different but adjacent slices. Consequently,
an \otelo\ pseudo-spectrum could have $N$ or slightly more data points, except for local anomalies 
related to the gap between detectors.

\begin{figure}[h]
\centering
\includegraphics[angle=0,width=\linewidth]{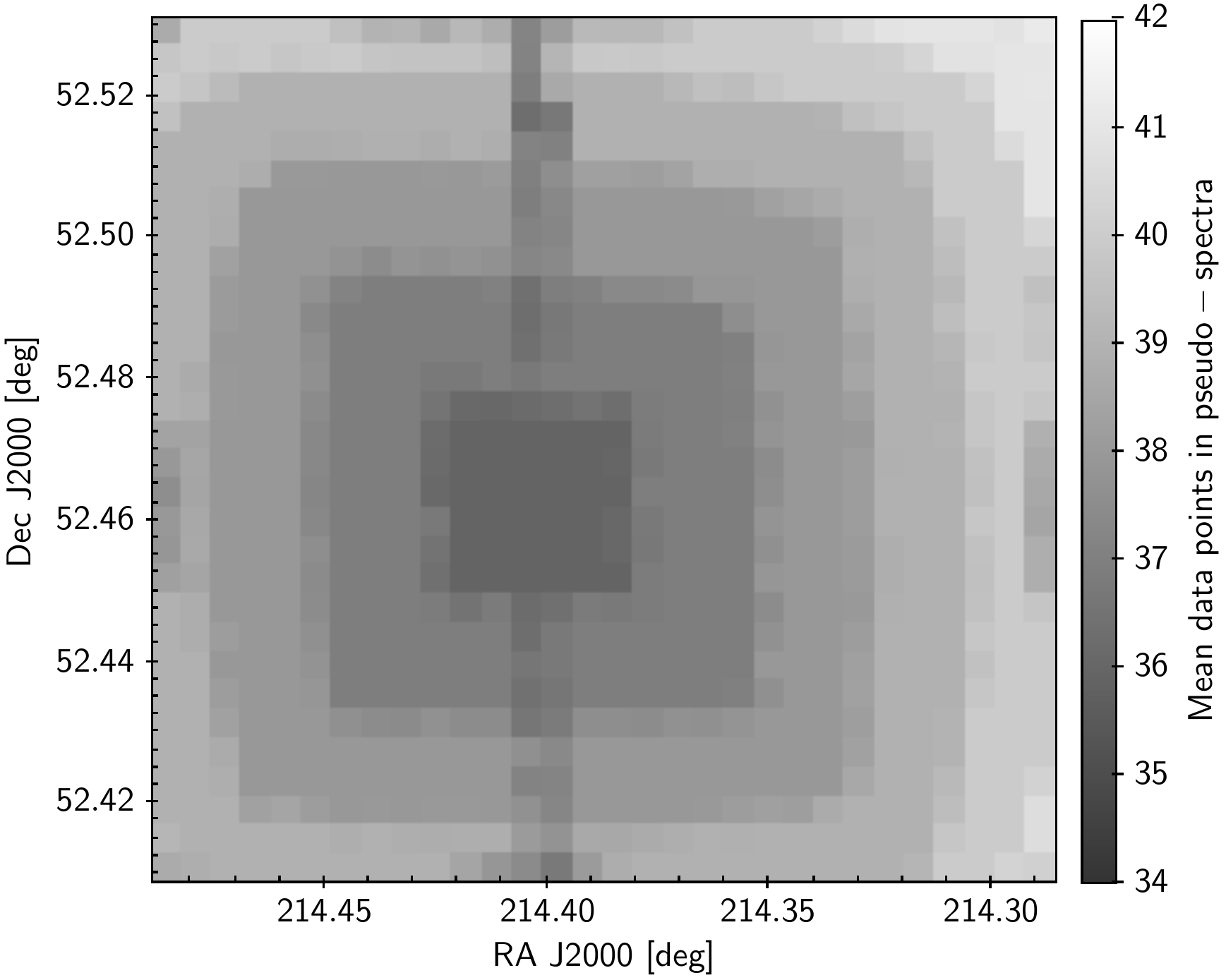}
\caption[Distribution of data point number in pseudo-spectra]{Spatial distribution of the mean number 
of data points in \otelo\ pseudo-spectra. The departure from concentric rings around the projected
position of the RTF optical centre is an effect of the cross-shaped dithering pattern of individual RTF 
observations. We note that, for sources located in the vicinity of the CCD gap, the number of data
points could be even smaller than the nominal number of slices: only 55 catalogued sources are
affected by this contingency.}
\label{nslices}
\end{figure}

Taking  these effects into account, we used a custom code to tackle the pseudo-spectra assembly.
The f$_{\rm i}$($\lambda_{\rm ob}$) vectors of each source 
are sorted in wavelength and, taking the first one as initial guess, the algorithm searches 
for accumulation points of data in $\lambda$-windows or cells in such a form that the maximum wavelength 
difference of a given \textit{n}-tuple of data is smaller than the scan step and finds the 
optimal, equal-spaced wavelength sequence for each source. Each element of this sequence is 
a wavelength label of the resulting 
pseudo-spectrum. The \textit{n}-fluxes associated with each wavelength are then combined using a weighted
mean scheme, using the inverse square of the flux error as a weighting factor. Finally, the
instrumental flux of the resulting pseudo-spectrum is converted into physical flux density units
and then formatted. Figure \ref{nslices} illustrates the dependence of the 
number of data points of a pseudo-spectrum as a function of the distance of the sources
to the optical centre. 

\section{The OTELO catalogue}
\label{sec:catalogue}

\begin{figure}[h]
\centering
\includegraphics[angle=-90,width=\linewidth]{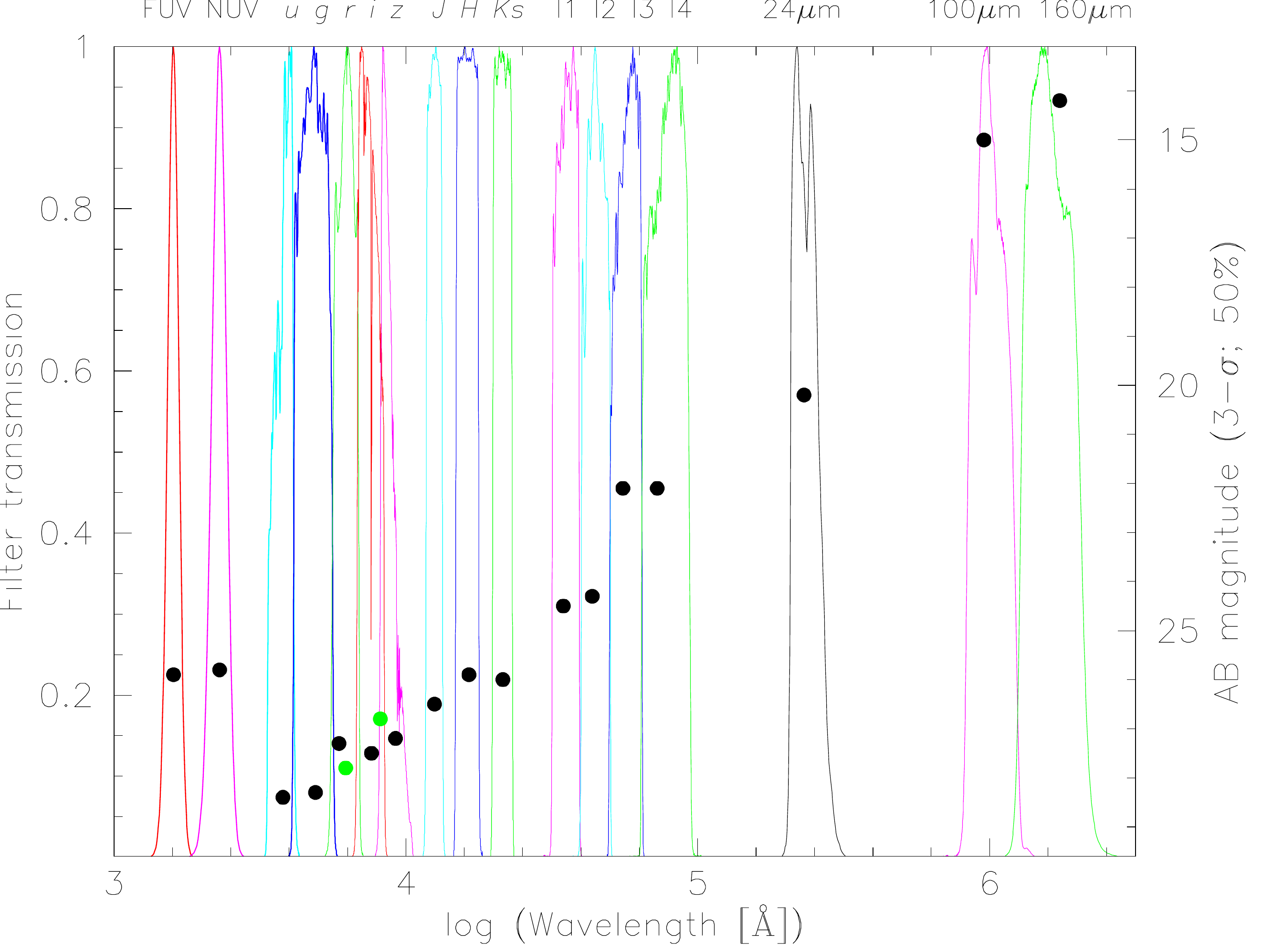}
\caption[Bands included in the final OTELO's multi-wavelength catalogue]{Bands included
in the final \otelo\ multi-wavelength catalogue. From right to left: \galex\ FUV and NUV,
optical \bandu, \bandg, \bandr, \bandi, \bandz\ from \cfhtls; \bandj, \bandh, \bandk\ from
\wirds, I1, I2, I3, I4 corresponding to
channels 1 to 4 of \spitzer/\irac, and far-infrared 24$\,\mu$m from \spitzer/\mips,
100$\,\mu$m \& 160$\,\mu$m from PEP/\herschel. The profiles of \herschel/\spire\ are not
shown. As a reference, black dots represent limiting magnitude (3$\sigma$,
$\sim$50\% completeness) compiled from the literature or, in the worst case, measured directly
in the corresponding field data. Green dots represent the limiting magnitude of
\acs606\ and \acs814\ data, whose profiles are not represented here for the sake of clarity.}
\label{bands}
\end{figure}

Apart from the pseudo-spectra, the main product of the \otelo\ survey is a raw source catalogue
extracted from \otelodeep\,
suitably complemented with X-ray, UV, optical, NIR, MIR, FIR, and spectroscopic data.
The process that leads to this subject is divided into two steps: we prepared a {\it core} 
catalogue, composed of ancillary data directly measured in images whose PSF FWHM does 
not exceed that of the \otelodeep\ by more that a factor $\sim$2, regardless the image scale factor, and 
with similar limiting fluxes. Apart from essential photometric data, the {\it core}
catalogue contains identification and position coordinates of all sources measured in the
\otelodeep\ image. Secondly, we cross-correlated the {\it core} catalogue 
with complementary data on the region surveyed to obtain the \otelo\ multi-wavelength catalogue. 
Figure \ref{bands} shows the main bands included in the latter and the respective limiting magnitudes.
The following sections are devoted to explaining this process. 

\subsection{Optical and NIR ancillary data}
\label{sec:ancillary}

Ancillary data that meet the definition of the {\it core} catalogue are composed of
optical images from the \cfhtls\ survey (T0007 Release), \acs, and NIR data from the WIRcam 
Deep Survey (\wirds, Release T0002)\footnote{\tt http://terapix.iap.fr/rubrique.php?id\_rubrique=261}. 
The \cfhtls\ survey data correspond to the imprint of \otelodeep\ on the Deep-3 field (1 $\times$ 
1 sq.deg.; 0.186 \arcsec/pixel), composed of 24 $u$, $g$, $r$, $i$, and $z$ stacks that reach a limiting magnitude 
25 to 26 (AB; 80\% completeness in extended sources). \acs\ images in the F606W and F814W bands of 
the EGS were obtained as part of the GO programme 10134 \citep{davis07}. Data were reduced, 
mosaicked and pixel-resampled from native 0.03 to 0.1 \arcsec/pixel 
by A. Koekemoer.\footnote{\tt http://aegis.ucolick.org/mosaic\_page.htm} The \bandj, \bandh , and \bandk\
bands public data from \wirds\ is a sub-section of the \cfhtls\ deep fields. At 50\% completeness for
point sources the survey reaches a limiting magnitude between 24 and 25 (AB), making it one of 
the deepest homogeneous surveys in the NIR to date. Further detail can be found in \cite{bielby12}.

Native optical and NIR images and their weight maps used for this purpose were initially trimmed 
to the \otelodeep\ imprint plus a margin of 1\arcmin\ in each dimension. A pixel homogenization
to the \otelodeep\ image, conserving integrated flux per area unit, was carried out through
{\tt SWarp}. Using the reference catalogue and procedures mentioned in Section 
\ref{sec:astrometry}, we tweaked on the existing WCS calibration of each image. Image preparation 
concluded with their spatial registration and a final trimming of the \otelodeep\ image using 
{\tt IRAF wregister}. After that, an accurate mean PSF of each image was fitted by using the 
{\tt PSFEx} (v. 3.17.1) application \citep{bertin11}. 
Table \ref{core_catalogue_summary} contains the main properties of the images used as input to
the {\it core} catalogue. The mean PSF of the set oscillates between $\sim$0.7\arcsec\ and 1\arcsec. This
variation could be a critical issue when robust photometry across the bands involved is required. 

\begin{table*}[ht]
\vspace*{2mm}
\caption[Summary Core Catalogue]{\otelo\ survey: Properties of the image set used to
build the {\it core} catalogue. Filter profiles and limiting magnitudes are represented
in Figure \ref{bands}.
}
\vspace*{-5mm}
\label{core_catalogue_summary}
\begin{center}
\begin{tabular}{c c c c c c c}
\hline   \\
Survey & Filter & Filter & Filter & Limiting & Photometric & PSF \\
Image & Name & $\lambda_{\rm eff}$ & FWHM & Magnitude\tablefootmark{\ (a)} & zero-point& FWHM \\
 &  & [$\AA$] & [$\AA$] & [AB-mag] & [AB-mag] & [\arcsec] \\
\hline      \\[1pt]                       
\otelodeep\ & \otelocustom\ & 9175.0 & 229.4 & 27.8 & 30.504 & 0.87 \\
\cfhtls & \bandu & 3881.6 & 574.8 & 30.2 & 30.000 & 1.00 \\
\cfhtls & \bandg & 4767.0 & 1322.4 & 30.6 & 30.000 & 0.91 \\
\cfhtls & \bandr & 6191.7 & 1099.1 & 30.3 & 30.000 & 0.86 \\
\cfhtls & \bandi & 7467.4 & 1316.1 & 29.9 & 30.000 & 0.82 \\
\cfhtls & \bandz & 8824.0 & 998.4 & 28.9 & 30.000 & 0.77 \\
\acs\tablefootmark{\ (b)} & F606W & 5810.1 & 1776.5 & 29.2 & 26.486 & 0.87 \\
\acs\tablefootmark{\ (c)} & F814W & 7985.4 & 1876.7 & 28.6 & 25.937 & 0.90 \\
\wirds & \bandj & 12481.5 & 1547.9 & 27.4 & 30.000 & 0.86 \\
\wirds & \bandh & 16158.2 & 2885.7 & 26.8 & 30.000 & 0.79 \\
\wirds & \bandk & 21337.8 & 3208.6 & 26.8 & 30.000 & 0.81 \\
\hline  
\end{tabular}
\end{center}
\tablefoot{}
\tablefoottext{a}{Defined as 5$\times$ the median of background noise measured on the image.} \\
\tablefoottext{b, c}{Convolved to the \otelodeep\ PSF.} 
\end{table*}
\normalsize

\subsection{PSF-matched photometry}
\label{sec:dmodel}

A number of software utilities have been developed to obtain
homogeneous and reliable photometry data from multi-wavelength, combined ground- and 
space-based surveys with mixed bandwidths and variable PSF. Applications based
on real or model source profiles (including PSF models) constitute the state of the art in these 
kinds of tools, among which are included for optical/NIR data {\tt ColorPRO} \citep{coe06}, 
{\tt PyGFit} \citep{mancone13}, and {\tt T-PHOT} \citep{merlin16}. We explored different approaches 
to obtain reliable total fluxes and colours from the image set that contribute to the \otelo\ 
{\it core} catalogue in a quick and accurate fashion, using the \otelodeep\ image as
the origin of the source detection, and with a single photometric parameter.

\begin{figure}[h]
\centering
\includegraphics[angle=0,width=\linewidth]{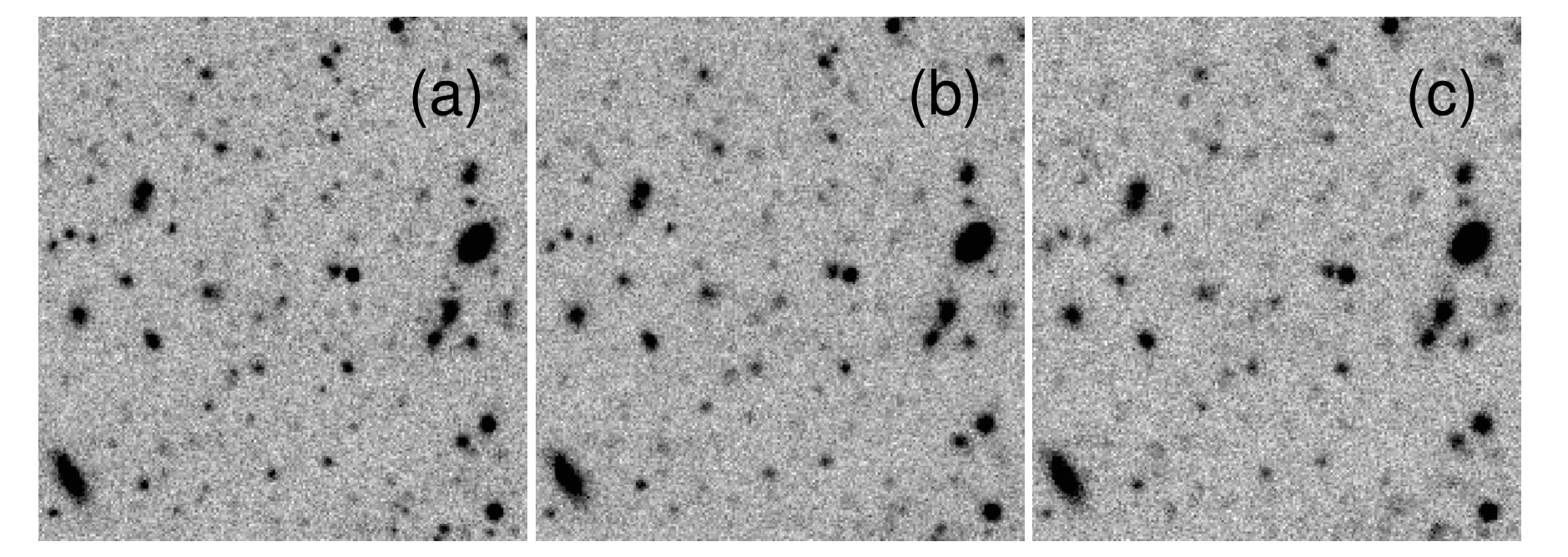}
\caption[Simulations z-band]{A detail (10\arcmin$\times$10\arcmin) of three mock images created with 
{\tt SkyMaker} from a \bandz-band simulated catalogue obtained using the {\tt Stuff} application.
Images differ only in the FWHM of their adopted PSF: (a) 0.7, (b) 0.9, and (c) 1.1 arcseconds These 
images were 
used to study the behaviour of the {\tt SExtractor} application to obtain photometric parameters, 
including PSF-model photometry, under controlled but conditions close to those of the images
used for the assembly of the \otelo\ {\it core} catalogue. More details in text.} 
\label{simulz}
\end{figure}

Inspired by the results of the analysis of the {\tt SExtractor} PSF-model photometry 
described by \cite{annunziatella13} and the simulation framework to model data of the Dark 
Energy Survey prepared by \cite{chang15}, we carried out our own tests to determine whether the
{\tt DETMODEL} parameter fulfils previous requirements within a reasonable error budget, compared 
with the most used ones. We built an extensive artificial source catalogue in the \bandz-band 
(hereafter, \ztrue) mimicking 
the corresponding real data for the {\it core} catalogue, as much far as the {\tt Stuff} 
(v. 1.26.0) application \citep{bertin09} allowed us. This condition includes identical pixel and image sizes, 
background noise level, effective gain, and photometric zero-point. Such a catalogue was used as 
input to the {\tt SkyMaker} (v. 3.10.5) software \citep{bertin09} to create three images that differ 
only in their PSF FWHM (i.e.\ 0.7\arcsec, 0.9\arcsec ,  and 1.1\arcsec, which cover the mean FWHM range of the 
real images considered here). Figure \ref{simulz} shows cutouts of these 
simulated images. We obtained the PSF model of each \bandz-band arbitrary image using {\tt PSFEx} and 
then recovered the artificial source fluxes using {\tt SExtractor} in {\it dual-mode} with the 
intermediate FWHM one as detection image.

\begin{figure*}[htb]
\centering
  \begin{tabular}{cccc}
    \includegraphics[angle=0,width=.23\textwidth]{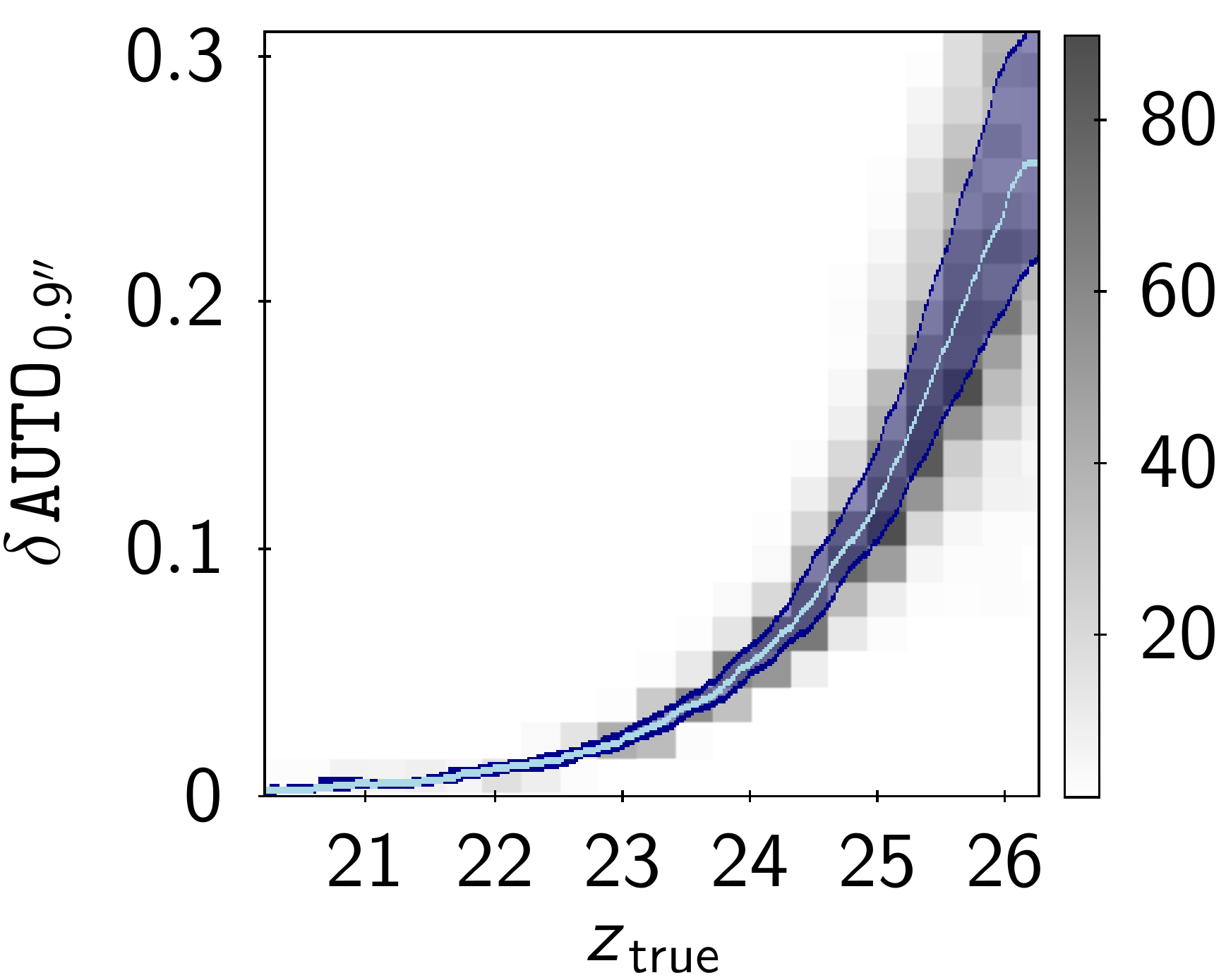} &
    \includegraphics[angle=0,width=.23\textwidth]{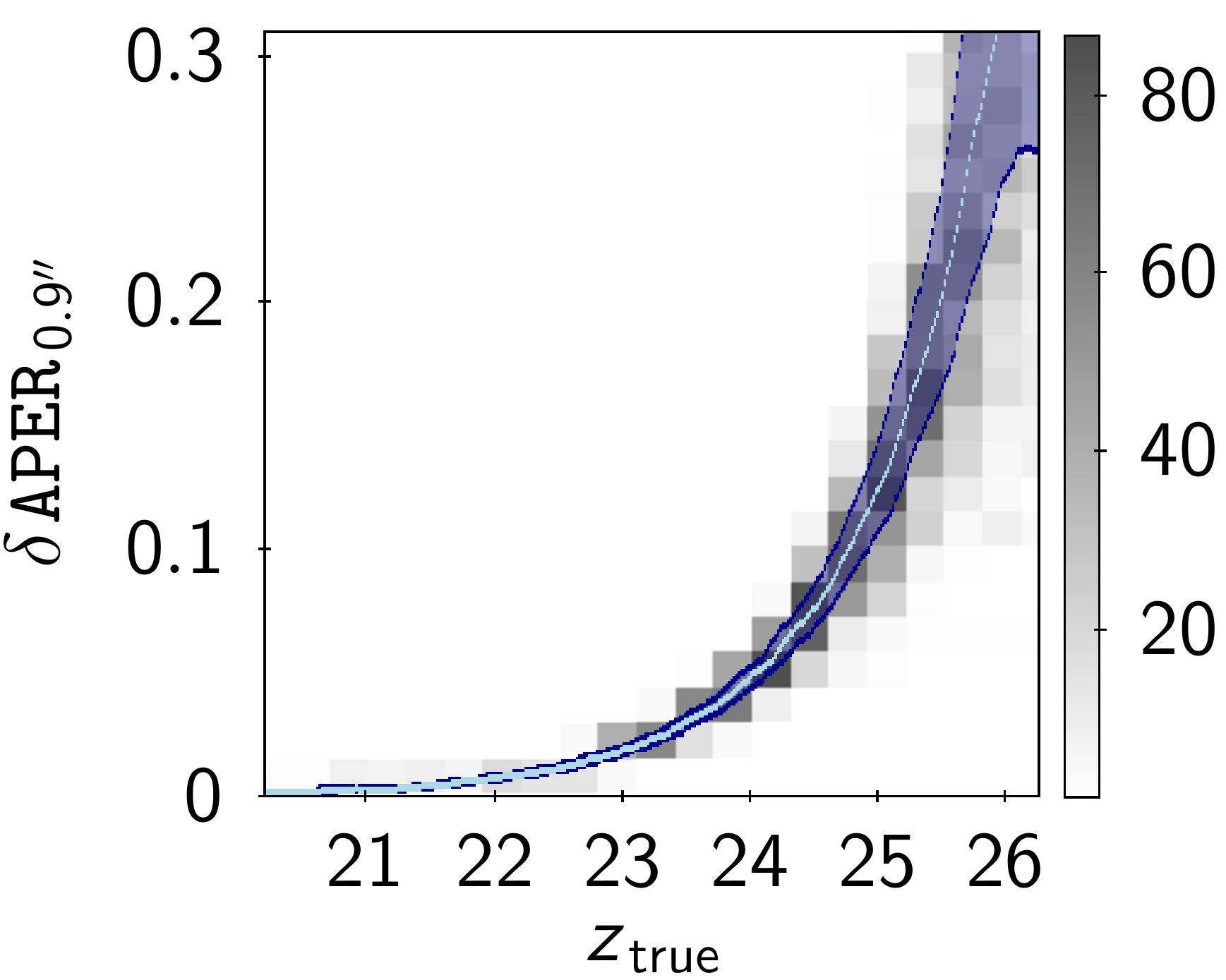} &
    \includegraphics[angle=0,width=.23\textwidth]{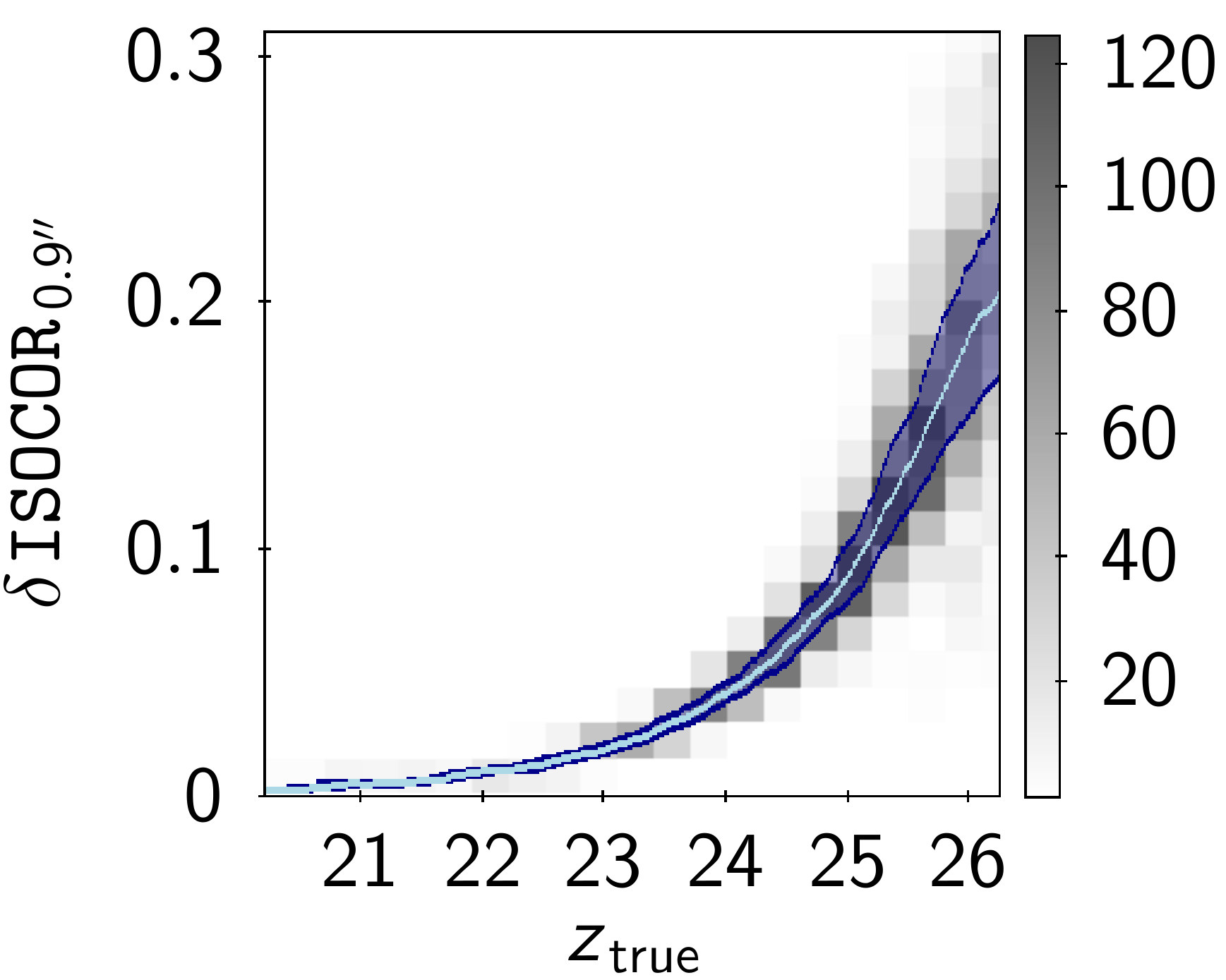} &
    \includegraphics[angle=0,width=.23\textwidth]{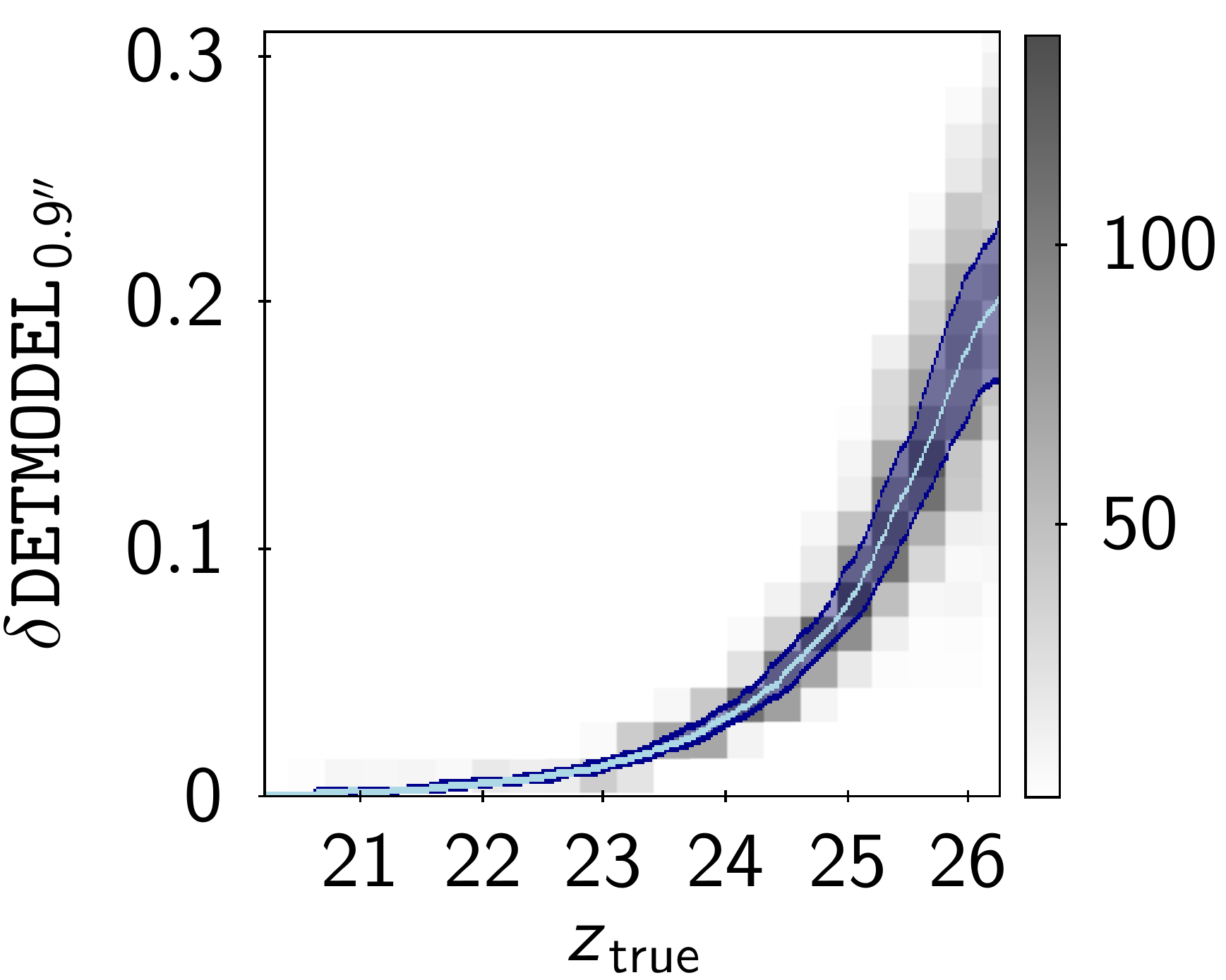} \\
  \end{tabular}
\caption[Magnitude error simulation]{Distribution of photometric errors from measurements
on the arbitrary \bandz-band image with mean FWHM=0.9\arcsec  for the set of photometric 
parameters ({\tt AUTO}, {\tt APER}, {\tt ISOCOR} and {\tt DETMODEL}) discussed in text
(and also presented in Figures \ref{magsim} and \ref{colorsim}). Light blue tracks
represent the running median of the distributions. Very similar trends (not shown) were also found 
for the two extreme mean FWHMs (0.7\arcsec\ and 1.1\arcsec).
}
\label{magerrsim}
\end{figure*}

Photometric error distributions from the simulated detection image (i.e.\ with a mean FWHM=0.9\arcsec)  
are pictured in Figure \ref{magerrsim}. The {\tt DETMODEL} and {\tt ISOCOR} parameters give a more favourable
balance against {\tt AUTO}, and even 3 \arcsec\ in diameter aperture ({\tt APER}), photometry
distributions. The error distribution corresponding to mean FWHMs of 0.7 and 1.1 arcseconds are not
represented because they resemble the one plotted.

\begin{figure*}[htb]
\centering
  \begin{tabular}{ccc}
    $\overline{\sf FWHM}$ = {\sf 0.7}\arcsec & $\overline{\sf FWHM}$ = {\sf 0.9}\arcsec & $\overline{\sf FWHM}$ = {\sf 1.1}\arcsec \\
    \includegraphics[angle=0,width=.31\textwidth]{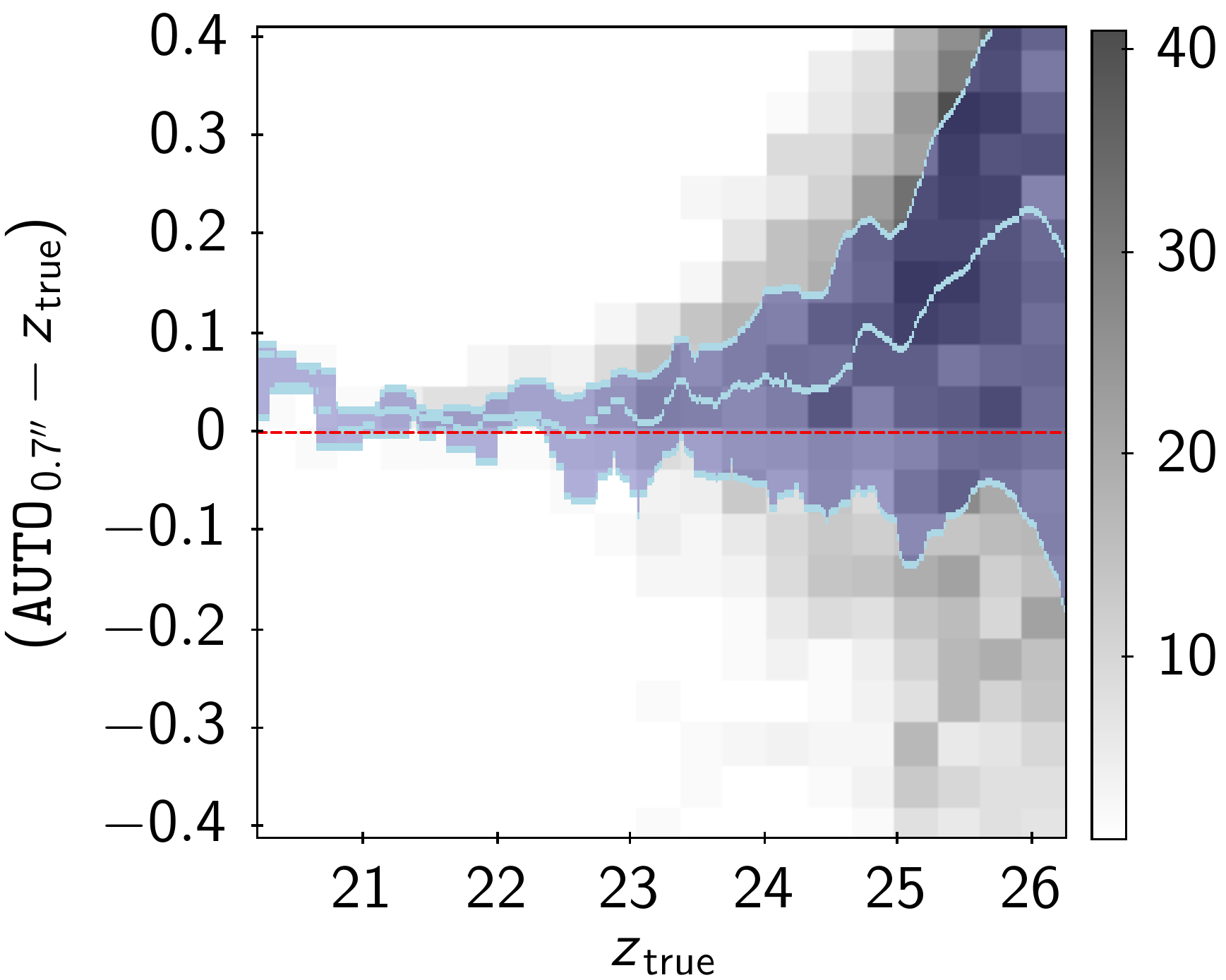} &
    \includegraphics[angle=0,width=.31\textwidth]{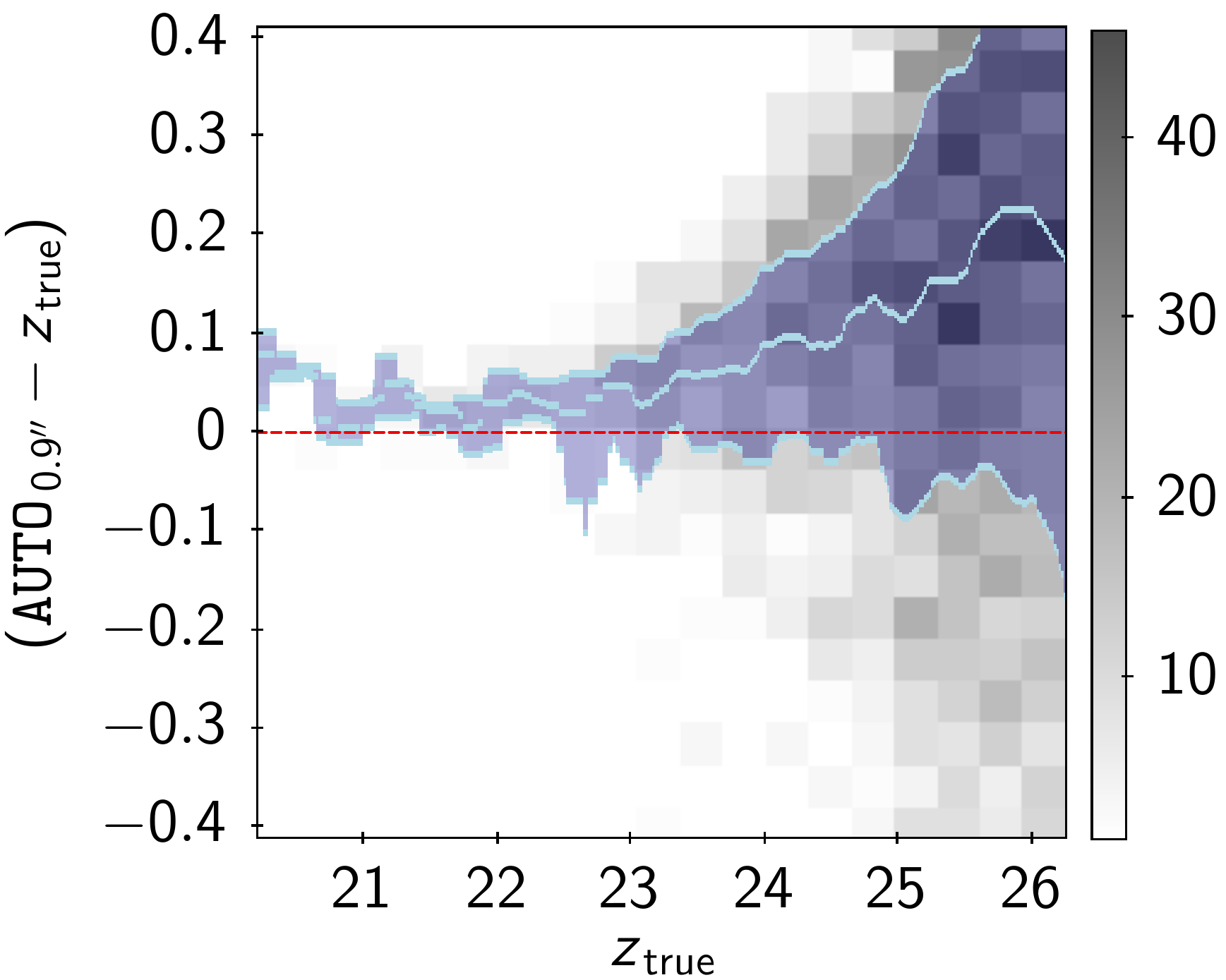} &
    \includegraphics[angle=0,width=.31\textwidth]{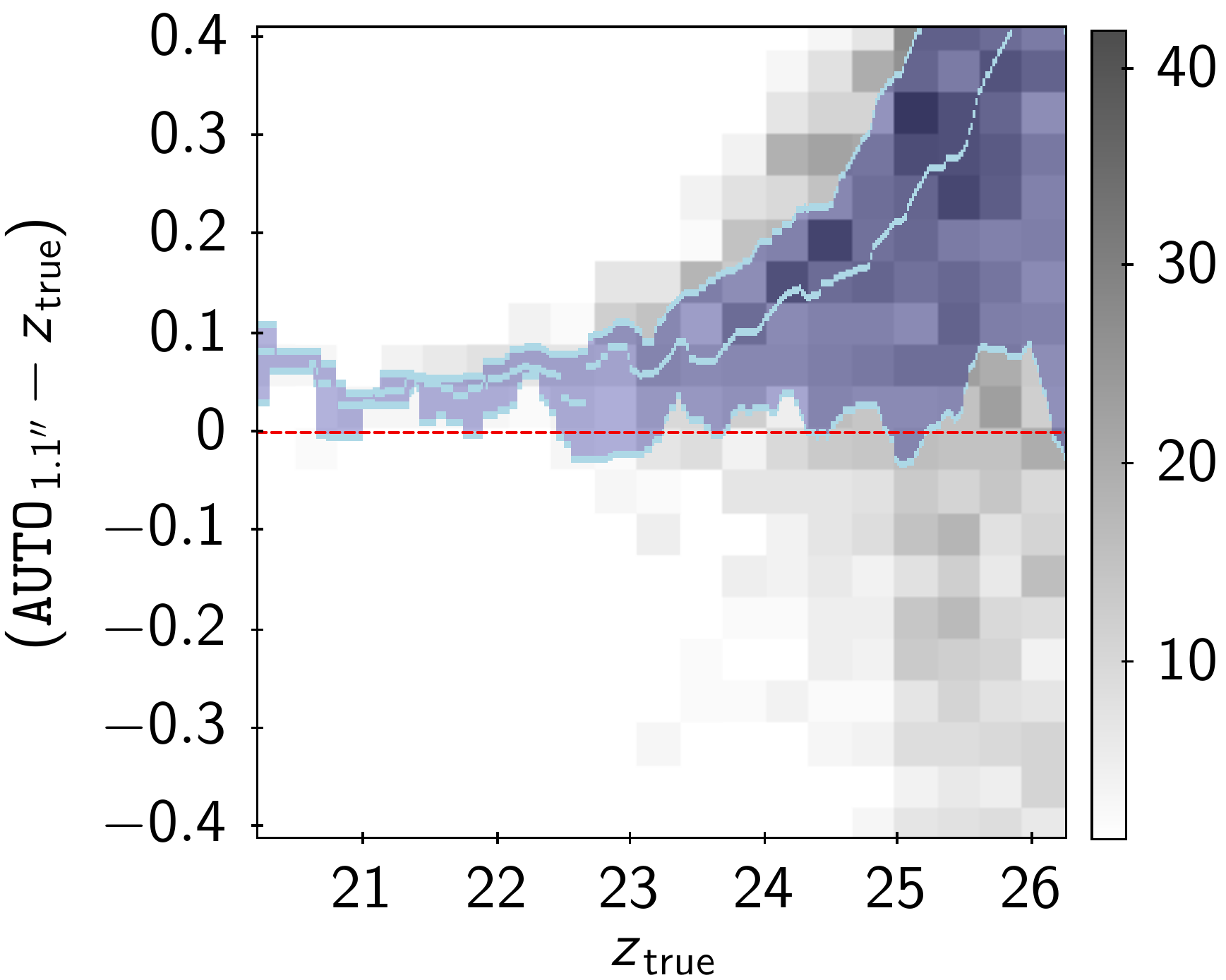} \\
    \includegraphics[angle=0,width=.31\textwidth]{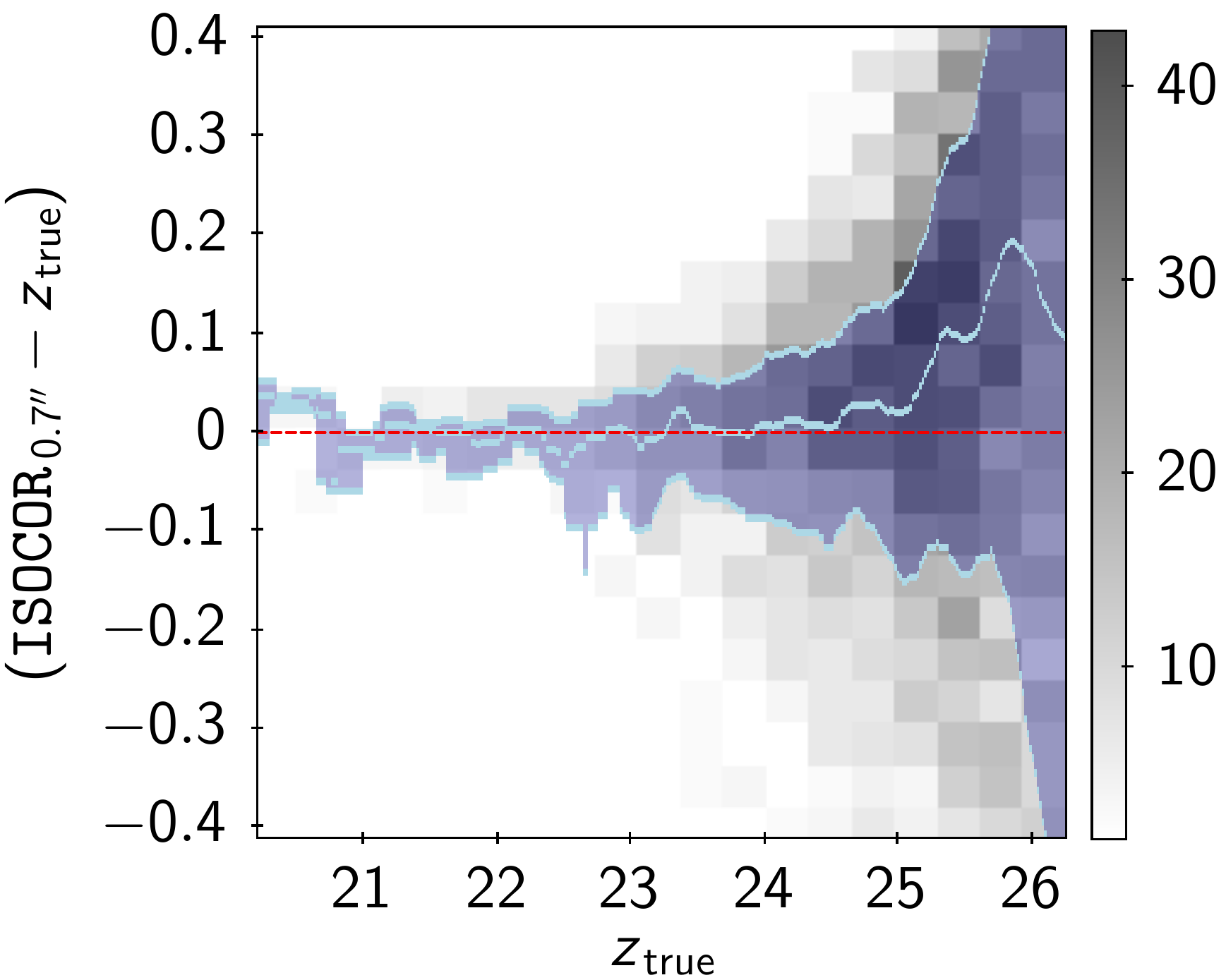} &
    \includegraphics[angle=0,width=.31\textwidth]{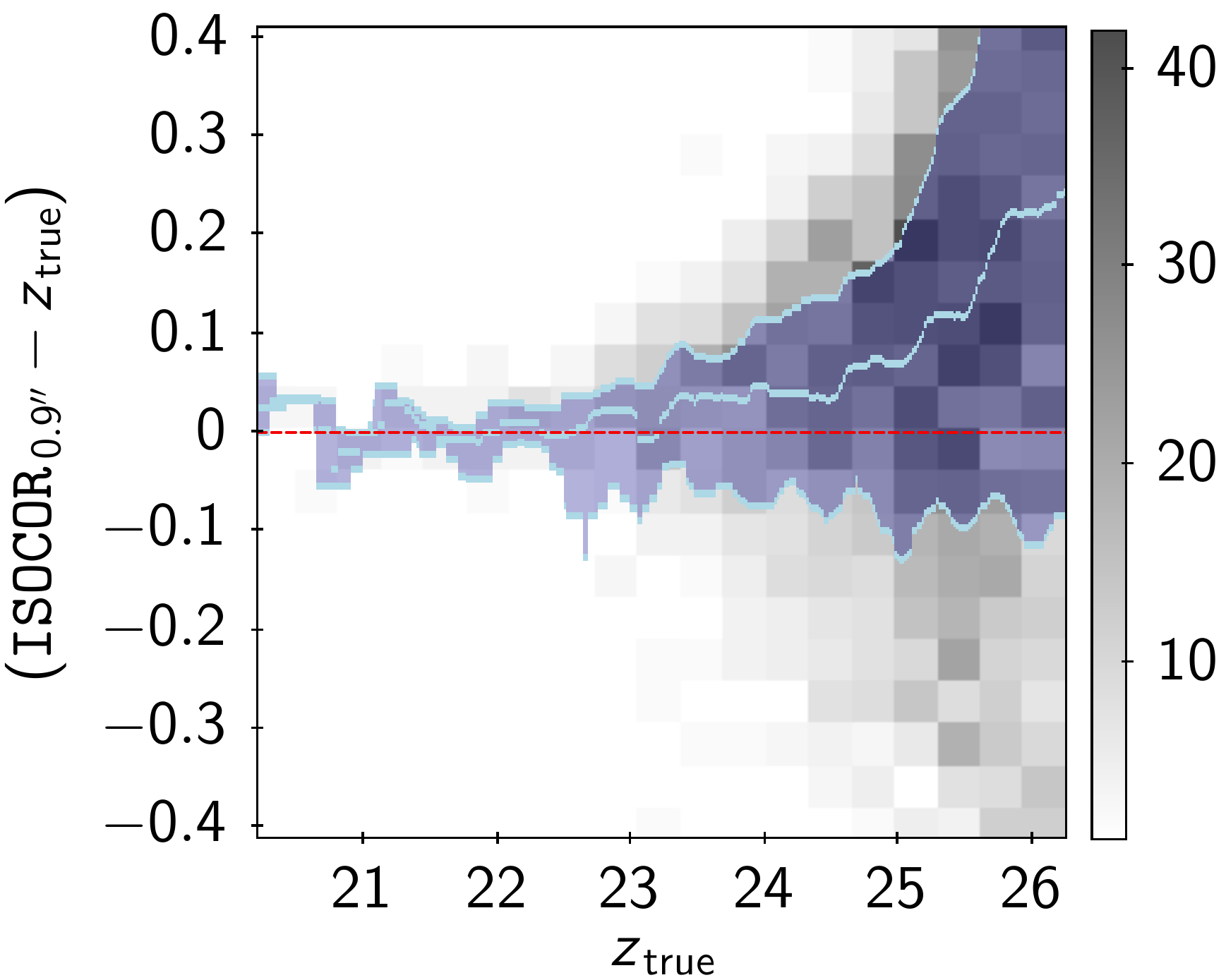} &
    \includegraphics[angle=0,width=.31\textwidth]{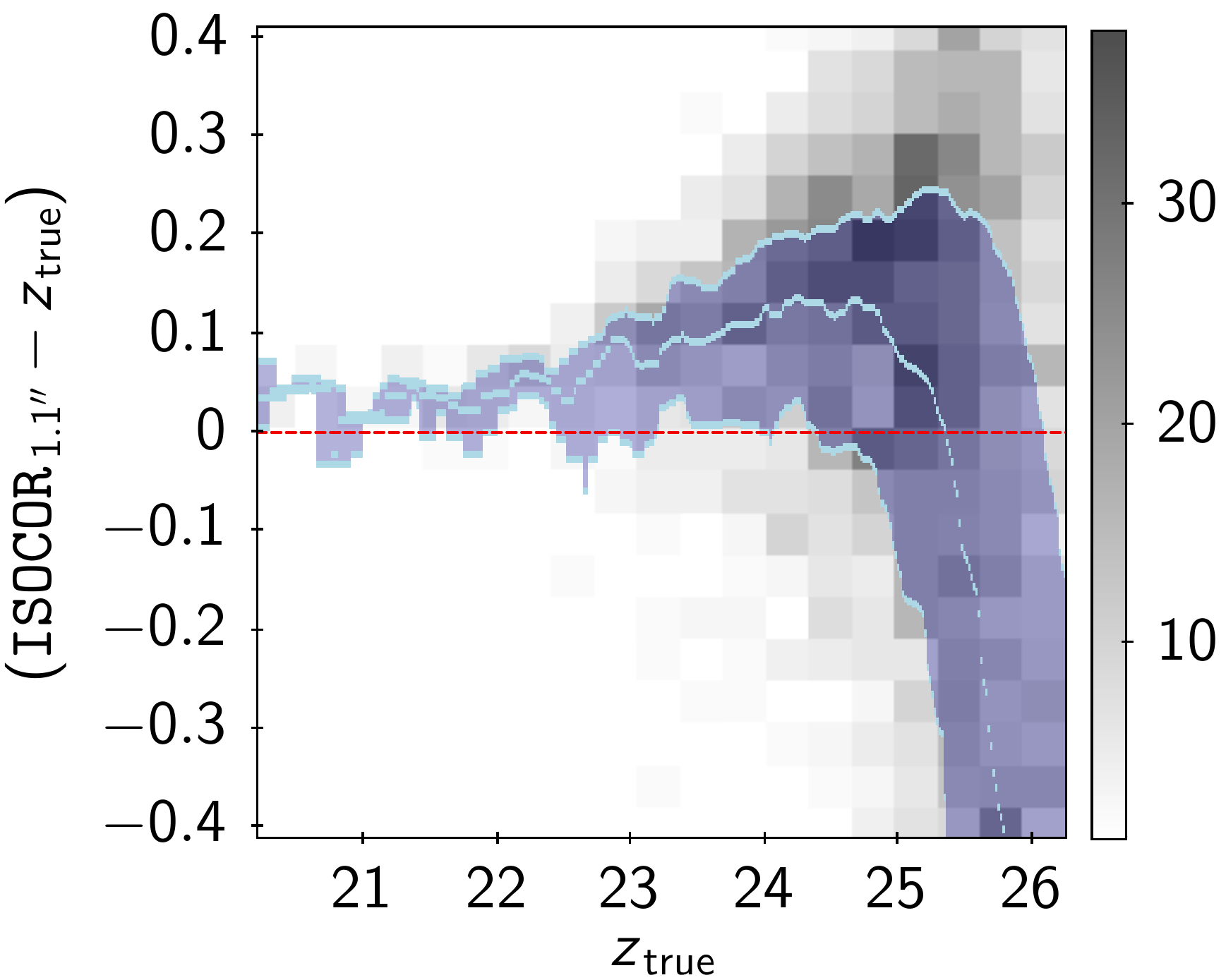} \\
    \includegraphics[angle=0,width=.31\textwidth]{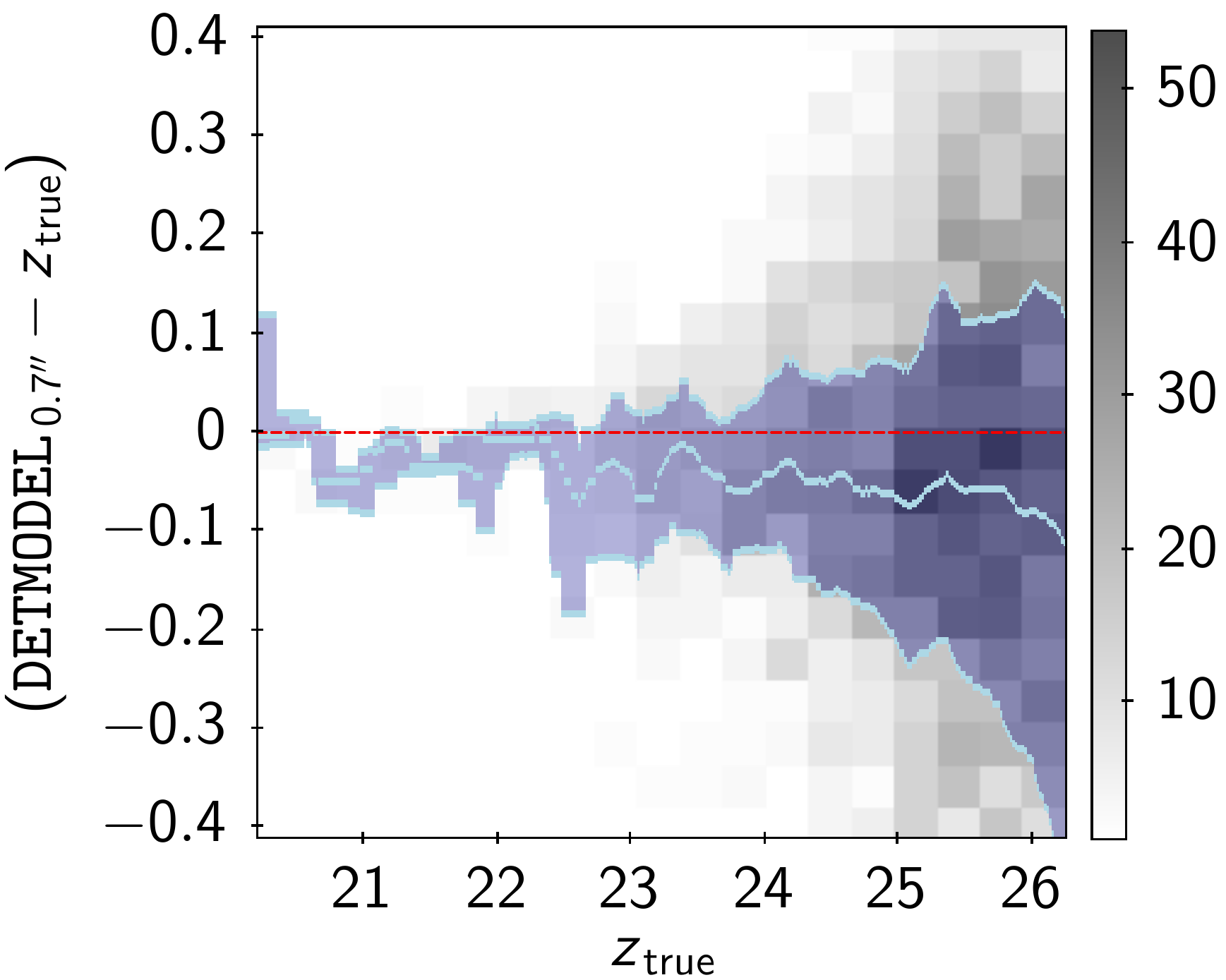} &
    \includegraphics[angle=0,width=.31\textwidth]{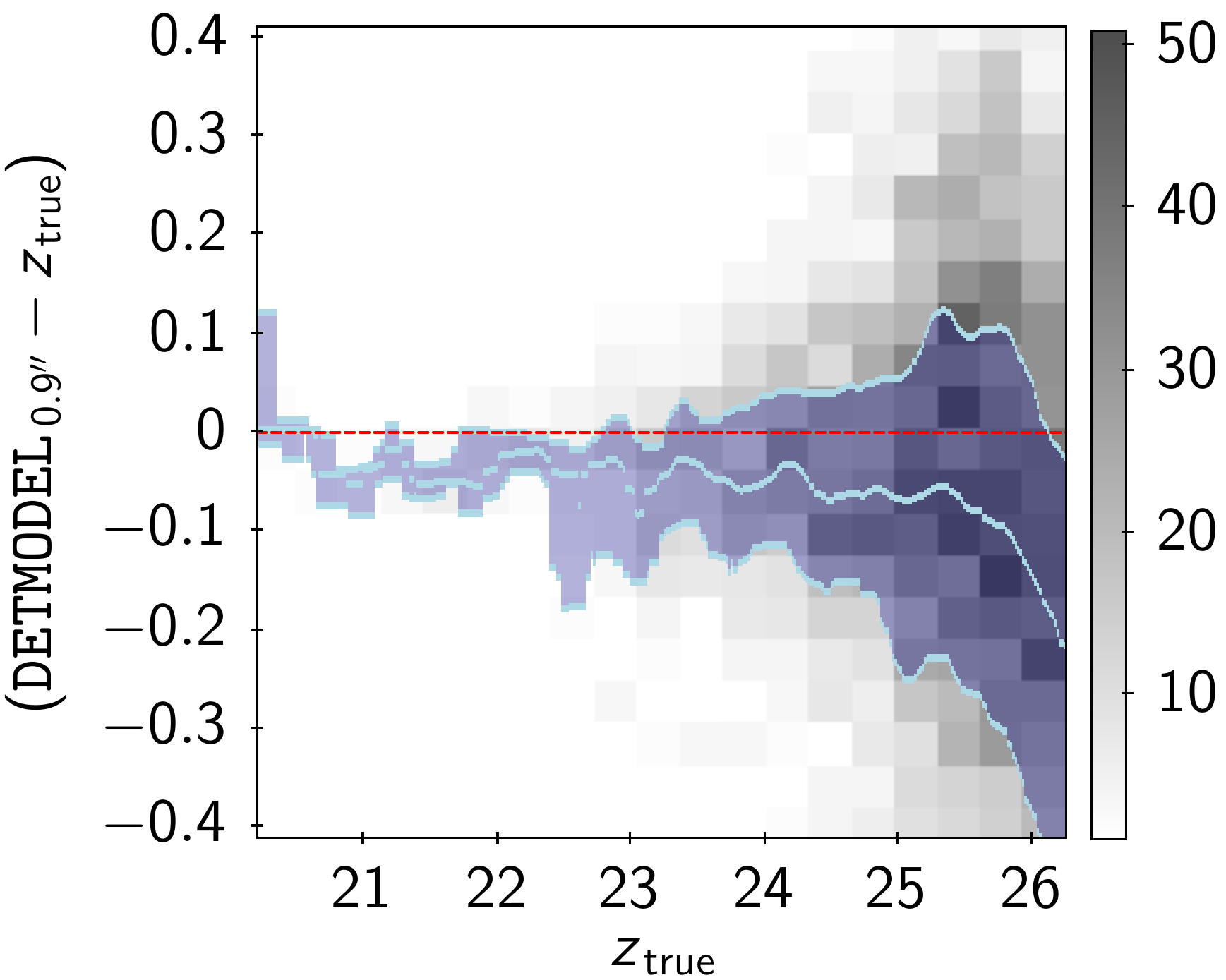} &
    \includegraphics[angle=0,width=.31\textwidth]{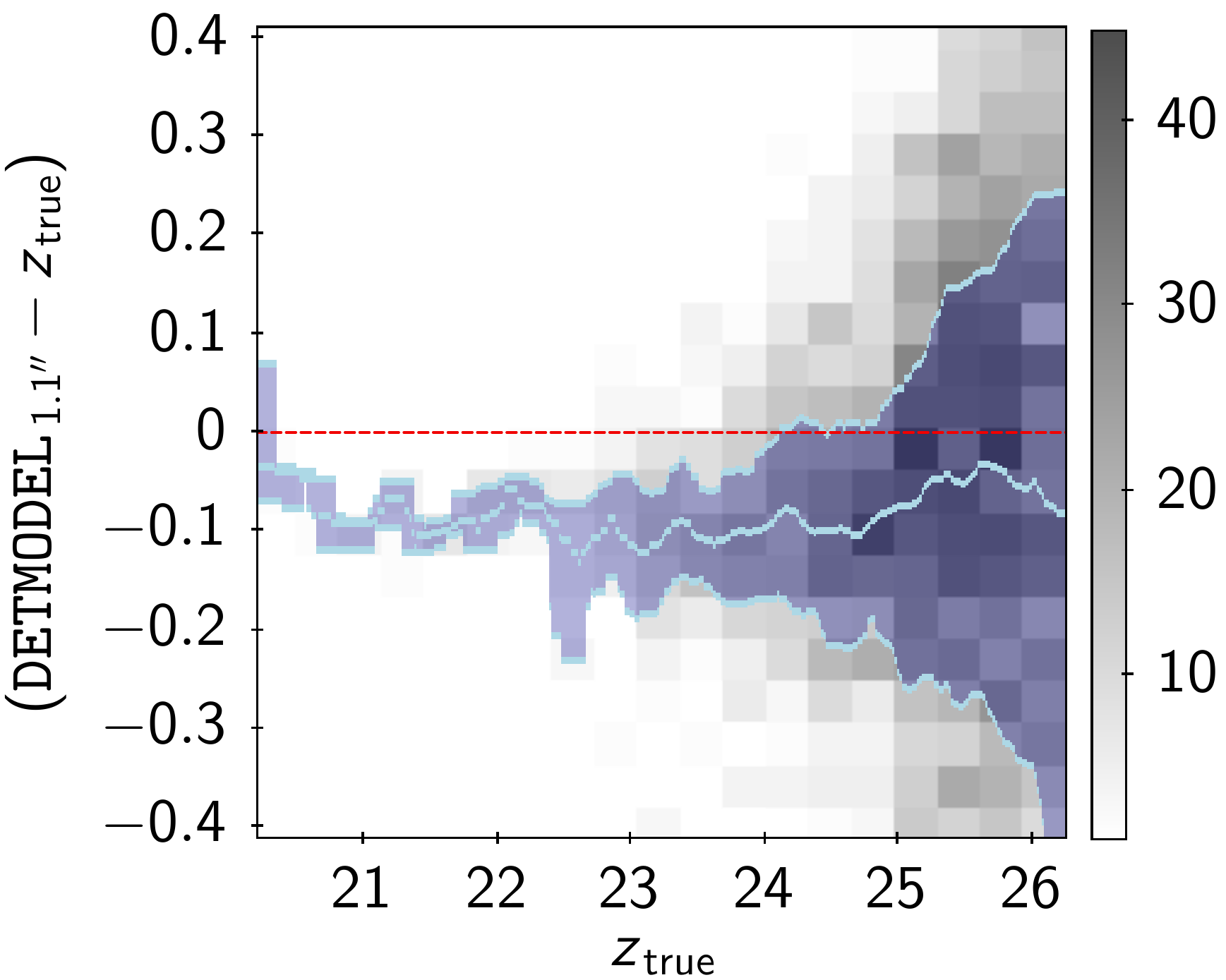} \\
  \end{tabular}
\caption[DET\_MODEL simulation]{Distribution of measured-to-true magnitude recovered using
{\tt SExtractor}, from three arbitrary \bandz-band images that only differ from the PSF FWHM (i.e.\
0.7, 0.9, and 1.1 arcseconds, from left to right). Photometric parameters used ({\tt AUTO}, 
{\tt ISOCOR}, and {\tt DETMODEL}) are arranged from top to bottom. Light blue tracks 
represent the running 25\%, 50\% (median), and 75\% quantiles. Scale bars represent the
background cells counts.
}
\label{magsim}
\end{figure*}

Unreliable detections (non-zero {\tt FLAGS}) in any of the input catalogues were discarded. 
The consolidated catalogue was cross-matched in turn
with the input one that contains the \ztrue\ photometry. Figure \ref{magsim} shows
magnitude difference plots of Kron ({\tt AUTO}), which is the primary choice for a measurement
of the total brightness, isophotal (in this case, {\tt ISOCOR}), and 
{\tt DETMODEL} parameters when compared with \ztrue. These parameters
were selected from a larger set, and the represented ones showed the lowest dispersion, which
depends mainly on the measurement error. Attending to the overlapping running median 
plotted on the entire \ztrue\ and the dispersion profile, the {\tt DETMODEL} parameter 
is the best choice for total flux recovery for all three mean FWHMs.

\begin{figure*}[htb]
\centering
  \begin{tabular}{ccc}
    \includegraphics[angle=0,width=.31\textwidth]{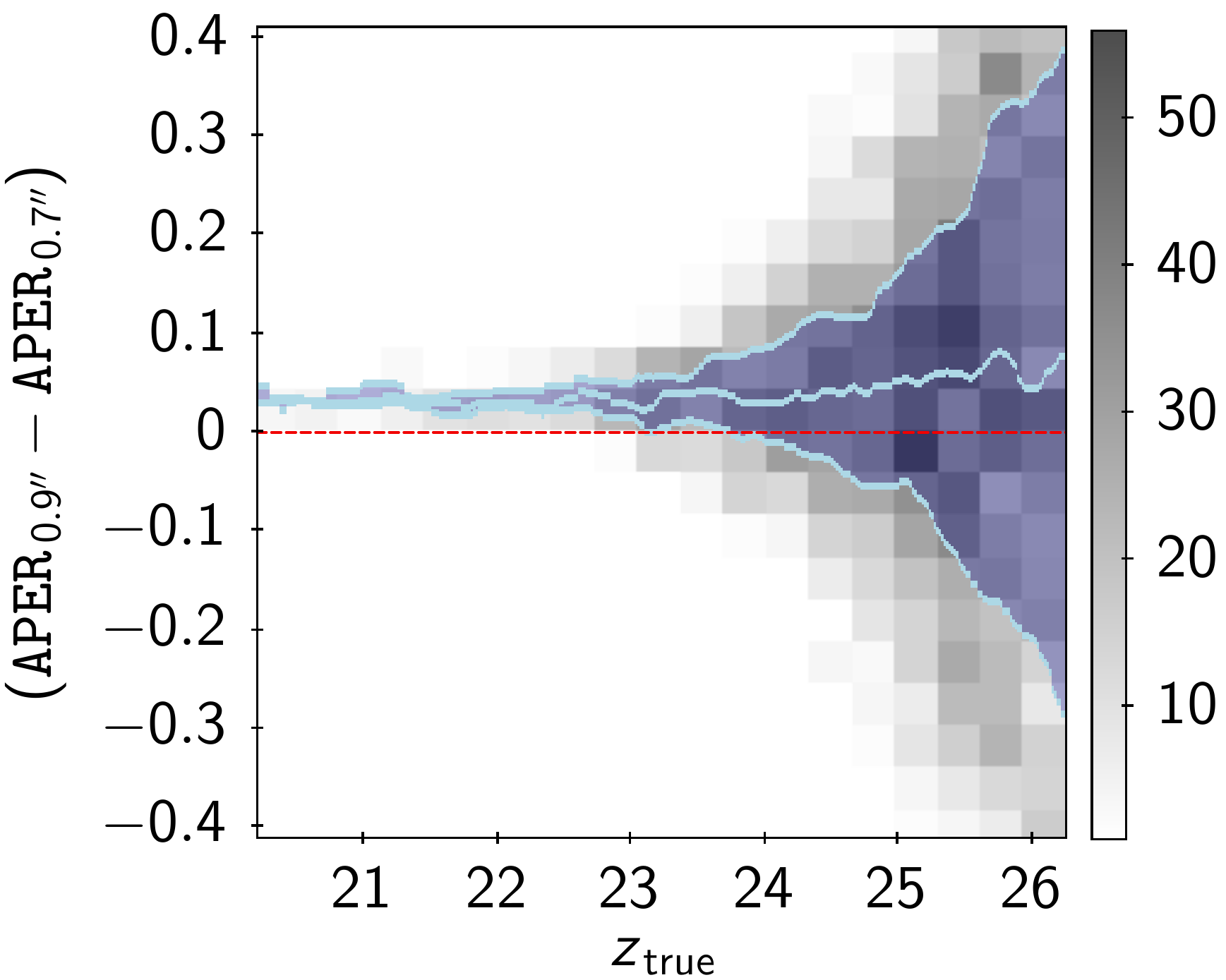} &
    \includegraphics[angle=0,width=.31\textwidth]{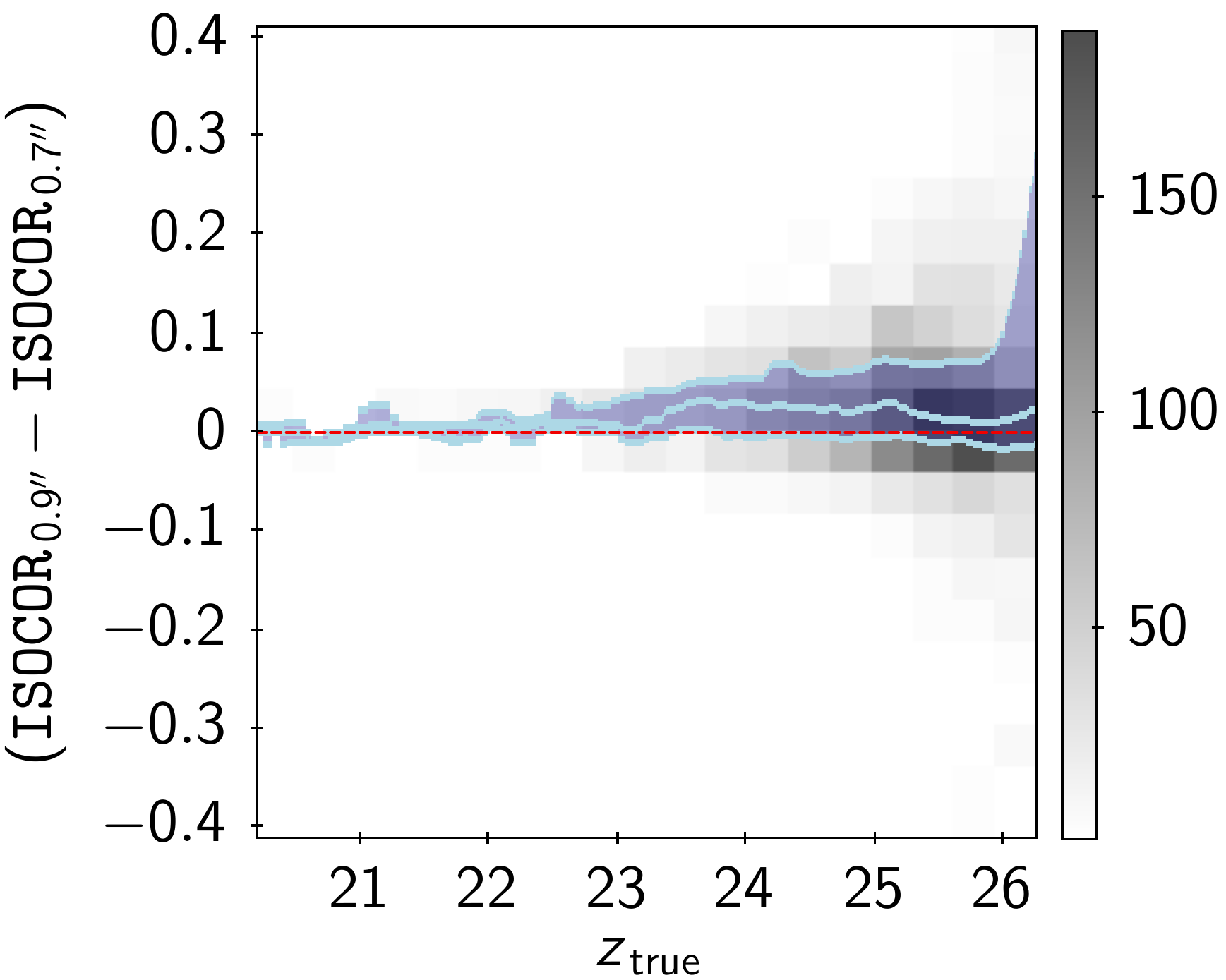} &
    \includegraphics[angle=0,width=.31\textwidth]{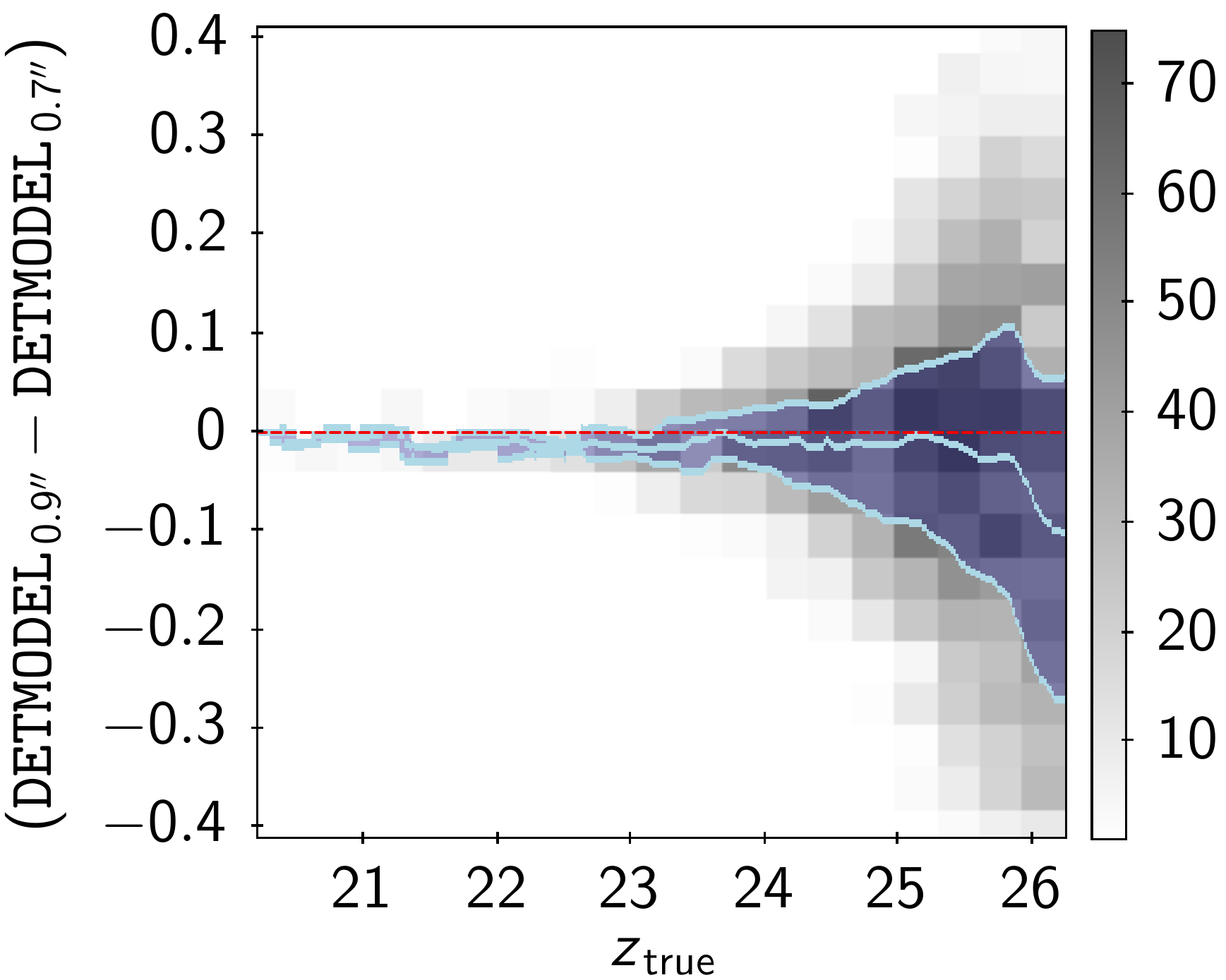} \\
    \includegraphics[angle=0,width=.31\textwidth]{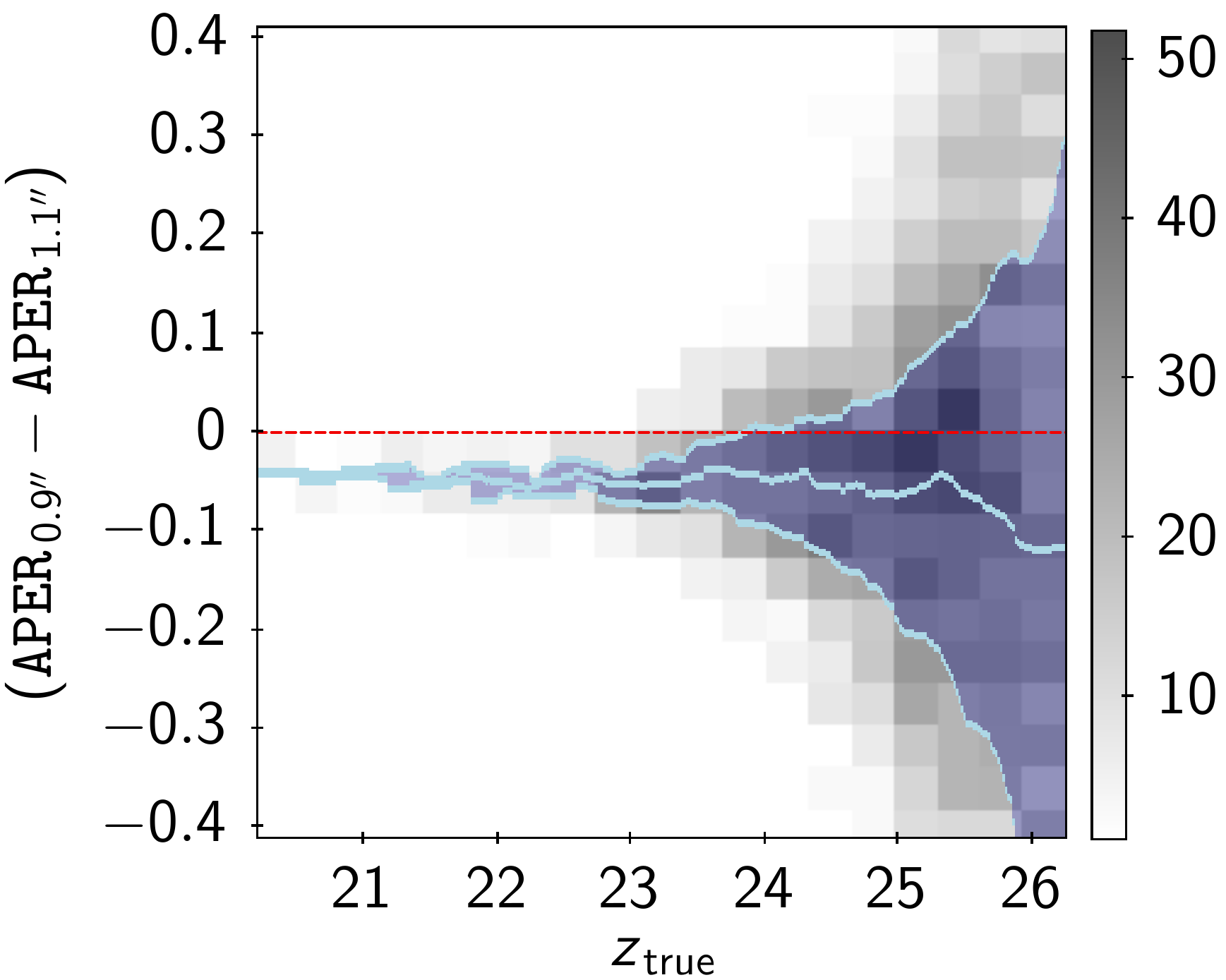} &
    \includegraphics[angle=0,width=.31\textwidth]{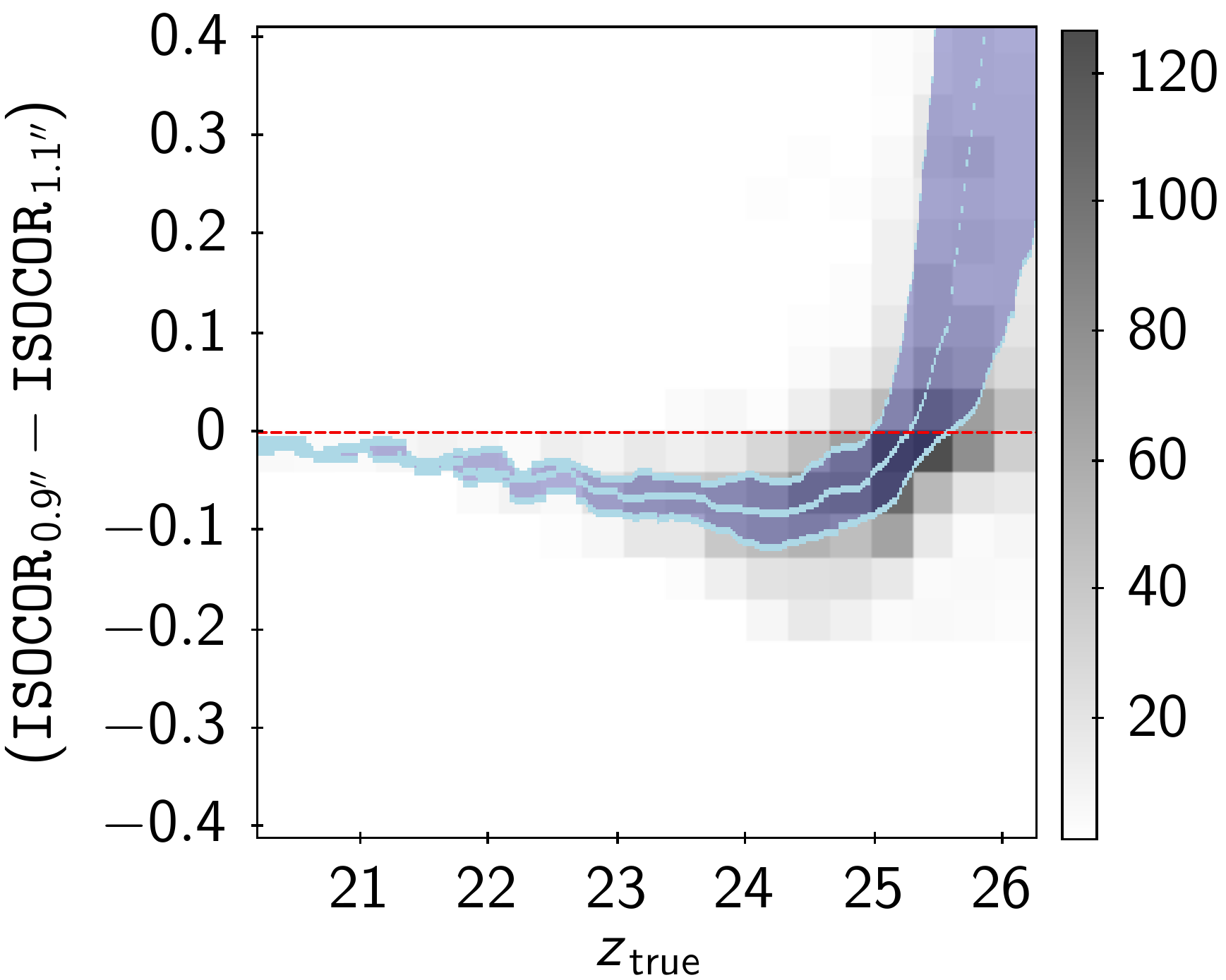} &
    \includegraphics[angle=0,width=.31\textwidth]{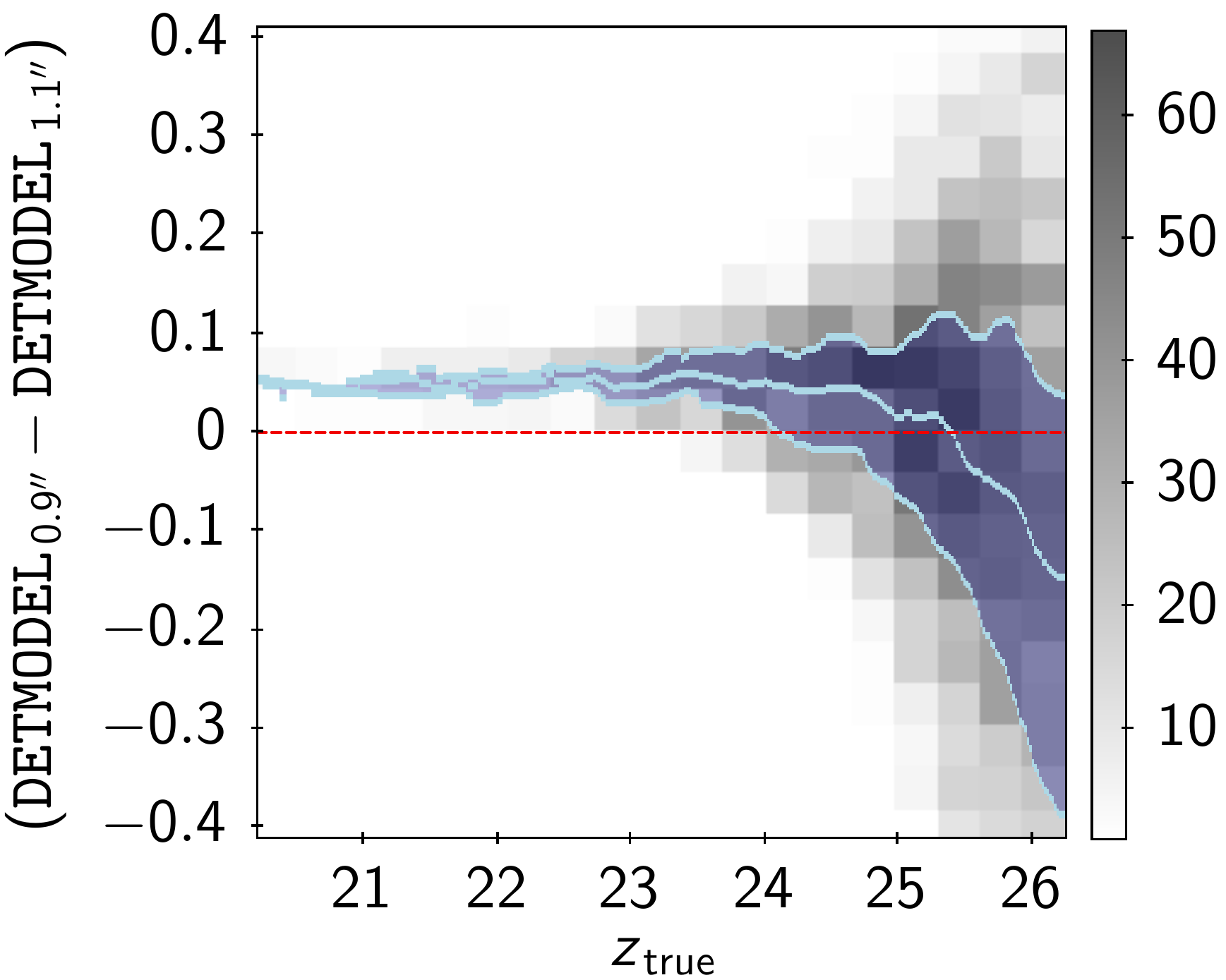} \\
  \end{tabular}
\caption[Colours simulation]{Distribution of measured colours recovered for paired combinations
of mean FWHM. Photometric parameters used ({\tt APER}, {\tt ISOCOR}, and {\tt DETMODEL}) are arranged 
from left to right. Light blue tracks represent the running 25\%, 50\% (median), and 75\% quantiles.
Scale bars represent the background cells counts.	
}
\label{colorsim}
\end{figure*}

Regarding colours (Figure \ref{colorsim}), we compared flux measured in image pairs with 
different mean FWHMs. The photometric parameters chosen were aperture ({\tt APER}), 
isophotal, and {\tt DETMODEL}. As in the case of Kron magnitudes for total flux parametrization, 
apertures (corrected) are the conventional choice to build colours, although \cite{benitez04} 
suggest that isophotal magnitudes provide the best estimate of `true' galaxy colours under the
same FWHM. The behaviour of {\tt DETMODEL} colours is slightly better than {\tt APER} colours,
and the {\tt ISOCOR} colours are even more consistent to zero differences but fail for
faint sources. Thus, the compromise solution for single-parameter photometry in the present 
case is provided by {\tt SExtractor} using the {\tt DETMODEL} approach. Obviously, this does not 
preclude complementary measures with other photometric parameters.

\subsection{The \otelo\ {\it core} catalogue}
\label{sec:core}

Based on the results of the simulations described above, we carried out the construction of the
\otelo\ {\it core} catalogue by forcing {\tt DETMODEL} photometry in all the images, using
the \otelodeep\ image as the detection map in {\it dual-mode} with {\tt SExtractor}. The 
catalogue contains logical
and equatorial coordinates of the sources measured on the \otelodeep\ image, flux measurements
in \otelodeep, \bandu, \bandg, \bandr, \bandi, \bandz, \bandj, \bandh, \bandk, \acs606\ and \acs814\ 
bands using the {\tt AUTO}, {\tt ISOCOR},
{\tt APER}, and {\tt DETMODEL} photometric parameters, as well as the light-spread fitting parameter
{\tt SPREAD\_MODEL} (see Section \ref{sec:stargalaxy}), flux radii, peak surface brightness, isophotal 
area and generic flags for each source in those bands.

\subsection{Complementary catalogues}
\label{sec:addancillary}

The {\it core} catalogue of \otelo\ was cross-matched with public value-added data
to obtain the final \otelo\ multi-wavelength catalogue.
For this purpose, catalogues in X-ray, UV, mid- and far infrared were used (see Table
\ref{table_catalogues}). \par

These complementary catalogues vary in both PSF and astrometric uncertainties, the
latter being the result of different treatments depending on the data source. The differences of 
the PSF in size and shape at the various wavelengths considered have a noticeable influence
on the accuracy of the matches, namely source confusion and multiple matches. The final \otelo\ 
catalogue is the result of the cross-match of the source lists described below through an
algorithm that takes into account not only the relative position of the match candidates 
but also their magnitude distribution and that of the background sources.

\begin{table*}[ht]
\vspace*{2mm}
\caption[Catalogues used for the construction of OTELO's multi-wavelength catalogue]{Catalogues used for the construction of OTELO's multi-wavelength catalogue.} 
\vspace*{-5mm}
\label{table_catalogues}   
\small                                 
\begin{center}\begin{tabular}{c c c c c}        
\hline   \\                     
Domain / Content & Survey / Mission & Spectral region & Astrometry Error & Mean PSF FWHM\tablefootmark{\ (a)} \\    
 &  &  & [\arcsec] & [\arcsec] \\  
\hline      \\[1pt]                        
    X-Rays & \chandra\ &   0.5-10 keV & 0.7 & $<$0.5 \\
    Ultraviolet & \galex\ &   Far-UV: 1350-1780 $\AA$ &  0.6 & 4.5 \\
    & & Near-UV: 1770-2730 $\AA$ & & 5.1 \\
    Mid-Infrared & \spitzer\ (\irac) & 3.6, 4.5, 5.8 \& 8.0$\,\mu$m &  0.37 & 1.9, 1.8, 2.1 \& 2.8 \\
    Far-Infrared (I) & \spitzer\ (\mips) \& \herschel\ (\pacs)  &  24, 100 \& 160$\,\mu$m  &  1.0  & 6.4, 7.0 \& 11.2 \\
    Far-Infrared (II)  & \herschel\ (\spire) & 250, 300 \& 500$\,\mu$m &  0.5 & 18.2, 24.9 \& 36.1 \\
    Photo-z & \cfhtls\ T0004 Deep-3 &  &  0.26  & -- \\
    Spec-z & \deep\ Galaxy Redshift Survey &  &  0.50 & -- \\
\hline                                             
\end{tabular}
\end{center}
\tablefoot{}
\tablefoottext{a}{\chandra\ data from http://cxc.harvard.edu/proposer/POG/html/index.html. Remaining values were extracted from \cite{aniano11}.} \\
\end{table*}
\normalsize

In X-rays, the catalogue from \cite{povic09} was employed. It contains 639 X-Ray sources 
in the Extended Groth Strip, selected from public \chandra\ data in five bands: {\tt full} (0.5-7 keV), 
{\tt soft} (0.5-2 keV), {\tt hard} (2-7 keV), {\tt hard2} (2-4.5 keV), and {\tt vhard} 
(4-7 keV). When cropped to the \otelo\ field, 74 sources are left. The AEGIS-X catalogue 
\citep{laird09} was also checked to include non-redundant X-ray emitters. In the \otelo\ field, it 
contains 50 sources with fluxes in four bands: {\tt full} (0.5-10 keV), {\tt soft} (0.5-2 keV), 
{\tt hard} (2-10 keV), and {\tt ultra-hard} (5-10 keV). Both X-ray catalogues were first 
cross-matched with search radii from 1 to 2.5 arcseconds. In this range, 42 sources had a match, 
of which more than 90\% were closer than 0.5\arcsec\ from their counterpart. Based on that, 
a new X-ray catalogue with nine bands was constructed, including those 42 sources plus the 
remaining 32 sources from \cite{povic09} and the 8 sources from \cite{laird09}.
 
In the UV, data from the Galaxy Evolution Explorer (\galex\, \citealt{martin05}), as part of the 
AEGIS survey, were used (\citealt{bianchi14,morrissey07}). In total, 5185 \galex\ sources 
fell in the \otelo\ field.
 
In the infrared, we used data from the \herschel\ and \spitzer\ Space Observatories 
(see \citealt{pilbratt10} and \citealt{werner04}, respectively). We employed the first 
full public data release from the \pacs\footnote{Photoconductor Array Camera and 
Spectrometer.} Evolutionary Probe (PEP) survey of \herschel\, which includes data in the 
Extended Groth Strip \citep{lutz11}. This catalogue uses the 24$\,\mu$m 
MIPS\footnote{Multiband Imaging Photometer.} band of Spitzer as a prior to select the 
100 and 160$\,\mu$m \pacs bands. A total of 553 objects from this catalogue fell within 
\otelo's FoV. According to \cite{lutz11}, the astrometry precision of the catalogue is 
sub-arcsecond, hence we adopted a maximum position error of 1.0\arcsec\ for those 
sources.

We also took advantage of the third data release of the \herschel\ Multi-tiered 
Extragalactic Survey (HerMES, \citealt{oliver12}), which makes use of the 
SPIRE\footnote{Spectral and Photometric Imaging Receiver.} instrument on board the 
\herschel\ Space Observatory. This catalogue employs the 24$\,\mu$m MIPS band as a prior to 
select the 250, 350, and 500$\,\mu$m bands. It contains 822 sources in \otelo's field with an 
astrometrical precision of 0.5\arcsec\ \citep{roseboom10}.
 
As for \spitzer, we initially employed the \irac\footnote{InfraRed Array Camera.} 3.6 
$\,\mu$m-selected catalogue of the Extended Groth Strip from \cite{barmby08}, which 
contains the four \irac\ bands (3.6, 4.5, 5.8, and 8$\,\mu$m) and 2374 objects in our field 
with a precision in astrometry of 0.37\arcsec. However, this catalogue does not include 
the lower left corner of our field and has extremely large errors in magnitude for the 
faintest sources. We therefore added the catalogue made by \cite{barro11}, which comprises 
2317 sources in our field selected over the 3.6$\,\mu$m and 4.5$\,\mu$m \irac\ images, 
measured with aperture photometry. We cross-matched these two catalogues with our own 
independently, and when both had a match we favoured the \cite{barro11} photometry.

Finally, we took advantage of two public redshift catalogues to add this information 
to OTELO's multi-wavelength catalogue. One was the CFHTLS T0004 Deep3 photo-z catalogue 
\citep{coupon09}, with 7725 sources in our field obtained using optical and NIR data
only. We considered a maximum positional 
error of 0.26\arcsec\ for all the sources. The other source of redshift data was 
the catalogue corresponding to the 4th data release 
of the \deep\ Galaxy Redshift Survey \citep{newman13}, which contains 517 sources in 
OTELO's field. Those targets were selected from a broad-band photometric catalogue 
obtained with the CFHT and had absolute errors of 0.5\arcsec, as defined by the 
USNO-A2.0 catalogue used for the astrometry \citep{coil04}. The imprint of the spatial 
distribution of sources in each of these catalogues is shown in Figure \ref{catalogues}.

To correlate these catalogues with our own, we used the methodology first developed 
by \cite{deruiter77}, later improved by \cite{sutherland92}, which defines a 
likelihood-ratio ($LR$) to distinguish between true counterparts and false identifications.
 This approach has been successfully used to match radio and X-Ray sources to optical 
or infrared ones \citep[see for example][]{ciliegi03,luo10}. Given a non-optical source, 
\cite{deruiter77} described the $LR$ as the ratio between the probability of finding 
its true optical counterpart at a certain distance, and the probability of finding 
instead a background source at that distance. They assumed that background sources 
followed a Poisson distribution and only took into account the radial distance of the 
optical to the non-optical sources and the positional error of both.

\cite{sutherland92} later introduced magnitude information to improve the $LR$ 
technique. They calculated not only the probability that the true counterpart lay at a 
given distance from the non-optical source but also that its magnitude lay in a certain 
interval. In this work we have followed that approach and used the procedure developed 
by \cite{perezmartinez16}, defining the $LR$ as:

\begin{equation}
LR= \frac{q(m) f(r)}{n(m,r)},
\end{equation}

\noindent $q(m)$ being the magnitude distribution of the true counterparts, $f(r)$ the probability 
distribution function of a true counterpart being at a distance $r$ of the object and 
$n(m,r)$ the surface density of background objects with magnitude $m$. 
For each catalogue to be matched, the procedure gets the best candidate counterpart
for each \otelo\ source and an estimate of the reliability of the association, $R_c$. First, 
the $LR$ of each candidate is calculated as per the previous
expression, keeping those with $LR$ above certain threshold. The election of this threshold
is key to the final result and is obtained iteratively by maximizing the sum of the reliability 
and completeness of the cross-matched catalogue. The reliability $R$ of the catalogue
cross-correlation is the average of the individual reliabilities $R_{\rm c}$ of each counterpart,
defined as the ratio between the $LR$ of the current candidate over the sum 
of the others plus a completeness correction factor:

\begin{equation}
R_{\rm c}=\frac{LR}{\Sigma\, LR + (1-Q)},
\end{equation}

\noindent where the sum is over all the candidates found for a given source, and 
$Q=\int_{\,-\infty}^{\,m_{\rm \,lim}} q(m)\ dm$ is the fraction of the true counterparts we 
are able to detect, obtained again by the iterative calculation of its magnitude distribution.
The catalogue completeness, $C$, is defined as the ratio of the sum of the reliabilities of 
all the sources over the total number of objects in the non-optical catalogue:

\begin{equation}
C=\frac{\Sigma\, R_{\rm c}}{N}.
\end{equation}

Depending on the density of objects in the catalogue, a safely broad radial search in a 
radius of 5\arcsec\ was performed. From all the sources found at that distance, only those with a good $LR$ were retained after computing it for all of them with the methodology 
explained. In this way, we were able to select the best counterpart and calculate the
reliability and the overall completeness of the result. A summary of the likelihood-ratio
matching parameters and results obtained after applying this procedure is shown in Table 
\ref{LR_catalogues}.

\begin{table}[ht]
\vspace*{2mm}
\caption[Results of the cross-correlation of catalogues]{Results of the cross-correlation
of catalogues.  $LR\text{-}th$: value of the likelihood-ratio used as threshold, $R$: reliability of
the matching, $C$: completeness, $N_{\rm X}$: number of sources in the non-optical catalogue and
$N_{\rm matches}$: number of reliable matches found. }
\vspace*{-5mm}
\label{LR_catalogues}
\centering   \small
\begin{center}\begin{tabular}{c c c c c c}
\hline   \\
 Catalogue & $LR\text{-}th$ & $R$ & $C$ & $N_{\rm X}$ & $N_{\rm matches}$  \\
\hline      \\[1pt]
    X-Rays & 0.069 & 0.810 & 0.553 &  82 & 56 \\
    Ultraviolet &  0.105 & 0.907 & 0.753 & 5185 & 4223\\
    Mid-Infrared & 0.01 & 0.947 & 0.850 &  2374 &  2128\\
    Far-Infrared (I) & 0.022 & 0.940 & 0.855  & 553 &  503\\
    Far-Infrared (II) &   0.023 & 0.962 & 0.876 & 822 & 749 \\
    Photo-z & 0.281 & 0.901 & 0.568 & 7725 & 4860\\
    Spec-z & 0.542 & 0.992 & 0.886 & 517 & 461\\
\hline
\end{tabular}\end{center}
\end{table}
\normalsize

In the cases of
the \chandra\ and \cfhtls\ D3 data, the crowdedness of the latter and the sparsity of the 
former affect the completeness and the reliability of the results by producing more
sources below the acceptance threshold $LR\text{-}th$ and higher -tuples of multiple matches
with similar $LR$. In general, the reliability of the cross-match is well above 0.90,
except for the X-ray case ($R=0.810$), with a mean completeness of 0.76 (median 0.85).

\subsection{Photometric redshifts}
\label{sec:photoz}

The finding of photometric redshifts for \otelo\ sources is the first exploitation of
the {\it core} plus ancillary data catalogue. Redshift estimates are mandatory for
creditable labelling of the emission lines detected in \otelo\ pseudo-spectra
and useful for a first classification of the sources based on SED fitting. In order to 
obtain them, we took advantage of the \lephare\ code  (\citealt{arnouts99,ilbert06}), 
adopting the $\chi^2$ minimization approach to find the best fit between the observed 
flux of an object and different SED templates.

\begin{figure}[h]
\centering
\includegraphics[angle=-90,width=\linewidth]{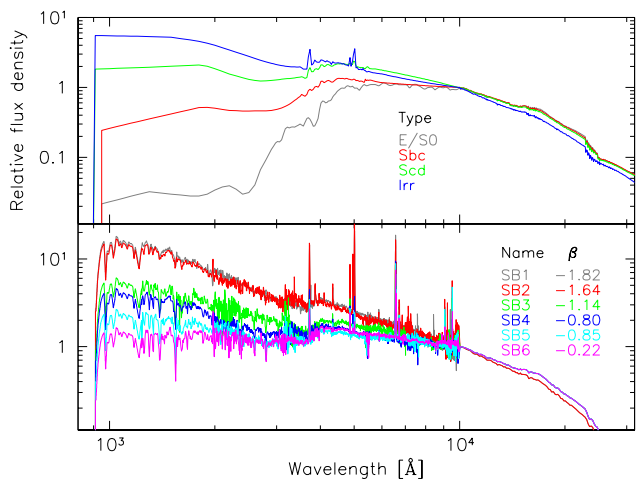}
\caption[Galaxy templates]{Galaxy templates used in photometric redshift estimations.
Upper panel show the basic set of Hubble types E/S0, Sbc, Scd, and Irr from \cite{coleman80}, 
whilst the lower panel contains the collection of star-forming galaxy templates from 
the \cite{kinney96} Atlas. Their measured UV slopes, $\beta$ (2200-3100 $\AA$), are
also given. In both cases the continuum was extrapolated in UV and IR as explained
in \cite{arnouts99}.}
\label{galtemplates}
\end{figure}

Three different libraries were used for the UV and optical range: one for galaxies, one 
for stars and one for AGN/QSOs. The galaxy library is composed of ten SED templates: 
4 representative of Hubble types (E, Sbc, Scd, Im), observed by \cite{coleman80}, 
and six representative of starburst galaxies, built by \cite{kinney96}. As a survey 
biased to ELS finding, star-forming systems with a broad span in 
UV-slope ($\beta$) should be included in the \otelo\ distinctive galaxy template set. All these galaxy SEDs 
are shown in Figure \ref{galtemplates}. The 
AGN/QSO templates were selected from the SWIRE\footnote{Spitzer Wide-area InfraRed Extragalactic survey.} 
library, created by \cite{polletta07}, and include templates of two Seyfert galaxies, three type-1 QSO, 
two type-2 QSO and three composite galaxies (starburst+AGN). As for the star library, it 
consisted mainly of the 131 templates calibrated by \cite{pickles98}, covering all the 
usual stellar spectral types (O-M) and luminosity classes, plus four white dwarf templates 
from \cite{bohlin95} and 26 brown dwarfs representative of stellar spectral types M, L 
and T from the SpeX Prism 
library\footnote{\label{spex}\tt http://pono.ucsd.edu/$\sim$adam/browndwarfs/spexprism}. 
In order to fit the infrared (IR) part of the spectra from 5$\,\mu$m and to calculate infrared 
luminosities, the \cite{chary01} library (CE01), consisting of 105 templates with different 
luminosities, was also used.  The extinction law of \cite{calzetti00} was adopted, with 
values of extinction E($B$-$V$) ranging from 0 to 1.1 in steps of 0.1. The redshift 
range was defined from 0.04 to 10, in intervals of 0.05.

In summary,
for each problem source \lephare\ provides photometric redshift solutions for each template
set in the  UV and optical range. Apart from redshift, the solution includes the best fitted template 
and the $\chi^2$-statistic. Each \lephare\ launch starts with the creation of the filter library, 
which includes, in our case, all the filters described in section \ref{sec:catalogue}: FUV and NUV 
from \galex, \bandu, \bandg, \bandr, \bandi, and \bandz\ from \cfhtls, \bandj, \bandh, and \bandk\ 
from \wirds, the four \irac\ bands, 24$\,\mu$m from \mips, 100 and 160$\,\mu$m from \pacs, and finally 250, 350 and 
500$\,\mu$m from \spire, and -optionally, as explained below- the \oteloint\ data. Apart from redundancy 
with the \bandi\ and \bandz\ 
bands of \cfhtls, resampled data in \acs606\ and \acs814\ do not cover the entire \otelo\ FoV and were excluded 
from photo-z estimates. Given this set of filters, magnitudes (AB) are then computed for
each template in the UV and optical range sets described above at each value of redshift 
and extinction in the corresponding input grids. A distribution of $\chi^2$ estimates
is obtained from a comparison of the observed and the model magnitudes for each one 
of the ten galaxy templates in order to find the best redshift solution.
If an additional local minimum in the $\chi^2$ 
distribution is significant, a secondary redshift solution and its corresponding 
template fit is also provided. The execution of the code continues with the replication of this procedure 
but using the AGN and star template sets. This analysis is independent of the previous step
and skips the finding of secondary solutions.
For AGN templates an additional redshift solution is given. For stars the expected
$\chi^2$-statistic is provided. Finally,
if the problem source have observed IR data, \lephare\ fits a CE01 template in the flux
dimension adopting the main photometric redshift solution previously found. Table \ref{otelo_catalogue} 
contains an extensive list of the photometric redshift parameters given in the \otelo\ 
multi-wavelength catalogue.

In order to maximize the analysis possibilities and the choice of a \zrel\ for each
suitable \otelo\ source, the code was launched twice: once for both the {\it core} 
plus ancillary data catalogue -- i.e.\ including the \oteloint\ data -- and again using only the 
broad-band data, as is traditionally done. 
In the first case, the contribution of emission lines was taken
into account when computing the template magnitudes. To that end, an extra filter was
added to the filter library (see Figure \ref{airy}), corresponding to the spectral responses 
of the tunable filters and representing the spectral window of \oteloint, when data from the
latter are included. 

\begin{figure*}[htb]
\centering
  \begin{tabular}{ccc}
    \includegraphics[angle=-90,width=.32\textwidth]{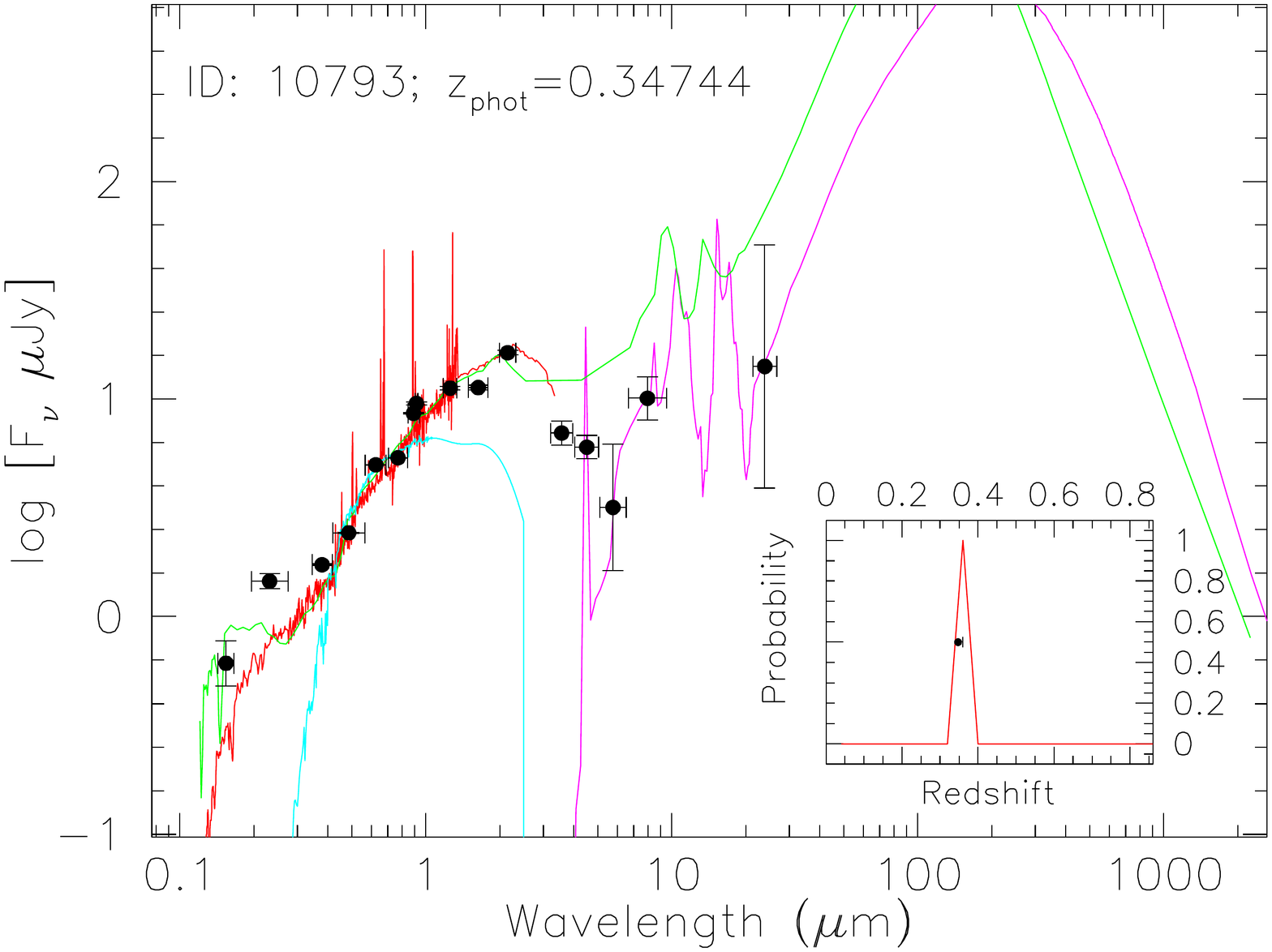} &
    \includegraphics[angle=-90,width=.32\textwidth]{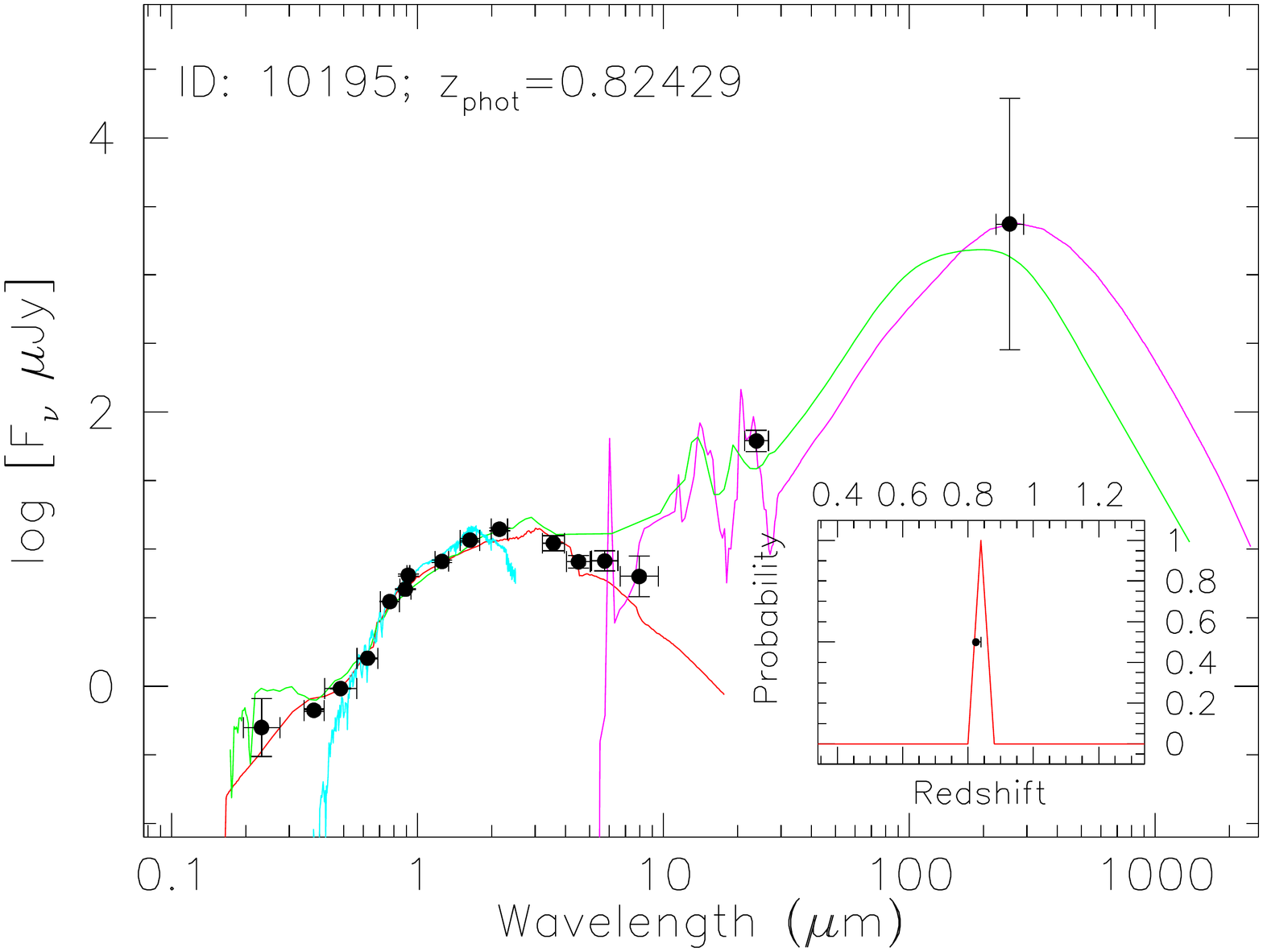} &
    \includegraphics[angle=-90,width=.32\textwidth]{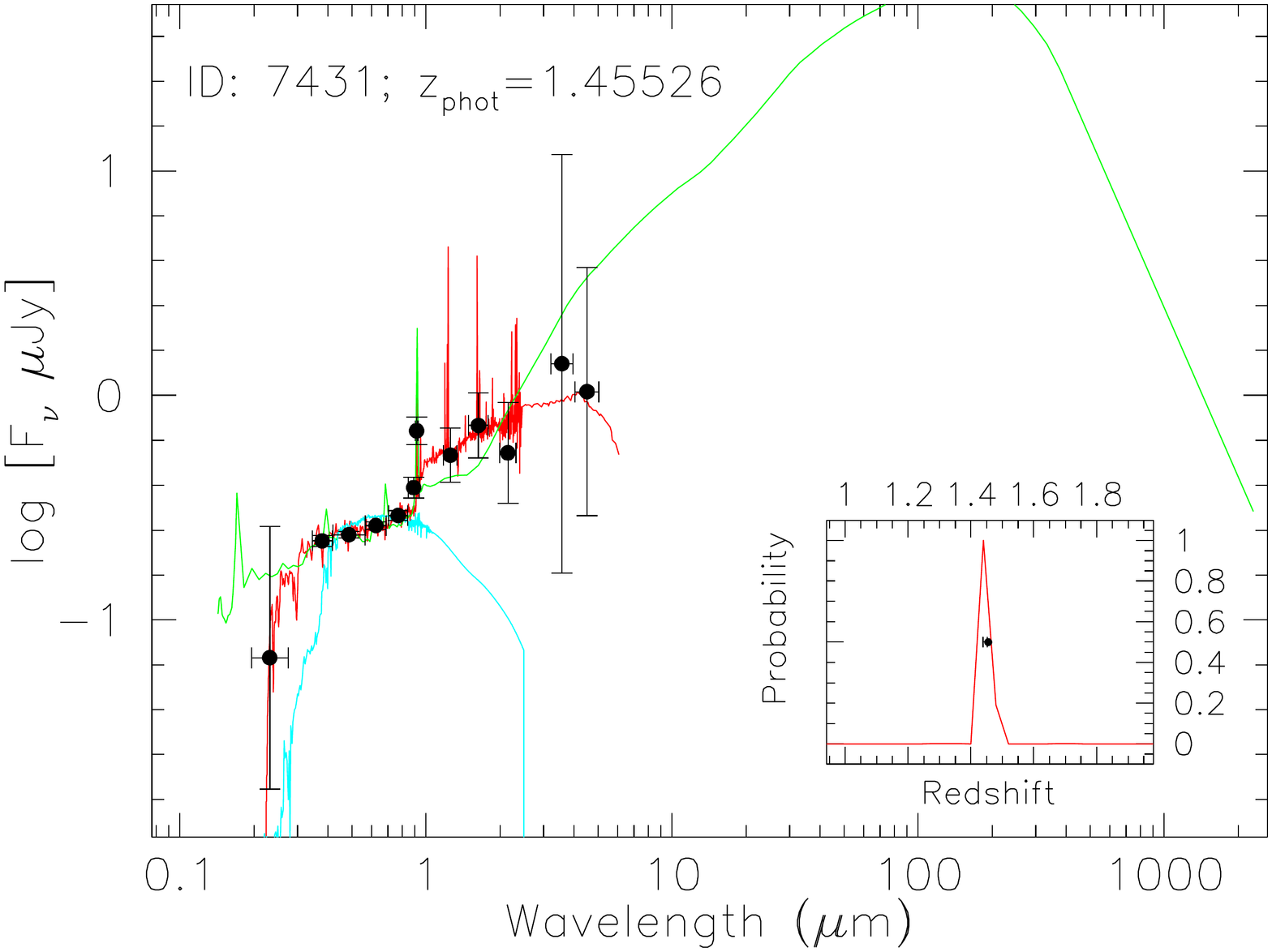} \\
  \end{tabular}
\caption[Examples of SED fitting performed by \lephare.]{Examples of SED fitting
performed by \lephare\ in order to obtain photometric redshifts. From left
to right, OTELO sources with ID 10793, 10195, and 7431 at z$\sim$0.35, 0.82
and 1.46, respectively, are shown. Black dots
are the observed fluxes of the source (including (\oteloint\ data), and
their errors are given by vertical bars. Horizontal bars represent the filter widths.
The red curve represent the best-fitting galaxy template from which the primary solution
of the photometric redshift is obtained.
The blue spectrum represent the stellar template that best fits the observed SED, while the
pink one corresponds to the best fit of the \cite{chary01} templates of the infrared
part of the SED (when applicable) for the photometric redshift obtained. The spectrum in green 
is the best alternative (QSO) photo-z solution using \cite{polletta07} templates.}
\label{lephare_fits}
\end{figure*}

Both solution sets, including either the \oteloint\ flux in the redshift estimate or not, are given 
in the multi-wavelength catalogue described in Section \ref{sec:catfeatures}. Unless otherwise mentioned,
in the remainder of this article we refer to the photometric redshift as that obtained using all the 
{\it core} plus ancillary data. Examples of the photometric redshift fits using \lephare\ for 
ELS candidates at different redshifts are shown in Figure \ref{lephare_fits}.

\begin{figure}[h]
\centering
\includegraphics[angle=-90,width=\linewidth]{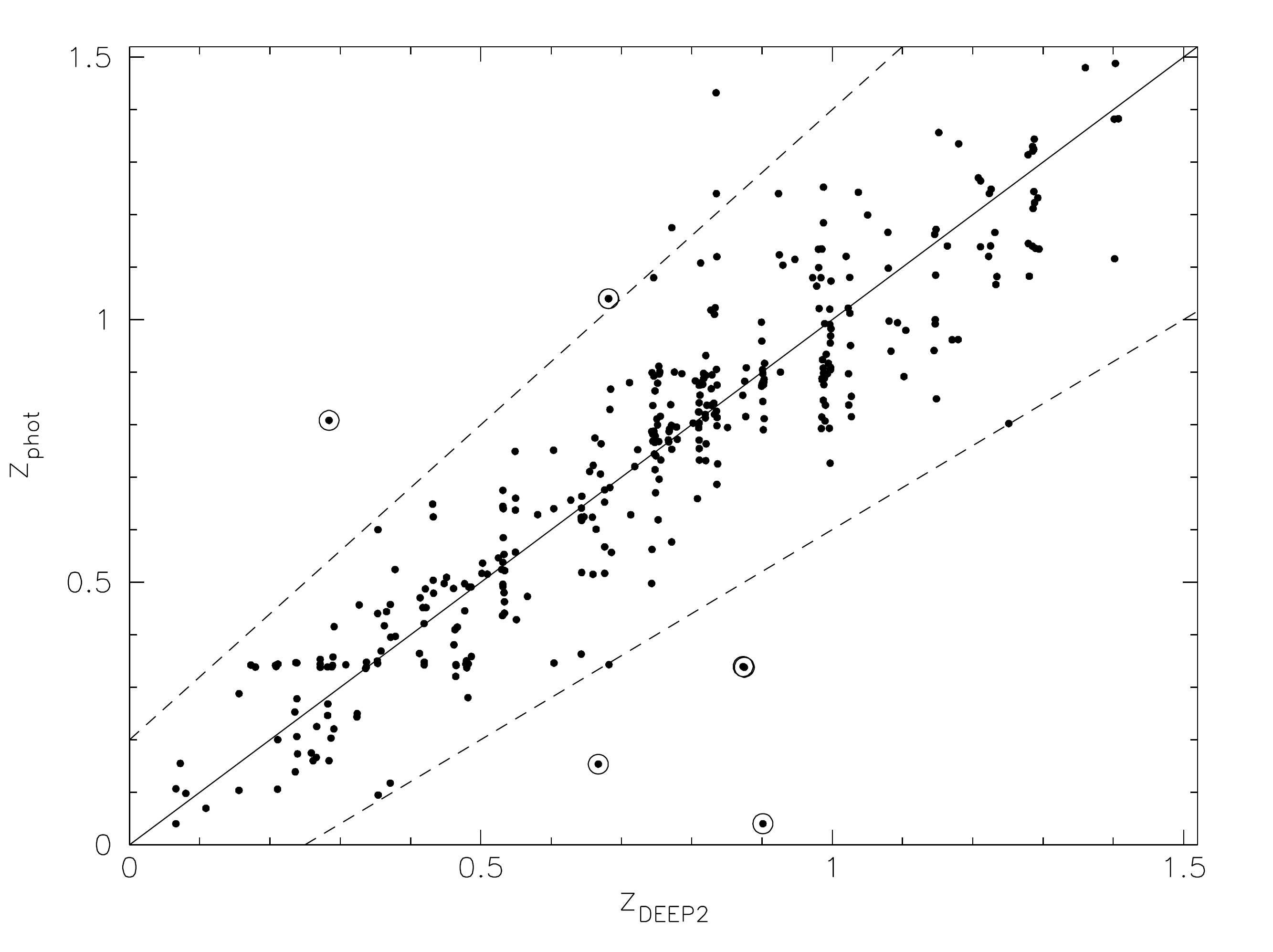}
\caption[Photometric redshifts versus spectroscopic redshifts from \deep]{Plots of
\zp\ as a function of \zs. Dashed lines indicate \disp\ = 0.2. Encircled dots 
correspond to multiple, unresolved sources, or objects with images on the border
of \otelodeep. See text for further details. 
}
\label{zdisp}
\end{figure}

\begin{figure}[h]
\centering
\includegraphics[angle=-90,width=\linewidth]{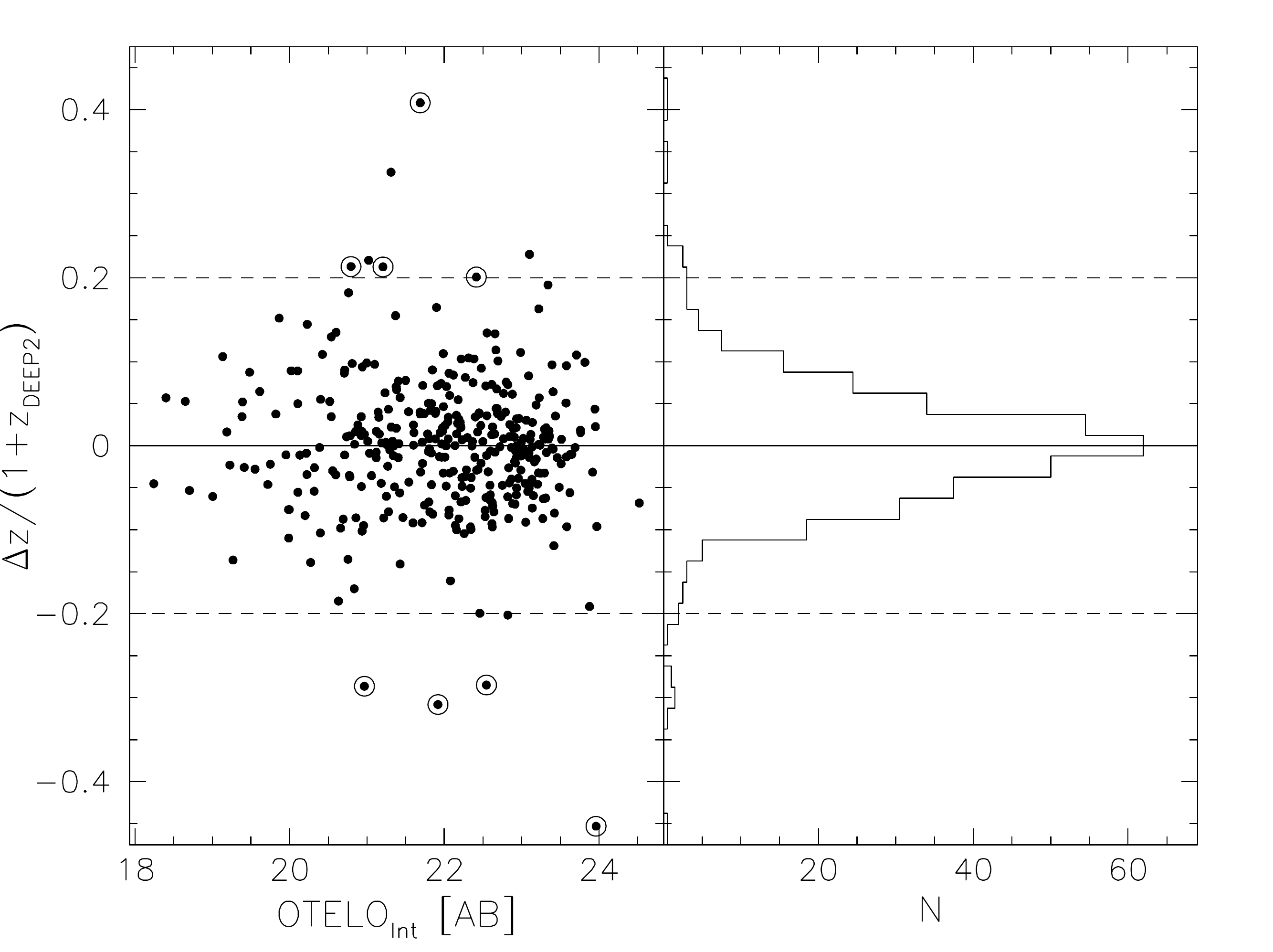}
\caption[Photometric redshifts dispersion distribution and versus \oteloint\ magnitude]{
Plot of \dispnobar\ versus \oteloint\ (left panel) and the distribution of the former (right).
Dashed lines and encircled points have the same meaning as Figure \ref{zdisp}.
}
\label{zspmag}
\end{figure}

To calibrate the accuracy of the computed best-fitted photo-z, we compared them 
to the spectroscopic redshifts up to \zred\ $\sim1.5$ from the \deep\ catalogue (Fig.           
\ref{zdisp}), described in section \ref{sec:addancillary}, which are flagged as 
secure/very secure redshifts. A total of 368 \otelo\ sources
match with \deep\ data entries. The best fit for 17 of these sources corresponds
to QSOs/AGN instead of galaxy templates. Then, assuming that the true redshift, \zreal\ = \zs, 
we found that the redshift accuracy \disp\ is better than 0.2 (0.1) 
for 96\% (87\%) of all selected sources with spectroscopic redshift (see Fig. 
\ref{zspmag}). Nearly identical results are reached if photo-z solutions are estimated
without data from \otelodeep. As noted from this figure, there does not seem to be a 
dependence of accuracy scatter on the apparent magnitude of the source.

We checked one by one all 13 fof the 368 sources that appear outside of the region defined 
by \disp\ $<$ 0.2 in Figures \ref{zdisp} and \ref{zspmag}. This examination went through
the analysis of the SED fit quality, as well as the corresponding pseudo-spectra and 
source cutouts from \acs606\ and \acs814\ imaging data at native spatial resolution. Such images
show evidence of multiple components, unresolved in ground-based data, for seven out of 13 objects.
It is likely that the composed, uneven SED of these sources are poorly represented by photo-z 
solutions based on the selected templates. On the other hand, two of these outliers correspond to 
sources directly located on the border of the \otelo\ FoV. Outliers belonging to these categories
are shown encircled in the mentioned figures, and all but two of them have unusually large
$\chi^2$ values ($>10^3$) from galaxy template fitting when the median of its distribution is 
10.62.

Despite the fraction of inaccurate SED fits ($\sim$4\%), the level of accuracy achieved, 
if extrapolated to all the sources with a fair $\chi^2$-statistic, is good enough to meet 
the primary goal of the redshift estimates;  that is, to obtain a \zrel\ that serves to 
guide the emission-line identification in otelo\ pseudo-spectra. Photo-z distribution 
of preliminary ELS candidates are given in Section \ref{sec:elgselection}. 

\subsection{Data integration and the multi-wavelength catalogue}
\label{sec:catfeatures}

Matched {\it core} (Section \ref{sec:core}) and complementary (Section \ref{sec:addancillary}) catalogues 
were integrated with the results of the photometric redshift analysis described above, yielding the 
\otelo\ multi-wavelength catalogue, whose entries are listed in Table \ref{otelo_catalogue}. The 
catalogue includes solutions with ({\zpY}) and without 
({\zpN}) photometry measured on the \otelodeep\ image, together with secondary solutions if found. As 
described in Section \ref{sec:photoz}, the catalogue also contains redshift solutions of QSO/AGN models. 
For objects with detected FIR emission, total IR luminosity is given. Photo-z data for each template set 
include in addition the best model fitted and the $\chi^2$-statistic.
Spectroscopic redshifts ({\tt zspec}) from \deep\ survey \citep{newman13} with a quality 
flag ${\rm Q}\geqslant3$ were also incorporated. 

\begin{table*}[h]
\vspace*{2mm}
\caption[OTELO catalogue contents]{Contents of the \otelo\ multi-wavelength catalogue.}
\vspace*{-5mm}
\label{otelo_catalogue}
\small
\begin{center}\begin{tabular}{l l}
\hline   \\
 Parameter & Description   \\[0.5mm]
\hline      \\[1pt]
{\tt idobj} & OTELO object number in raw catalogue \\
{\tt x, y} & Logical coordinates of the object in \otelodeep\ image \\
{\tt ra, dec} & Equatorial coordinates (J2000.0) of the object in \otelodeep\ image \\
{\tt dmodelX}\tablefootmark{\ (a)} & Magnitude from detection model-fitting in {\tt X}-band of the {\it core} catalogue \\
{\tt edmodelX} & RMS error on detection model-fitting magnitude \\
{\tt isoareaX} & Isophotal area \\
{\tt fradiX-n} & source radius at {\tt n}-per cent of flux, with {\tt n}=20, 50, 80 \\
{\tt spreadmX} & Spread parameter from model-fitting magnitude \\
{\tt espreadmX} & Spread parameter error from model-fitting magnitude \\
{\tt clasX} & {\tt SExtractor} star/galaxy classifier output \\
{\tt iflagX} & {\tt SExtractor} standard flag \\
{\tt totalZ}\tablefootmark{\ (b)} & Total magnitude in the {\tt Z}-complementary catalogue \\
{\tt etotalZ} & Total magnitude error in the {\tt Z}-complementary catalogue \\
{\tt els\_preliminary} & ELS preliminary candidate flag \\
{\tt zspec} & Spectroscopic redshift from \deep\ survey \citep{newman13} \\
{\tt zp\_reliable\_T04} & Most reliable photo-z from \cfhtls-T04 \citep{coupon09} \\
{\tt zp\_first\_T04} & Best photo-z estimate from \cfhtls-T04 \\
{\tt zp\_left\_68\_T04} & Minimum photo-z from $\Delta\chi^2$ = 1.0 (68\%) from \cfhtls-T04 \\
{\tt zp\_right\_68\_T04} & Maximum photo-z from $\Delta\chi^2$ = 1.0 (68\%) from \cfhtls-T04 \\
{\tt z\_BEST\_deepQ}\tablefootmark{\ (c)} & Best galaxy model photo-z estimate \\
{\tt z\_BEST68\_LOW\_deepQ} & Minimum photo-z from $\Delta\chi^2$  = 1.0 (68\%) \\
{\tt z\_BEST68\_HIGH\_deepQ} & Maximum photo-z from $\Delta\chi^2$  = 1.0 (68\%) \\
{\tt CHI\_BEST\_deepQ} & Lowest $\chi^2$  for galaxy templates \\
{\tt MOD\_BEST\_deepQ} & Galaxy model for best {\tt CHI\_BEST} \\
{\tt EBV\_BEST\_deepQ} & Reddening colour excess E($B$-$V$) \\
{\tt NBAND\_USED\_deepQ} & Number of bands used in photo-z fitting \\
{\tt z\_SEC\_deepQ} & Secondary photo-z solution \\
{\tt CHI\_SEC\_deepQ} & $\chi^2$ for secondary photo-z solution \\
{\tt MOD\_SEC\_deepQ} & Galaxy model for {\tt CHI\_SEC} \\
{\tt z\_QSO\_deepQ} & QSO model photo-z \\
{\tt CHI\_QSO\_deepQ} & $\chi^2$ for {\tt Z\_QSO} \\
{\tt MOD\_QSO\_deepQ} & Galaxy model for {\tt CHI\_QSO} \\
{\tt CHI\_STAR\_deepQ} & $\chi^2$ for {\tt MOD\_STAR} \\
{\tt MOD\_STAR\_deepQ} & Fitted star model \\
{\tt LUM\_TIR\_BEST\_deepQ} & IR luminosity from {\tt MOD\_FIR} (when apply) \\
{\tt LIB\_FIR\_deepQ} & FIR library used (when apply) \\
{\tt MOD\_FIR\_deepQ} & FIR model fitted (when apply) \\
{\tt CHI\_FIR\_deepQ} & $\chi^2$ for FIR template fitting (when apply) \\
\hline
\end{tabular}
\end{center}
\tablefoot{}
\tablefoottext{a}{where {\tt X} is D (\otelodeep), \bandu, \bandg, \bandr, \bandi, \bandz\ (CFHT filters), \acs606, \acs814, \bandj, \bandh\ or \bandk.} \\
\tablefoottext{b}{where {\tt Z} is \chandra, FUV or NUV from \galex, 3.6, 4.5, 5.8, 8.0, or 24$\,\mu$m from \spitzer, or 100, 160, 250, 300, 500$\,\mu$m from \herschel.} \\
\tablefoottext{c}{where {\tt Q} is {\tt Y} or {\tt N}: including/not including \oteloint\ data in photo-z solutions, respectively.} \\
\end{table*}
\normalsize

Following \cite{coupon09}, we can define an estimate of the photo-z uncertainty $\delta\,$\zp\ by

\begin{equation}
\begin{aligned}
&\delta\,({\scalebox{.9}{\tt z\_BEST\_deepQ}}) = \\
&\lvert {\scalebox{.9}{\tt z\_BEST\_LOW\_deepQ}} - {\scalebox{.9}{\tt z\_BEST\_HIGH\_deepQ}} \rvert / 2,
\end{aligned}
\end{equation}

\noindent where ${\tt z\_BEST\_[LOW,HIGH]\_deepQ}$ are the low and high redshift values included in the
\otelo\ multi-wavelength catalogue, corresponding
to the 68\% confidence interval of the photo-z probability distribution function (PDF) obtained
with \lephare, and {\tt Q} = {\tt Y,N}, depending whether or not the \oteloint\ data are included in photo-z solutions,
respectively. Thus, based on the photo-z accuracy given above, a useful
criterion to discard possible photo-z outliers could be defined by

\begin{equation}
\begin{aligned}
\label{eq:zunc}
\delta\,({\scalebox{.9}{\tt z\_BEST\_deepQ}})\leqslant 0.2\ (1+{\scalebox{.9}{\tt z\_BEST\_deepQ}}).
\end{aligned}
\end{equation}

Thus, if we choose \zp\ = {\tt z\_BEST\_deepY}, from the 9709 sources with non-null photo-z solutions, a total
of 6600 have an uncertainty $\delta\,$\zp\ $<$ 0.2 (1+\zp). This constraint is used in what follows, for example, to
debug preliminary ELS selections.

\subsection{\otelo\ public data release}
\label{sec:datarelease}
From extensive data described in previous sections, the \otelo\ Team has selected the following value-added
data for the first public data release: 

\begin{itemize}
\item {A basic catalogue containing (i) the identification of the source, (ii) equatorial coordinates, 
(iii) the optical and NIR photometric data from the {\it core} catalogue (Section \ref{sec:core}), (iv) the best 
photometric redshift solutions and uncertainties (Section \ref{sec:catfeatures}), and (v) a morphology classification 
flag (Nadolny et al. 2019; in prep.).}
\item {A supplementary catalogue containing (i) the identification of the source, (ii) the X-ray, UV, MIR and FIR 
photometric data and additional redshift data (i.e.\ \cfhtls\ and \deep), all obtained from complementary catalogues
as described in Section \ref{sec:addancillary}.}
\item  {A public version of the {\em Web-based GUI} described in \otelotwo, containing (i) the identification
data of the source, (ii) a set of 8\arcsec$\times\,$8\arcsec\ image cutouts of the source from ground-based
and HST imaging data available, (iii) all the photometric redshift solutions included in Table \ref{otelo_catalogue} 
and their corresponding SED (as represented in Figure \ref{lephare_fits}, and (iv) the pseudo-spectrum of the source.} 
\end{itemize}

Both catalogues and the Web-based visualisation tool, along with the corresponding Explanatory Supplement, will be 
publicly available from the \otelo\ survey URL (see Section \ref{sec:intro}) from the second half of 2019.

\section{\otelo\ demographics}
\label{sec:demography}

\subsection{Star--galaxy separation}
\label{sec:stargalaxy}

The \otelo\ survey could be used as a true sensitive probe of Galactic thick disk and halo 
structures \citep{alfaro03}. 
A reliable segregation of stars present in the \otelo\ field is not only interesting in itself because of the 
existence of e.g. potential cool and ultra-cool sources, but also for the analysis purposes that
follow in the next sections of this work. To this aim, the standard approach takes advantage of a combination of
the source image geometry and optical/NIR colour criteria.  

Apart from the usual {\tt CLASS\_STAR} star/galaxy separator of {\tt SExtractor}, we also obtained the
{\tt SPREAD\_MODEL} classifier for each band, which is a by-product of the PSF-model photometry
that quantifies the differences between PSF-like and resolved objects  through a linear discriminant
\citep{desai12}. In a comparison of different methods used for 
star/galaxy separation, \cite{annunziatella13} concluded that the {\tt SPREAD\_MODEL} largely improves
the classification of extended and point-like sources down to such faint apparent brightness as $B$=26 for stellar
ones, above the performance of the {\tt CLASS\_STAR} parameter or software applications such as {\tt DAOPHOT}
and {\tt ALLSTAR}. 

\begin{figure}[htb]
\centering
\includegraphics[angle=0,width=\linewidth]{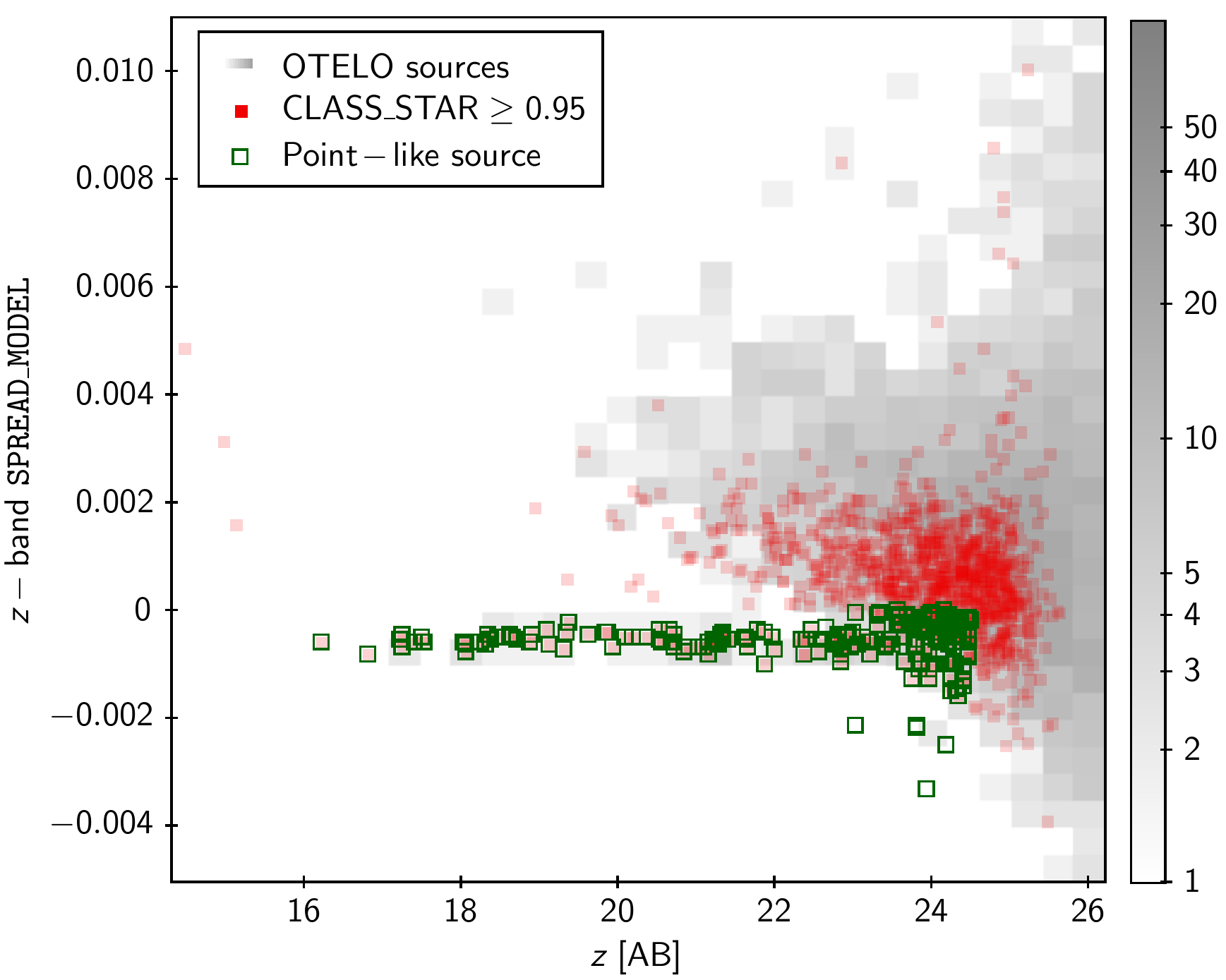}
\caption[Star-galaxy separation]{Example in the  \bandz-band of the {\tt SPREAD\_MODEL} performance in
the star--galaxy separation. Red squares represent sources with {\tt CLASS\_STAR} $\ge$ 0.95. Point
sources selected (open squares) are brighter than AB=24.5 \bandz-band magnitude and limited by a 
{\tt SPREAD\_MODEL} value below which the star-like sequence is clear.
}
\label{spread}
\end{figure}

Figure \ref{spread} shows the distribution of the {\tt SPREAD\_MODEL} compared with the 
{\tt CLASS\_STAR} parameter  for the \otelo\ sources detected in \bandz-band. Even though any photometric band can 
be used for this task, the choice of the \bandz-band utilizes data that show the smallest mean PSF FWHM 
(see Table \ref{core_catalogue_summary}). As expected, from this plot it is clear that {\tt SPREAD\_MODEL} 
gives a sharp sequence of stellar-like sources around zero, and that extended sources display increasingly
positive values. Remarkably, the vast majority of sources that could have been selected as point-like 
using the {\tt CLASS\_STAR} $\geqslant$ 0.95 criterion are clearly deviate from the star-like sequence up
to AB$\sim$24. At this value both sequences begin to merge. On this basis, we adopted AB=24.5 \bandz-band 
magnitude as a flux limit for star/galaxy separation and a {\tt SPREAD\_MODEL} parameter threshold below 
zero to obtain a collection of 197 point sources. 

\begin{figure}[htb]
\centering
\includegraphics[angle=0,width=\linewidth]{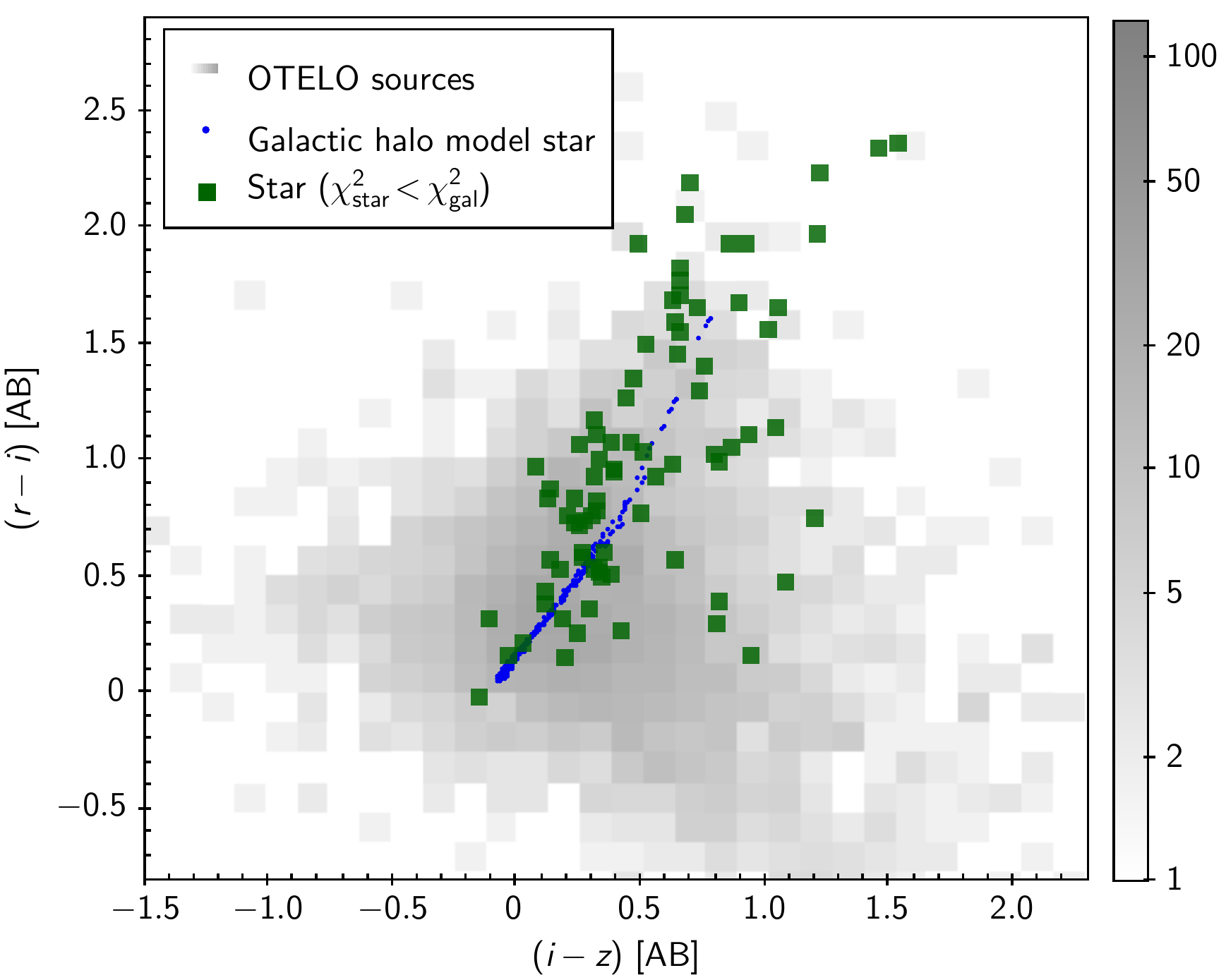}
\caption[Optical colour--colour diagram]{($r-i$) versus ($i-z$) colours distribution of the final \otelo\
catalogue. Green squares represent sources classified as stars and the blue dots trace a sequence of
typical Galactic halo stars from \cite{allende14}. See text for details.
}
\label{optical_colors}
\end{figure}

As mentioned in Section \ref{sec:photoz}, an extensive set of AGN and stellar templates was used during
\lephare\ runs to obtain best SED fits of all \otelo\ sources, independently of the main photo-z 
solution obtained through galaxy templates. Following \cite{yang14}, we used the $\chi^2_{star} < \chi^2_{galaxy}$ 
as a valid criterion to depurate the star selection based on the {\tt SPREAD\_MODEL}. The 81 point-like sources 
that met this condition are represented in Figure \ref{optical_colors} and constitute a fair sample of 
Galactic source candidates subject to possible spectroscopic follow-up. More than 90\% 
of these sources have colours congruent with those of a sequence of Galactic halo stars obtained
from models of \cite{allende14}. A similar fraction of the star candidates are located in the expected 
region of the {\it BzK}-equivalent diagram presented in Section \ref{sec:colourdiagrams}.

The best templates fitted to the star candidates are mainly distributed between spectral types K and
M (14 and 46 sources, respectively), of which 45 correspond to sub-dwarfs. Most of the templates
fitted belong to the Pickles Atlas \citep{pickles98}, and five of the M-type stars were well-fitted by SpeX 
Prism\footnote{Supra at \ref{spex}.} templates.   

A catalogue simulation run of the Besan\c con model\footnote{\tt http://model.obs-besancon.fr} of the Galaxy 
\citep{robin03}, without kinematics, predicts for the \otelo\ field a total of 112 stars between magnitudes 
16.0 and 24.5 (\bandz-band of \cfhtls) in a solid angle of 0.014 deg$^2$ around l=96$^\circ$.46 and 
b=60$^\circ$.03. According to this, the expected stars on the \otelo\ field would be distributed between 
spectral types F3 and M7, with a luminosity class corresponding to dwarfs or sub-dwarfs, except two that 
would belong to the red sub-giants. A detailed study of the \otelo\ Galactic component with new statistical
estimates is a part of a forthcoming contribution. 

\subsection{Preliminary ELS selection}
\label{sec:elgselection}

According to the science goals of \otelo, once obtained the calibrated pseudo-spectra
corresponding to each entry of the multi-wavelength catalogue, it is necessary to select the
ELS candidates. The creation of robust lists of ELS segregated by emission line(s) necessarily 
involves a sequence of procedures, from which the first two are described in this paper.
The completion of these tasks contributes towards quantifying the potential and intrinsic merits 
of the survey. Further analysis of the preliminary ELS selection is a part of the scope 
of \otelotwo.

Preliminary ELS selection starts with  blind processing of the individual pseudo-spectra.
A straightforward application reads each calibrated pseudo-spectrum, determines a constant 
pseudo-continuum flux density ($f_{\rm c}$) and then estimates median absolute deviation 
($\sigma_{\rm MAD}$) of the flux vector (f[i], i=1 to N, and N = number of the RTF scan slices).
This is a more robust measure of the pseudo-continuum fluctuations than the simple standard 
deviation when deal with data with strong fluctuations (i.e.\ the absorption/emission 
features in a pseudo-spectrum). Assuming that the fluctuations of the pseudo-continuum are 
normally distributed, $\sigma_{\rm MAD}$ can be simply converted to standard deviations 
$\sigma_{\rm c}$ around the median multiplying by a factor that depends on the quantile 
function, $\Phi^{-1}$. 

An emission/absorption feature in a given pseudo-spectrum is considered noticeable if (i) at 
least two consecutive slices of the pseudo-spectrum are above/below a value defined by 
$f_{\rm c}$ +/- 2 $\times$ $\sigma_{\rm c}$, or (ii) only one slice flux is above/below this 
value, but with an adjacent point above/below $f_{\rm c}$ +/- $\sigma_{\rm c}$
and the other above/below $f_{\rm c}$. An example of the former is the [NII]$6584\AA$ emission line in the 
synthetic pseudo-spectrum represented in Figure \ref{pseudospectra}.

Both criteria are complementary and not mutually exclusive (i.e.\ they are partially correlated). 
Together, they are consistent with the condition behind the source detection threshold in the \otelodeep\ 
image (Section \ref{sec:sextraction}). If the noise around the pseudo-continuum approximates the Gaussian
distribution, the independent probabilities that any feature in a given pseudo-spectrum with N $\geq$ 36 data 
points satisfies criteria (i) or (ii) because it is a random event are, respectively, 0.018 and 0.051. Therefore, 
the joint probability of having false-positive emission/absorption line allocation in a pseudo-spectrum 
is at most $\sim$6\%.

\begin{figure}[htb]
\centering
\includegraphics[angle=0,width=\linewidth]{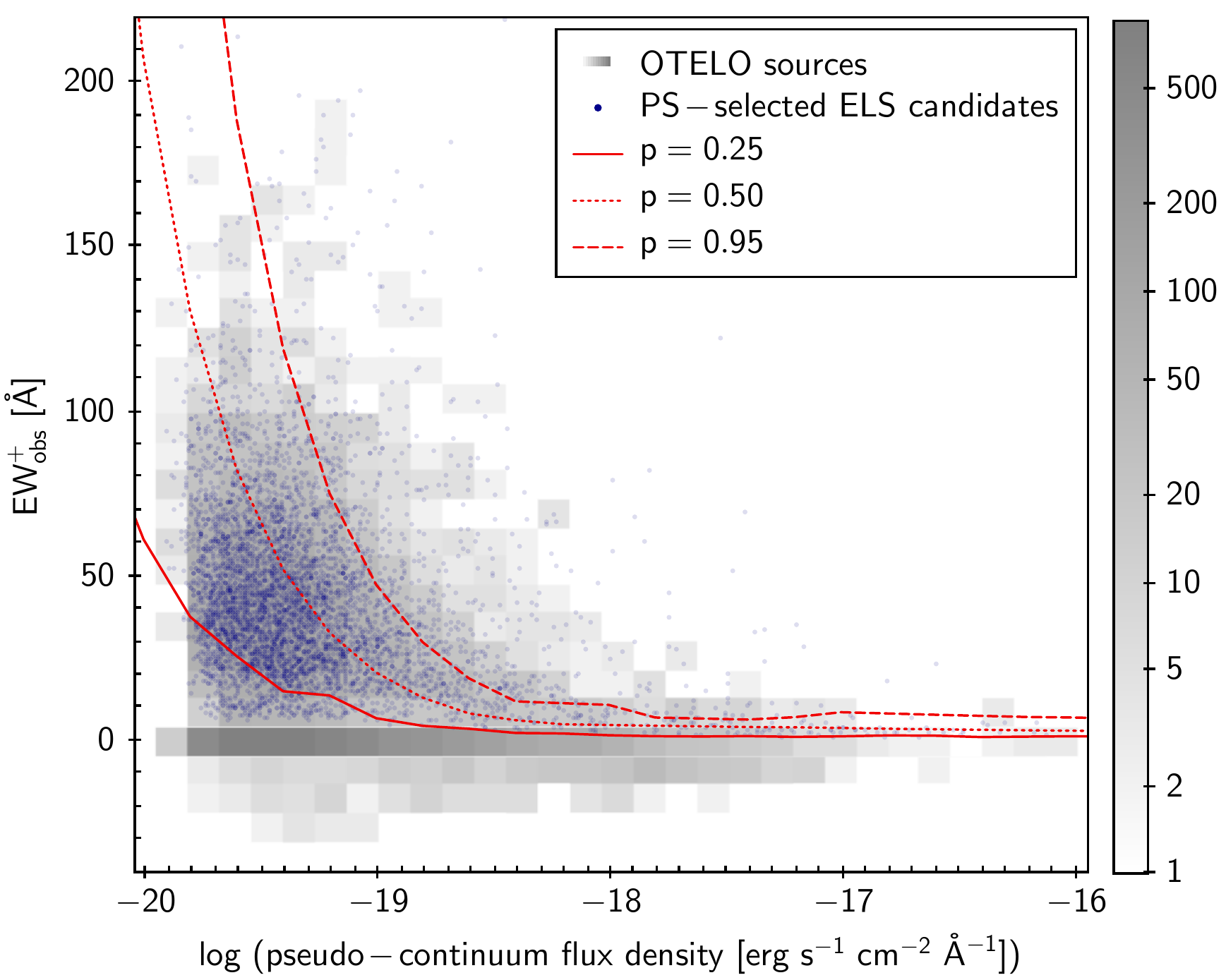}
\caption[EW--continuum flux density diagram for emitter selection from PS]{Upper-bound of
observed equivalent width (EW$^+_{\rm obs}$) as a function of the flux density of the pseudo-continuum
measured on the \otelo\ pseudo-spectra (PS). The preliminary ELS candidates, selected as described in the text,
are pictured as blue dots. The red lines represent the minimal
EW$^+_{\rm obs}$ at p=0.25, p=0.50 and 0.95, obtained from RTF data simulations 
described in \otelotwo. In this figure
(and hereafter) the background grid represents the binned sum of all \otelo\ sources in the space of
the variables represented.
}
\label{prelg}
\end{figure}

Using these criteria, an upper bound of the observed equivalent width (EW$^+_{\rm obs}$) in emission/absorption 
is measured in each pseudo-spectrum by integrating all the flux contributions above/below the pseudo-continuum.
All sources of \otelo\ survey in the EW$^+_{\rm obs}$ vs. $f_{\rm c}$ space are represented in 
Figure \ref{prelg}. A total 
of 5322 from 11237 sources met one or both selection criteria in terms of emission feature. This subset
constitutes the preliminary ELS candidates selected from pseudo-spectra. These sources are highlighted
in Figure \ref{prelg}. 

Assuming a Gaussian profile, the width of an emission line in a pseudo-spectrum of \otelo\
can be written as the sum in quadrature of line intrinsic and instrumental widths,
the latter being given by Eq. \ref{eq:effwidth} for \otelo\ RTF data, which is close to 19.4 \AA.
This condition is a good approximation as long as the line intrinsic width
$\ga \delta\lambda\,_{\rm FWHM}/2$ = 6 \AA\ (Section \ref{sec:observations}).

When the observed width of a given emission line reaches $\sim$50 \AA, that is $\sim$25\% of 
the \otelo\ spectral range (i.e.\ $\sim$230 \AA\ or the FWHM of the \otelocustom\ filter),
the convolved line begins to saturate the pseudo-spectrum, resulting in an overestimate of 
the pseudo-continuum. Under these circumstances, a likely ELS candidate is automatically 
ruled out if only the information given by the pseudo-spectrum is regarded. For instance, 
the width of AGN broad lines span from $\Delta v_{\rm FWHM} \sim$500 km s$^{-1}$ (a little more 
than the upper limit of the narrow line component) to $\ga10^4$ km s$^{-1}$, with typical values 
of $\sim$5000 km s$^{-1}$ \citep{peterson97}. This translates into 
25 \AA\ $<$ FWHM $\la$ 300 \AA\ (RTF-convolved) at the central wavelength of the \otelo\ scan. In addition,
RTF data simulations using synthetic pseudo-spectra \citep[][\otelotwo]{ramonperez17} support these 
effects, and confirm that potential ELS with intrinsic line widths above FWHM $\sim$60 \AA\ can be lost if 
an alternative recovery method is not enforced. \otelo\ data
can be used to complement the ELS selection from pseudo-spectra by adopting the colour-excess
technique traditionally used in narrow-band surveys.
 
There are several approaches to the colour selection of ELS \citep{pascual07}. The particular
case developed below corresponds to the scheme of using an intermediate-band filter (\otelocustom)
to measure the line+continuum flux (\oteloint), inside a broader one for continuum sampling 
(\bandz-band). The complementary search for ELS is then performed by selecting (\bandz-\oteloint) 
colour excess with respect to a zero point and above its uncertainties, as a function of the 
\oteloint. Comparing this colour around zero value (i.e.\ |(\bandz-\oteloint)| $<$ 0.5; 4337 sources) 
with the corresponding flux excess in \otelo\ pseudo-spectra, parametrized by the EW$^+_{\rm obs}$,
we found that the zero-point of the (\bandz-\oteloint) colour diagram does not have offsets above
0.008 mag, with a near zero dependence on the EW$^+_{\rm obs}$. Therefore, the (\bandz-\oteloint)
zero calibration is accurate enough to be used for ELS extraction.

Following \cite{bunker95}, isolines of colour-excess significance were computed using

\begin{equation}
\label{eq:colorsel}
($\bandz-\oteloint$) = -2.5 \log_{10}\ \left[1 - \delta\ \Sigma\ 10^{-0.4\ (zp-{\scalebox{.7}{\oteloint}})}\right],
\end{equation}

\noindent where $zp$ = 30.504 is the photometric zero-point of \oteloint\ data, $\delta$ is the sum in 
quadrature of the sky background in each band, and $\Sigma$ the colour excess significance. 
In this case, additional cuts or limits in colour (or equivalently observed EW) were 
disregarded. 

\begin{figure}[htb]
\centering
\includegraphics[angle=0,width=\linewidth]{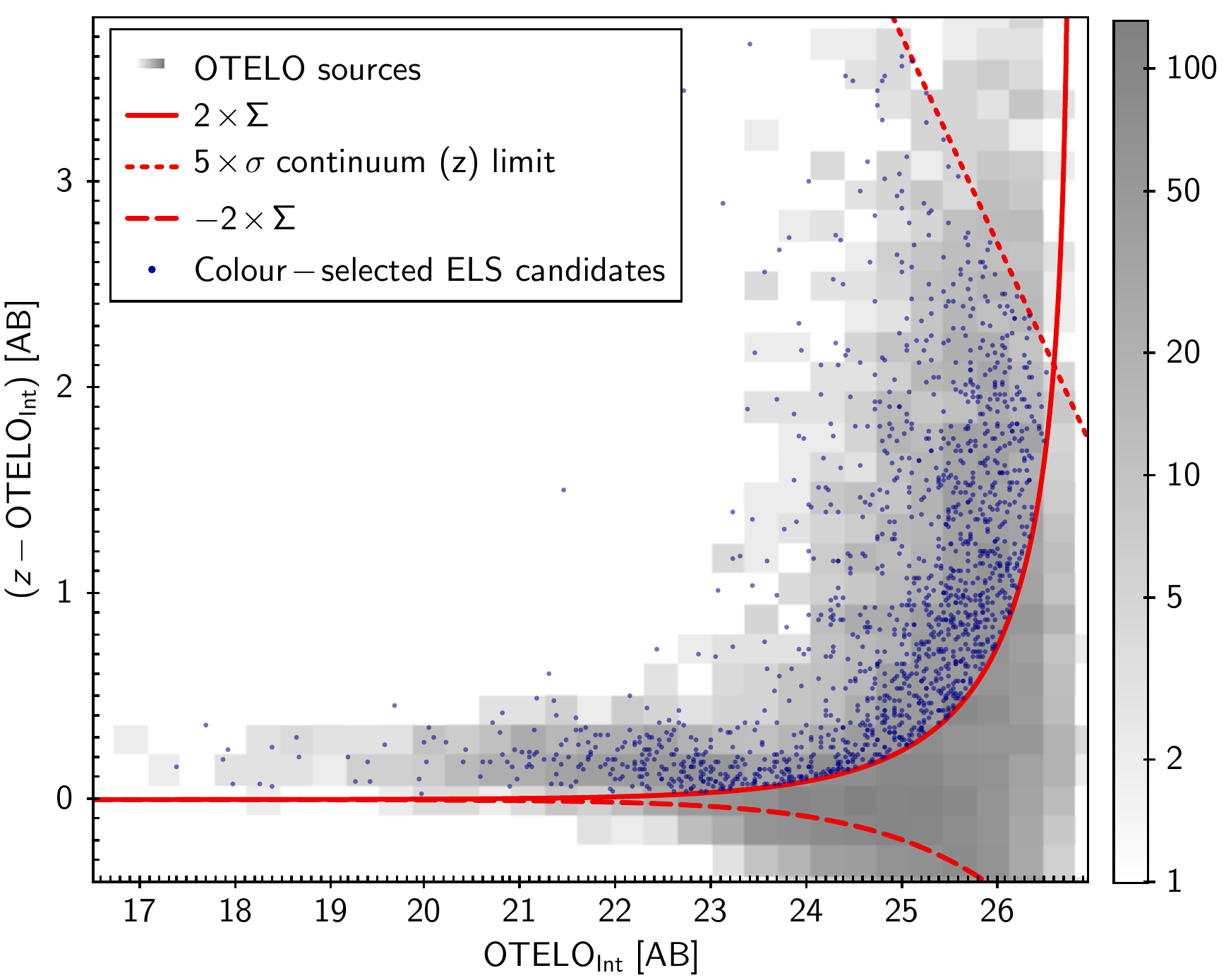}
\caption[Colour-magnitude diagram for emitter selection]{A (\bandz-\oteloint) colour-magnitude
diagram used for a selection of ELS complementary to the preliminary one based on pseudo-spectra.
The candidates are represented by blue dots. This selection is bounded by a 2-$\Sigma$ isoline of 
colour significance, and a 5-$\Sigma$ limit in \bandz\ (dotted line). All fluxes correspond to
the {\tt DETMODEL} measurements. \oteloint\ flux is measured on the \otelodeep\ image.
}
\label{zcolor}
\end{figure}

All sources with $\Sigma > 2$ and a signal-to-noise ratio on the continuum better than 5$\sigma$ were
selected, yielding a total of 3226 ELS candidates. From these sources, 1542
(48\%) were already selected by emission excess in pseudo-spectra as described above. Figure \ref{zcolor}
shows the constraints described besides the remaining 1684 colour-selected only ELS candidates.

If, therefore, the ELS candidate set that meets the selection criteria applied to pseudo-spectra is joined 
with the colour-selected ELS one without redundancies, a total of 7006 preliminary line emitters is obtained.
Regardless of the selection criterion used, all ELS candidates in the \otelo\ multi-wavelength catalogue
are identified with the corresponding flag, {\tt els\_preliminary} (see Table \ref{otelo_catalogue}).

\begin{figure}[h]
\centering
\includegraphics[angle=-90,width=\linewidth]{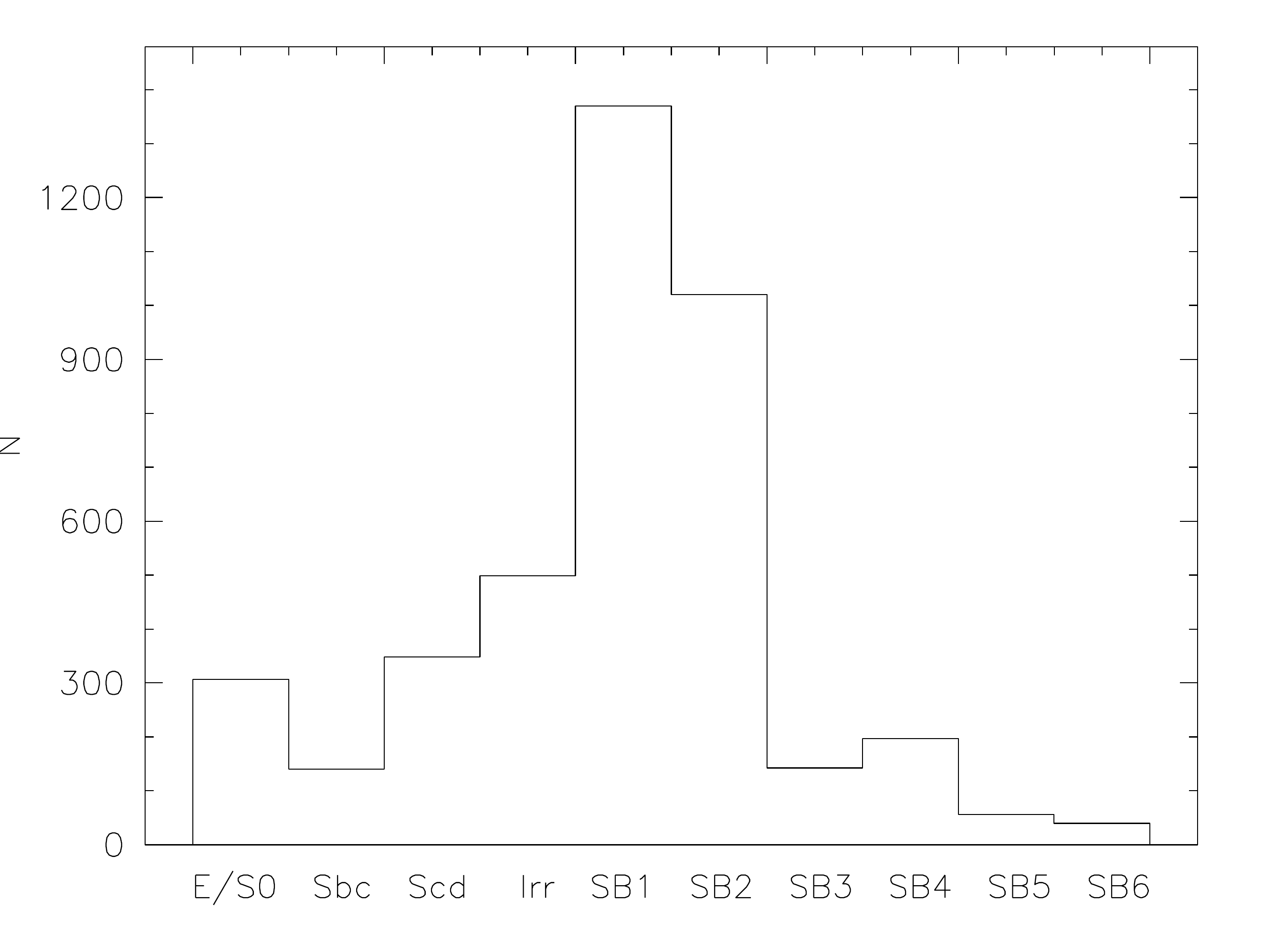}
\caption[Template distribution]{
Primary \zp\ template distribution the preliminary ELS candidates
of \otelo, as described in Section \ref{sec:elgselection}.
Names of templates are the same as in Figure \ref{galtemplates}. The templates
with the highest UV-slope (SB1 \& SB2) are the best fits for more than 55\% of
the ELS candidates that fulfil the condition in Equation \ref{eq:zunc}.
}
\label{templatedist}
\end{figure}

The sky footprint and spectral range of the  \otelo\ survey define comoving volumes corresponding to
noticeable emission lines. In order to segregate those preliminary ELS candidates by chemical species 
and go one step further in refining the number statistics of line emitters, it is necessary to cross-correlate 
the selection described above with photometric redshift data. Accordingly, from the 7k+ set of preliminary ELS,
a total of 4336 sources satisfy the photo-z constraint in Equation \ref{eq:zunc}. The frequency of best galaxy 
template fitting for these ELS candidates is shown in Figure \ref{templatedist}. The best fits for most than 
a half (55\%) of ELS correspond to the two hardest UV-slope starburst templates.

\begin{figure}[h]
\centering
\includegraphics[angle=-90,width=\linewidth]{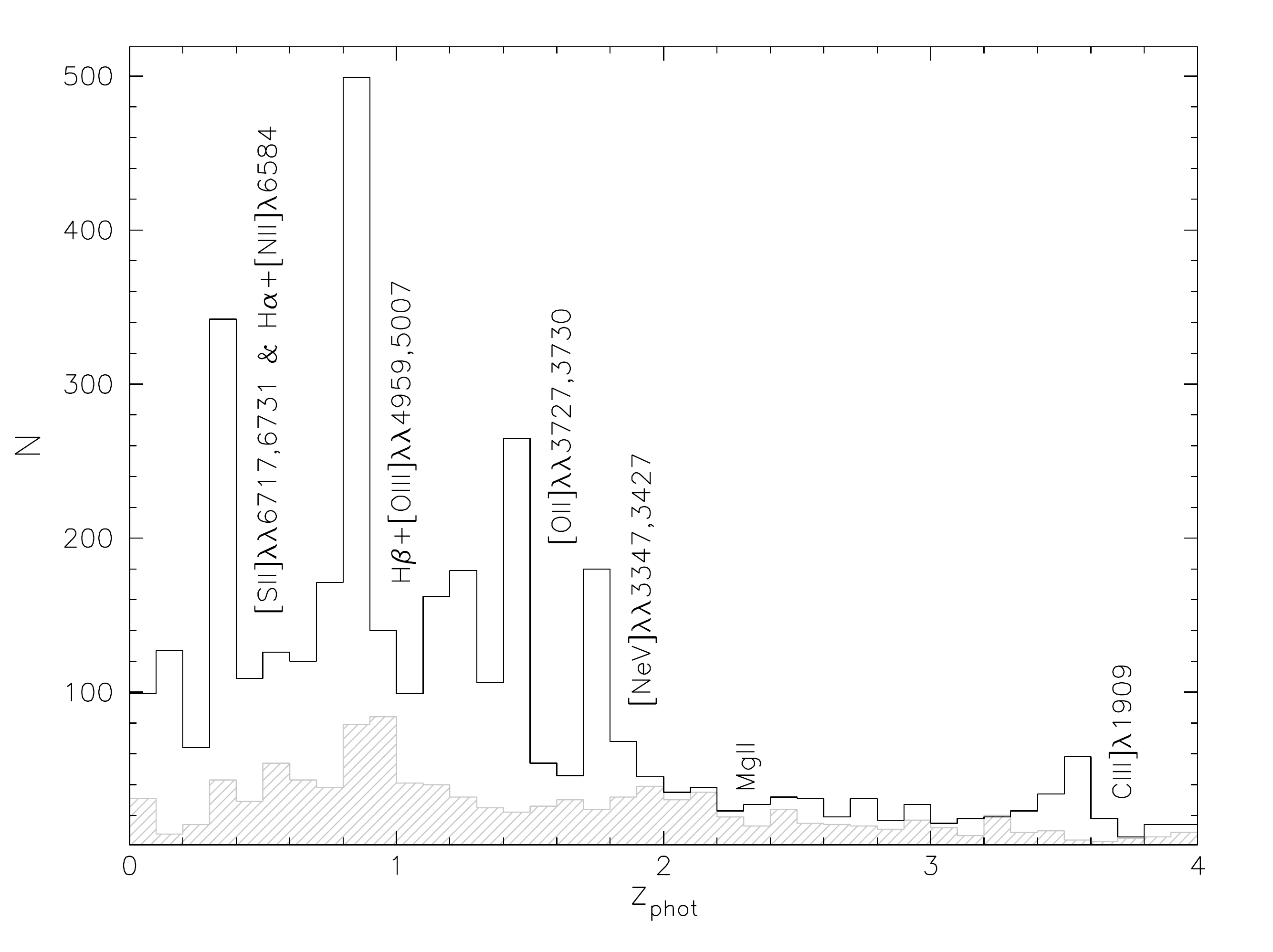}
\caption[Distribution of photometric redshifts.]{Distribution of
photometric redshifts up to \zred=4 obtained with \lephare\ for the preliminary ELS candidates
of \otelo, as described in Section \ref{sec:elgselection}.
As a comparison, the grey filled histogram corresponds to the photo-z of HDFN
galaxies \citep{fersoto99} under similar conditions to ours. The remarkable excess in the
histogram is labelled on the right side with the name of the corresponding redshifted emission 
lines.}
\label{zhist}
\end{figure}

The photo-z distribution of these sources up to redshift \zred=4 is provided by the histogram of
Figure \ref{zhist}, whose main spikes are associated to the bright emission lines scanned below this 
redshift. A total of 1516 candidates can be directly attributable to such spectral features, and a raw 
census of them is given in Table \ref{otelo_emitters_preliminary}, along with the redshift ranges and 
comoving volumes explored.

Regarding the high redshift (\zred\ $>$ 4) regime, a total of 611 from the 4336 ELS with acceptable 
uncertainty $\delta({\tt z\_BEST\_deepQ})$ are catalogued, and 296 of them are preliminary ELS
candidates for the species listed in Table \ref{otelo_emitters_preliminary_hiz}.  

\begin{table*}[h]
\vspace*{2mm}
\caption[OTELO preliminary emitters]{Counts of preliminary ELS candidates with acceptable uncertainty and
photo-z $\lesssim$ 4.}
\vspace*{-5mm}
\label{otelo_emitters_preliminary}
\begin{center}\begin{tabular}{c c c c}
\hline   \\
 Emission line(s) & Redshift range & Comoving volume [Mpc$^3$] & Raw ELS counts \\[0.5mm]
\hline      \\[1pt]
\sii & 0.33-0.38 & 1.769 $\times\ 10^3$ & 254 \\[0.5mm]
\ha\ + \nii & 0.36-0.42 & 2.463 $\times\ 10^3$ & 129 \\[0.5mm]
\oiii & 0.79-0.87 & 8.958 $\times\ 10^3$ & 416 \\[0.5mm]
\hb & 0.85-0.91 & 7.134 $\times\ 10^3$ & 186 \\[0.5mm]
\oii & 1.41-1.49 & 1.388 $\times\ 10^4$ & 261 \\[0.5mm]
\nev & 1.63-1.78 & 2.787 $\times\ 10^4$ & 209 \\[0.5mm]
\mgii & 2.21-2.32 & 1.995 $\times\ 10^4$ & 26 \\[0.5mm]
\ciii & 3.60-3.87 & 4.914 $\times\ 10^4$ & 35 \\[0.5mm]
\hline
\end{tabular}
\end{center}
\end{table*}
\normalsize

\begin{table*}[h]
\vspace*{2mm}
\caption[OTELO preliminary emitters - High-z]{Counts of high-\zred\ preliminary ELS candidates
with acceptable uncertainty.}
\vspace*{-5mm}
\label{otelo_emitters_preliminary_hiz}
\begin{center}\begin{tabular}{c c c}
\hline   \\
 Emission line(s) & Redshift range & Raw ELS counts \\[0.5mm]
\hline      \\[1pt]
\niii & 4.12-4.31 & 43 \\[0.5mm]
\oiiip & 4.39-4.58 & 22 \\[0.5mm]
\heii & 4.47-4.66 & 21 \\[0.5mm]
\civ & 4.79-4.99 & 10 \\[0.5mm]
\nv & 6.22-6.64 & 200 \\[0.5mm]
+ \lya & & \\
\hline
\end{tabular}
\end{center}
\end{table*}
\normalsize

Summarizing some contents of this Section, Table \ref{otelo_stats} condense the number statistics of
ELS after their preliminary examination. As a function of redshift, only a fraction of these ELS 
candidates would remain in the final
lists of true-positive emitters. Further refinements leading to these lists and the subsequent
analysis of their properties constitute the rationale of \otelo's forthcoming papers.

\begin{table*}[h]
\vspace*{2mm}
\caption[OTELO catalogue number statistics]{Essential number statistics of \otelo.}
\vspace*{-5mm}
\label{otelo_stats}
\begin{center}\begin{tabular}{c c}
\hline   \\
 Feature & Number \\[0.5mm]
\hline      \\[1pt]
Catalogue entries & 11237 \\
OTELO sources at 50\% completeness &  9862 \\
Sources with non-null photo-z solution (\zpY) & 9709 \\
Sources with $\delta$ (\zpY) $\leq$ 0.2 (1+\zpY)\ [Eq. \ref{eq:zunc}] & 6600 \\
Preliminary star candidates & 81 \\
Preliminary ELS candidates selected from pseudo-spectra & 5322 \\
Preliminary ELS candidates colour-selected only & 1684 \\
Preliminary ELS candidates with non-null photo-z solution & 4336 \\
Preliminary ELS candidates only detected in \bandj+\bandh+\bandk & 208 \\
Absorption line system candidates from pseudo-spectra & 483 \\
\hline
\end{tabular}
\end{center}
\end{table*}
\normalsize

\subsection{Colour--colour diagrams}
\label{sec:colourdiagrams}

Colour--colour plots provide validity tests for quality assurance of the multi-wavelength catalogue
and provide insights into the internal consistency among photometric redshifts, apparent brightnesses, 
diagnostic colours, and object types. Three different examples are discussed in the following paragraphs.

\begin{figure}[htb]
\centering
\includegraphics[angle=0,width=\linewidth]{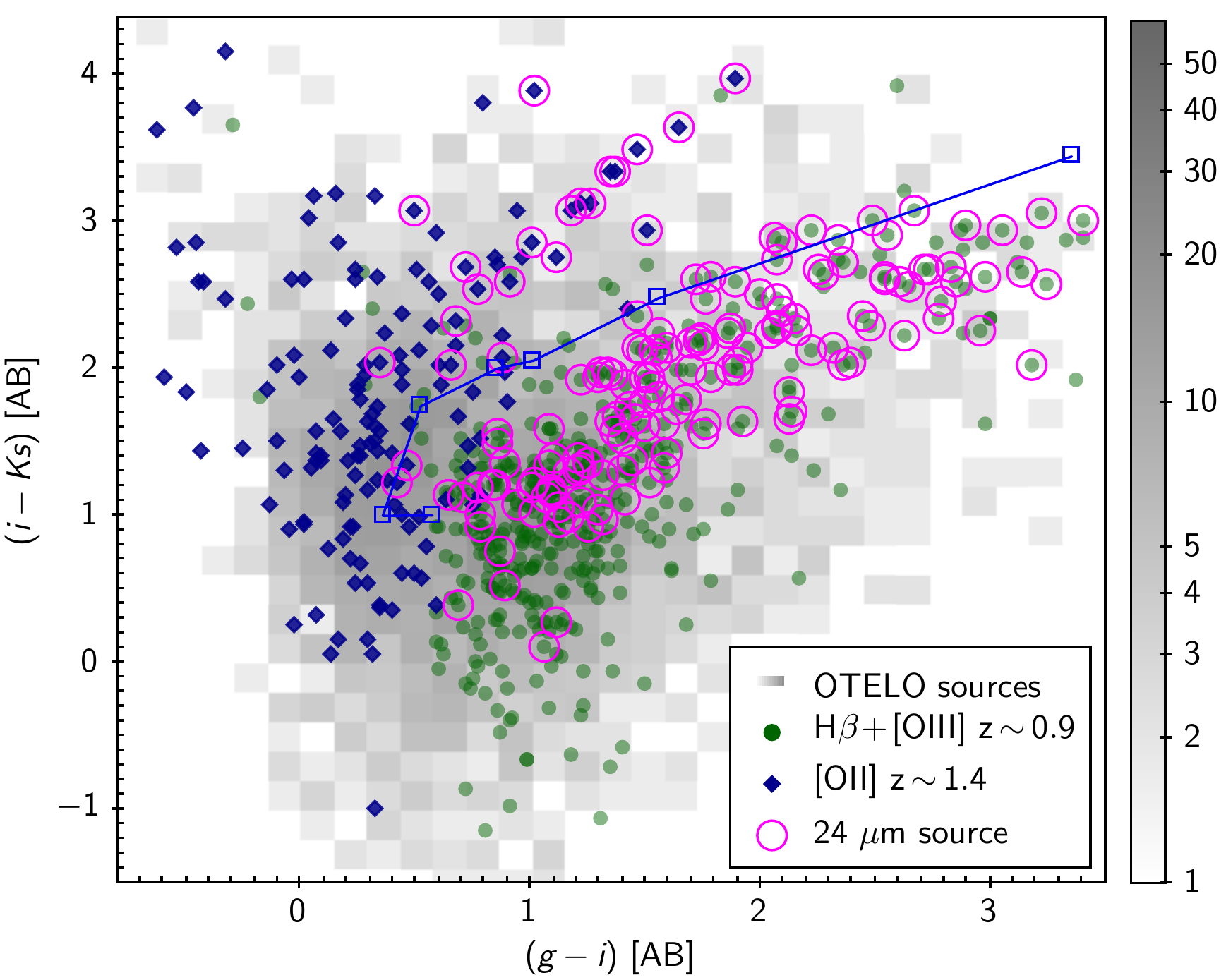}
\caption[Colour-colour diagram 0.9<z<1.4 ELS]{($i-K_S$) versus ($g-i$) colour-colour diagram showing the
\hb+\oiii\ (filled circles) and \oii\ (filled diamonds) emission-line candidates at \zred $\sim$0.9
and 1.4, respectively, selected as explained in the text. The grey background grid represents the OTELO sources
and the open circles represent those sources detected with \mips\ 24$\,\mu m$. As a reference, the blue line
represents a colour sequence of selected galaxy templates,
described in Section \ref{sec:photoz}, from E/S0 type to SB1 (open squares) at redshift \zred\ = 1.15
and E($B$-$V$)=0. The reddest segment of this sequence is concurrent with the intrinsic attenuation axis.}
\label{elg_color_giKs}
\end{figure}

Figure \ref{elg_color_giKs} represents the position ($i-K_S$) versus ($g-i$) colour--colour diagram of 
the \hb+\oiii\ and  \oii\
ELS candidates at \zred $\sim$0.9 and 1.4, respectively. Each subsample occupies well-defined regions with
a small mutual contamination (5\% of \hb+\oiii\ candidates are inside the region of the \oii\ ones
and vice versa, 9\%). As expected, an increasing number of detections in the \mips\ 24$\,\mu m$ band or IR-emitting
galaxies accompanies the intrinsic reddening. 

\begin{figure}[htb]
\centering
\includegraphics[angle=0,width=\linewidth]{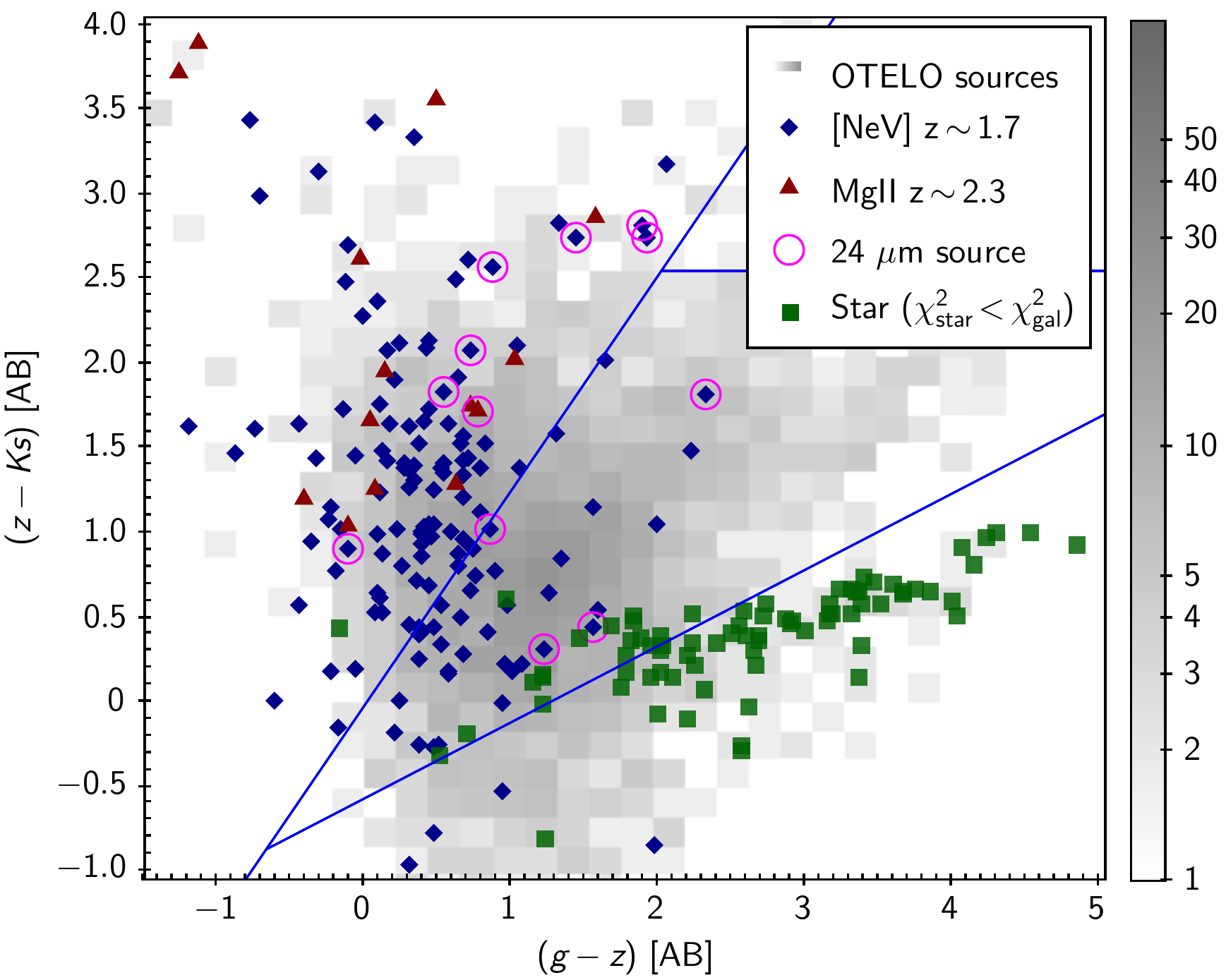}
\caption[Colour-colour diagram z>1.4 ELS (BzK type)]{($z-K_S$) versus ($g-z$) colour--colour diagram showing the
\nev\ (filled diamonds) and \mgii\ (filled triangles) emission-line candidates selected as explained in the text. 
This diagram is based on the adaptation \citep{arcila13} for \cfhtls\ and \wirds\ data of the {\it BzK} 
technique \citep{daddi04} used to select star-forming galaxies at 1.4 $<$ z $<$ 2.5. According to this,
the left side of the diagram would be occupied by the latter, and the central triangular region by passively evolved
galaxies at \zred\ $<$ 1.4. The colour cut between these regions matches with the reddening axis. The lowest region 
of the diagram would be populated by stars. The \otelo\ star candidates, obtained as described in 
Section \ref{sec:stargalaxy}, are symbolized by filled squares. The grey background grid represents the \otelo\ 
sources and the open circles represent those sources detected with \mips\ 24$\,\mu m$.
}
\label{elg_color_gzKs}
\end{figure}

According to Table \ref{otelo_emitters_preliminary} nearly 240 ELS candidates, distributed between
\nev\ and \mgii\ hosts, can be found at $\sim$1.5 $<$ \zred\ $<$ 2.5. In this case, a colour diagram
like the {\it BzK} plot \citep{daddi04} would be appropriate to confirm the selection and classification 
of star-forming galaxies, passive-evolved ones at \zred\ $<$ 1.4, and stars, whilst old galaxies at \zred\ 
$>$ 1.4 would occupy the reddest corner of the plot. Figure \ref{elg_color_gzKs}
show a {\it BzK}-equivalent diagram but using the \bandg\bandz\bandk\ filter system, following the offsets and
colour cuts established by \citep{arcila13} for \cfhtls\ and \wirds\ data. 

All \mgii\ ELS  and 77\% of  \nev\ candidates fall inside the region corresponding to star-forming
galaxies (s-\bandg\bandz\bandk) at \zred\ $>$ 1.4, and 85\% of the sources classified as 
{\it bona fide} stars in Section \ref{sec:stargalaxy} occupy the expected region in this diagram. 

The \nev\ are distributed in a colour sequence running along the colour cut between
the s-\bandg\bandz\bandk\ and passive-evolved galaxies at \zred\ $<$ 1.4. As other high ionization
lines, \nev\ is produced in the narrow-line region (NLR) of AGN \citep{vignali14}. It follows that
\otelo's \nev\ ELS population would mainly correspond to narrow-line AGN having Seyfert-like optical 
spectra with a star-forming component as found at \zred\ $<$ 1.2 by \cite{mignoli13} in the zCOSMOS-Bright 
Survey, or by \cite{vergani17} in the VIMOS Public Extragalactic Redshift Survey (VIPERS). In the latter 
case, the authors state that \nev-emitting galaxies 
populate the blue cloud, the green valley or the red sequence, depending on the stellar mass and the age of 
the stellar populations of the host galaxy. This fact could explain the distribution of \nev\ ELS candidates in 
Figure \ref{elg_color_gzKs}, which in turn may be understood by the existence of a causal relation between AGN 
activity and the mechanisms that regulate  star formation phenomena. All these contributions will provide a 
robust reference population for further evolutionary analysis of the star formation-AGN connection, using
the confirmed \nev\ emitters from \otelo.

\begin{figure}[htb]
\centering
\includegraphics[angle=0,width=\linewidth]{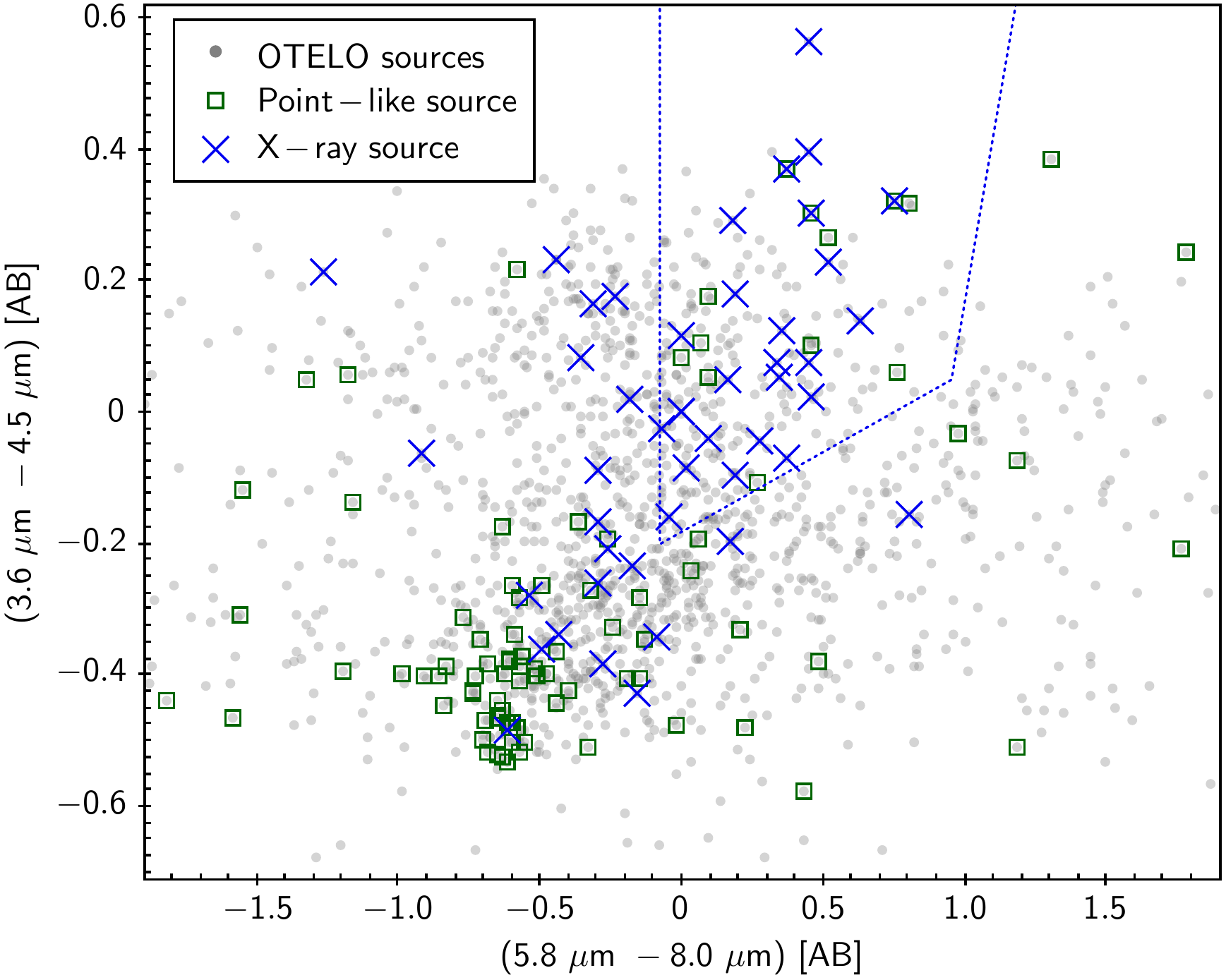}
\caption[MIR colour-colour diagram]{\irac\ data colour distribution of the final \otelo\
catalogue. Open green squares represent sources classified as point-like sources as explained in Section
\ref{sec:stargalaxy},
whilst blue crosses are the X-ray sources from the \cite{povic09} catalogue. The dotted line bounds the
region of MIR-selected AGN from \cite{stern05}.
}
\label{irac_colors}
\end{figure}

Going to redder broad bands in the multi-wavelength catalogue and using X-ray data, Figure \ref{irac_colors} 
shows remarkable \otelo\ sources in the \irac\ colour--colour plot. As expected, most of the point-like sources 
detected in the four MIR bands (96 from the total 191: Section \ref{sec:stargalaxy}) are grouped around the 
colour (3.6$\,\mu$m - 4.5$\,\mu$m) = -0.47, which corresponds to the zero-point of the Vega system. From 44 
sources detected in both X-ray and MIR bands, 24 are within the AGN selection polygon from \cite{stern05}, and 
11 of the remaining ones are close to its limits. Such sources could also qualify as AGN host galaxies. This 
selection box include 12 point-like sources that could be considered as clear QSO candidates.
 
\subsection{Galaxy number counts}
\label{sec:counts}

The contribution to the extragalactic background light at a given wavelength over large look-back time ranges
may be estimated from the galaxy number counts. Starting from early works of \cite{tinsley77} or \cite{tyson88}, up 
to the galaxy count statistics obtained for domains as dissimilar as X-ray \citep{harrison16} and millimetre 
\citep{aravena16}, the surface density of galaxies as a function of the apparent flux provides insights not only
into plausible scenarios that explain their evolution, but even as primary cosmological probe. In the
case of \otelo, galaxy number counts may be useful as a basic sensitivity test when compared
with similar surveys in depth and observed wavelength.

\begin{figure}[htb]
\centering
\includegraphics[angle=-90,width=\linewidth]{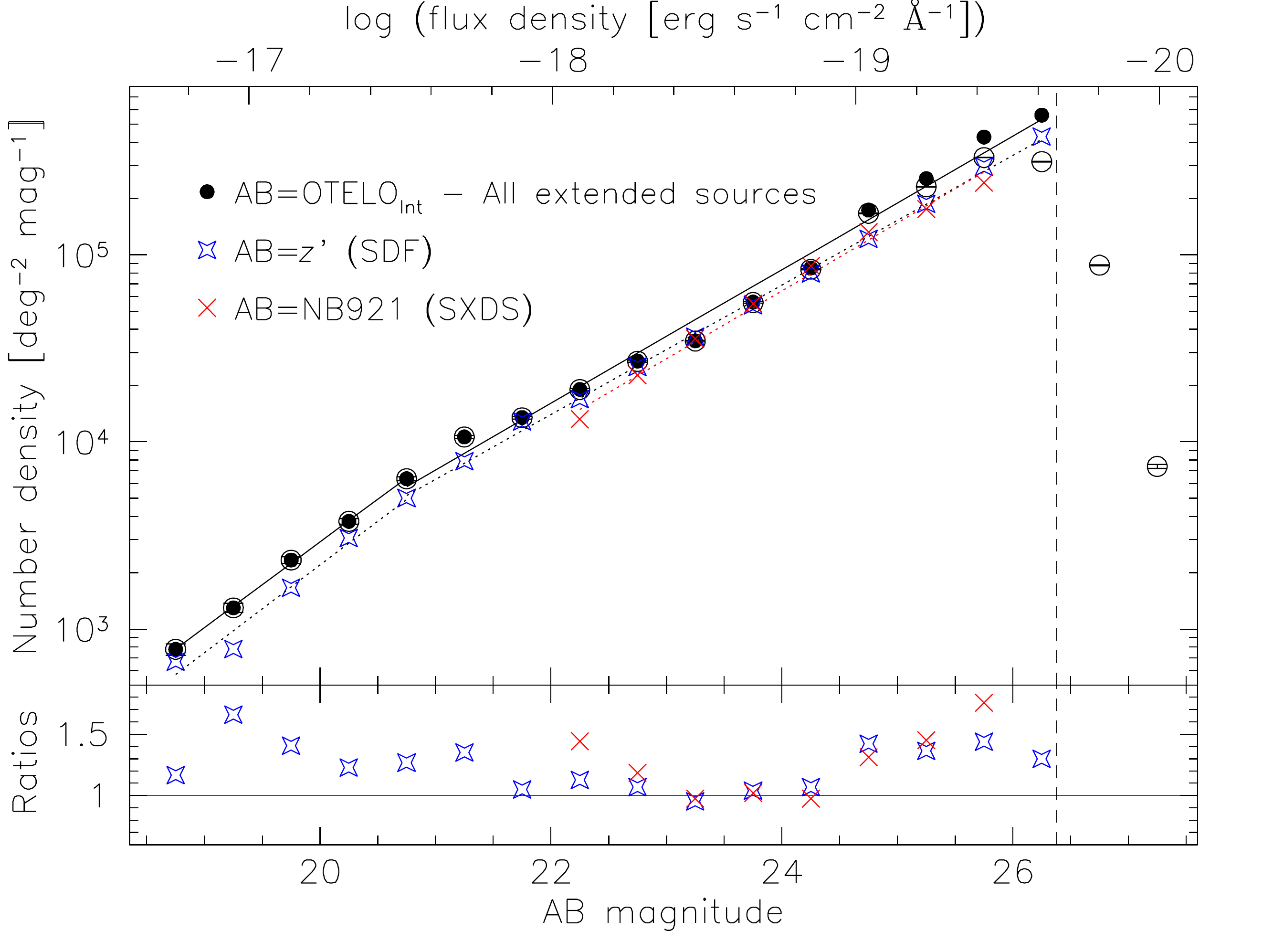}
\caption[Number counts OTELO-Deep]{Number counts of the \otelo\ raw extended sources
observed (open symbols) and corrected by completeness (filled symbols) according to the 
results presented in Section
\ref{sec:catalogue}. Error bars are confined to Poisson statistics and the dashed
line mark the 50\% completeness limiting flux, which was adopted as the significance limit for
the completeness correction. In comparison, blue stars and red crosses
mark the faintest total number counts (corrected) reported by \cite{kashikawa04} (\bandz'-band; SDF) and
\cite{ouchi10} (NB-921; SXDS), respectively.
}
\label{counts}
\end{figure}

Excluding stars, observed and completeness-corrected (Section \ref{sec:completeness}) number density counts for 
the raw extended source set (10839 entries of the \otelo\ multi-wavelength catalogue) is shown in 
Figure \ref{counts} (top panel). Instead of including galaxy count predictions from analytic or numerical models for 
comparison purposes, the completeness-corrected number counts from Subaru Deep Field (SDF) in the 
\bandz-band \citep{kashikawa04}, and from Subaru/XMM-Newton Deep Survey (SXDS) in NB921 narrow-band 
\citep{ouchi10}, are also represented. The number count statistics from \otelo\ are not corrected by cosmic variance 
effects (which should be noticeable mainly on the bright side of the number density distribution), or by possible 
false-positive contaminants (which increase markedly for fainter magnitudes than AB $\sim$ 25, as shown in 
Figure \ref{NBcompleteness}). On the other hand, a cosmic variance correction is assumed for the differential 
counts from both SDF and SXDS data, regarding the large area surveyed in contrast to \otelo. 
Under these hypotheses, the \otelo\ to Subaru-related number count ratios are shown in Fig. \ref{counts} (bottom panel).
On the bright side (AB $<$ 21.5), the number counts of \otelo\ show a mean excess of about 35\% with respect to 
SDF statistics that would be attributed to cosmic variance. Up to 98\% of this galaxy subsample is distributed 
in redshift below \zp=1. Following \cite{somerville04}, we consistently obtained an upper limit of the 
relative cosmic variance $\sigma_{\rm v} = b\, \sigma_{\rm DM} \simeq$ 0.38, with a bias $b \simeq$ 1.25 for a 
comoving number density of 0.0033 Mpc$^{-3}$, and a $\sigma_{\rm DM} \simeq$ 0.30 for the sampled comoving volume 
up to this redshift. On the faint end of the number count distribution (AB $>$ 24.5), a mean excess between 32\% 
and 37\% with respect to SDF and SXDS data, respectively, can be appreciated in Figure \ref{counts}. From testing 
different approaches for possible false-positive source rejection, such apparent excess in the number counts with 
respect to other surveys tends to dissapear. As evident, the Poissonian uncertainties assigned to the number counts 
of this survey (Fig. \ref{counts}, top) are certainly smaller than the differences which explanations are tried above. 

Depending on the magnitude range, the behaviour of all these NIR galaxy counts can be fitted 
by power-laws whose slopes are sensitive to varied effects that include those cited above. In the case of \otelo\ 
and SDF, the bright and faint ends of the galaxy count trends are well approximated by power-law pairs with an 
elbow at around AB = 21, which is consistent with previous studies in this spectral regime 
\citep[see for example][]{gardner93,prieto15,stefanon17}. The AB magnitude upper limit fit corresponds to $\sim$ 50\% 
completeness in all the sets represented. Number count slopes $\gamma$ from best-fitting in each case are 
given in Table \ref{counts_table}.

\begin{table}[ht]
\vspace*{2mm}
\caption[Number counts fits]{
Number count slopes from
data fitting in Figure \ref{counts}.
}
\vspace*{-5mm}
\label{counts_table}
\centering   \small
\begin{center}\begin{tabular}{c c c}
\hline   \\
Survey/band & Magnitude & Slope ($\gamma$) \\
  & range [AB] &  \\
\hline      \\[1pt]
\oteloint\ & 18.0-21.0 & 0.458 $\pm$ 0.008 \\
-- & 21.0-26.4 & 0.357 $\pm$ 0.012 \vspace{2 mm} \\
SDF / $z'$ & 18.0-21.0 & 0.468 $\pm$ 0.046 \\
SDF / $z'$ & 21.0-26.5 & 0.346 $\pm$ 0.004 \vspace{2 mm} \\
SXDS / NB921 & 22.0-26.0 & 0.363 $\pm$ 0.013 \\
\hline
\end{tabular} \end{center}
\end{table}
\normalsize

\section{Discussion and conclusions}
\label{sec:conclusions}

The \otelo\ survey is the deepest emission--line survey to date, being unique in terms of minimum detectable 
flux and emission--line equivalent width. \otelo\ demonstrates the power of narrow-band imaging by using low
spectral resolution tunable filters as implemented in the OSIRIS instrument at the 10m GTC.

The first pointing of the survey was  a 7.5\arcmin$\times\,$7.4\arcmin\ ($\sim$0.015 square degrees) area in 
the EGS field. We scanned the spectral window from 9070 to 9280 \AA, free of strong sky emission lines, with
a total exposure time of 108 h. The first data product of the survey consists of 36 TF slices, according 
to the central wavelength, composed in turn of six  narrow-band images of 12 \AA\ bandwidth, with an overlap 
between contiguous images of 6 \AA.

In this paper we present the final \otelo\ catalogue. \otelo\ sources have been obtained after using {\tt SExtractor} 
on a Deep image built by combining all science frames. Specific algorithms have been developed to subtract sky 
rings, ghosts, and spurious artefacts from the RTF images. Astrometric calibration has been done carefully in order
to align the individual science frames, resulting in an internal RMS better than 0.03\arcsec. The flux calibration 
has been obtained using two spectroscopic standard stars within the \otelo\ field.

From the 11237 raw entries, the \otelo\ catalogue contains 9862 objects at a 50\% completeness AB-magnitude of 26.38,
as measured in the \otelodeep\ image. The catalogue is complemented with ancillary catalogues coming from X-rays, UV,
optical, mid- and far infrared. It is also cross-matched with \cfhtls\ and \deep\ surveys to obtain either 
spectroscopic or complementary photo-\zred\ information. 

Our photometric redshifts have been determined through the \lephare\ code, using libraries for normal and starburst 
galaxies, Seyferts, QSOs, and  stars, including T brown dwarfs. The accuracy of the photometric   
compared to spectroscopic redshifts is better than $\vert\Delta$z$\vert$/(1+z) $\leq$ 0.2. Therefore, the \otelo\ 
catalogue has 9709 sources with non-null photo-z solutions, and 6600 of them have an uncertainty 
$\delta\,$\zp\ $<$ 0.2 (1+\zp). This is a crucial step, as redshift information is needed to identify emission 
lines in the \otelo\ survey. 

From the total number of sources detected, 81 are stellar candidates, probably with spectral types ranging from F3 
through M7, with the luminosity class of dwarfs or subdwarfs, according to the Besan\c con 
model\footnote{\tt http://model.obs-besancon.fr} of the Galaxy \citep{robin03}. 
On the other hand, the number counts distribution of the \otelo\ extended sources is consistent with similar 
data given in literature whenever the effects of the relative cosmic variance and the fraction of possible 
false-positive contaminants are considered.

From the analysis of the pseudo-spectra, 5322 ELS candidates with line widths smaller than FWHM $\sim$ 60 \AA\ 
are detected. Using a colour-excess technique, a total of 3226 candidates were segregated, from which 1542 were 
already detected using the pseudo-spectra. The total number of raw emitter candidates found is then 7006. It is 
important to note that the colour-excess technique complements the selection using TF pseudo-spectra, but that the 
colour-excess technique alone would miss more than 50\% of the candidates to ELS. Also, another main result is that 
16\% of the objects (1812) from the raw catalogue are ELS candidates directly attributable to the strongest emission 
lines in the optical. This ELS population consists of objects emitting [SII], H$\alpha$+[NII], [OIII], H$\beta$, [OII], \mgii, 
[NeV], and CIII], and [SII], at redshifts from 0.33 to 3.9. Other emission lines can be also identified by using
photometric redshift distributions, such as HeII and \lya. Finally, another 483 objects are candidates to be 
absorption-line systems. These population of emitters and absorbers at different redshifts will be studied in 
forthcoming papers. The first release of the \otelo\ value-added products will be public on the second half of 2019.

\begin{acknowledgements}

This  work  was  supported  by  the  Spanish  Ministry  of  Economy  and
Competitiveness  (MINECO) under  the  grants  
AYA2013\,-\,46724\,-\,P,
AYA2014\,-\,58861\,-\,C3\,-\,1\,-\,P,
AYA2014\,-\,58861\,-\,C3\,-\,2\,-\,P,
AYA2014\,-\,58861\,-\,C3\,-\,3\,-\,P,
AYA2016\,-\,75808\,-\,R,
AYA2016\,-\,75931\,-\,C2\,-\,2\,-\,P,
AYA2017\,-\,88007\,-\,C3\,-\,1\,-\,P and
AYA2017\,-\,88007\,-\,C3\,-\,2\,-\,P.

Based on observations made with the Gran Telescopio Canarias (GTC), installed in the 
Spanish Observatorio del Roque de los Muchachos of the Instituto de Astrof\'isica de 
Canarias, on the island of La Palma.

This study makes use of data from AEGIS, a multi-wavelength sky survey conducted with the
Chandra, GALEX, Hubble, Keck, CFHT, MMT, Subaru, Palomar, Spitzer, VLA, and other telescopes
and supported in part by the NSF, NASA, and the STFC.

Based  on  observations  obtained  with  MegaPrime/MegaCam,  a  joint  project  of  CFHT  and
CEA/IRFU, at the Canada-France-Hawaii Telescope (CFHT), which is operated by the National 
Research Council (NRC) of Canada, the Institut National des Science de l'Univers of the 
Centre National de la Recherche Scientifique (CNRS) of France, and the University of 
Hawaii.  This work is based in part on data products produced at Terapix available at 
the Canadian Astronomy Data Centre as part of the Canada-France-Hawaii Telescope Legacy 
Survey, a collaborative project of the NRC and CNRS. 

Based on observations obtained with WIRCam, a joint project of the CFHT, Taiwan, Korea, Canada,
France, at the Canada-France-Hawaii Telescope (CFHT) which is operated by the National 
Research Council (NRC) of Canada, the Institute National des Sciences de l'Univers of the 
Centre National de la Recherche Scientifique of France, and the University of Hawaii.  
This work is based in part on data products produced at TERAPIX, the WIRDS (WIRcam Deep 
Survey) consortium, and the Canadian Astronomy Data Centre. This research was supported by 
a grant from the Agence Nationale de la Recherche ANR-07-BLAN-0228.

Jos\'e A. de Diego thanks the Instituto de Astrof\'isica de Canarias for it
s support through 
the Programa de Excelencia Severo Ochoa and the Gobierno de Canarias for the Programa de 
Talento Tricontinental grant.

\'A. Bongiovanni thanks the anonymous Referee for her/his
feedback and suggestions, and Terry Mahoney (at the IAC's Scientific Editorial Service)
for his subtantial improvements of the manuscript.
   
\end{acknowledgements}

\bibliographystyle{aa}
\bibliography{otelo1} 

\end{document}